\documentclass[prd,preprintnumbers,onecolumn,floatfix,footinbib,nofootinbib,superscriptaddress]{revtex4}
\usepackage{amsmath,amssymb}
\usepackage{graphicx}
\usepackage{mathtools}
\usepackage{epsfig}
\usepackage{color}
\usepackage{bm}
\usepackage{enumerate}
\newcommand{\trans}[1]{\ensuremath{{\bm #1}_{\rm T}}}

\newcommand{\bma}{\ensuremath{b_{\rm max}}}

\newcommand{\lid}{\ensuremath{{}^6{\rm LiD} \;}}

\newcommand\MSbar{\ensuremath{\overline{\text{MS}}}}

\begin{document}
\title{Limits on TMD Evolution From Semi-Inclusive Deep Inelastic Scattering\\ at Moderate $Q$}

\preprint{YITP-SB-14-01}

\author{C.~A.~Aidala}
\email{caidala@umich.edu}
\affiliation{Department of Physics, \\ University of Michigan, \\ Ann Arbor, MI 48109, USA}
\author{B.~Field}
\email{bryan.field@farmingdale.edu}
\affiliation{Department of Physics, \\ Farmingdale State College, \\ 2350 Broadhollow Road, \\ Farmingdale, NY 11735-1021, USA}
\author{L.~P.~Gamberg}
\email{lpg10@psu.edu}
\affiliation{Science Division, \\ Penn State University-Berks,\\  Reading, PA 19610, USA}
\author{T.~C.~Rogers}
\email{rogers@insti.physics.sunysb.edu}
\affiliation{C.N.~Yang Institute for Theoretical Physics, \\ Stony Brook University, \\
  Stony Brook NY 11794, USA}
\date{May 27, 2014}

\begin{abstract}
In the QCD evolution of transverse momentum dependent 
parton distribution and fragmentation functions, 
the Collins-Soper evolution kernel includes both a perturbative short-distance 
contribution as well as a large-distance nonperturbative, but strongly universal, contribution.  
In the past, global fits, based mainly on larger $Q$ 
Drell-Yan-like processes, have found 
substantial 
contributions from nonperturbative regions 
in the Collins-Soper evolution kernel.  
In this article, we investigate semi-inclusive deep inelastic 
scattering measurements in the region of relatively small $Q$, of the order of a few GeV, 
where sensitivity to nonperturbative 
transverse momentum dependence may 
become more important or even dominate the evolution.
Using recently available deep inelastic scattering 
data from the COMPASS experiment, we  provide estimates of the 
regions of coordinate space that
dominate in TMD processes when  
the hard scale is of the order  of
only a few GeV. We find that distance scales that are
much larger than those commonly 
probed in large $Q$ measurements become important, 
suggesting that the details of nonperturbative effects in 
TMD evolution are especially significant in the region of 
intermediate $Q$.  We highlight the strongly universal nature of the nonperturbative component of
evolution, and its potential to be tightly constrained by fits from a wide variety of observables that 
include both large and moderate $Q$.  
On this basis, we recommend detailed treatments of the nonperturbative component of the 
Collins-Soper evolution kernel for future TMD studies.
\end{abstract}
\maketitle

\section{Introduction and Motivation}
 \label{ref:intromot}

This paper is intended to be part of an ongoing project dedicated to constraining the contributions from intrinsic,
nonperturbative parton transverse momentum in inclusive high-energy hadronic collisions within a consistent 
transverse momentum dependent (TMD) factorization formalism.
Efforts to improve constraints on nonperturbative input in TMD factorization are 
becoming increasingly relevant, both for studies of QCD bound state structure as well as for general 
perturbative QCD (pQCD) calculations of transverse momentum dependence in inclusive high energy interactions 
that extend to very low or zero transverse momentum. (See, for example,~\cite{Boer:2011fh,Balitsky:2011jrs} and references therein.)  

The TMD factorization theorem establishes objects like TMD parton distribution functions (PDFs) and 
TMD fragmentation functions (FFs), which contain nonperturbative information, as universal. 
Constraints on PDFs and FFs, obtained 
from measurements or from nonperturbative calculational techniques, can be combined with perturbative
calculations to produce first principles predictions for future experimental measurements.  
A steady accumulation of further measurements
also improve the accuracy and precision of fits to the nonperturbative contributions, which then become input for increasingly precise 
predictions, which can in turn be tested against still further experiments.  This 
process culminates 
in a repeating cycle of testing and refitting and further 
testing. Successful convergence toward increasingly accurate predictions and more tightly constrained
nonperturbative input is an important test of small-$\alpha_s$ perturbative QCD.  Moreover, it justifies the
interpretation of the nonperturbative TMD PDFs and TMD FFs as descriptions of truly intrinsic bound state 
properties of the colliding hadrons, and relates them to fundamental quark and gluon degrees of freedom.  
In this regard, the situation with TMD functions closely mirrors 
that of standard collinear factorization, wherein the intrinsic nonperturbative collinear properties of the colliding hadrons
are encoded in the collinear PDFs. 

Specifically, a TMD factorization theorem separates a transversely differential cross section into a
perturbatively calculable part and several well-defined universal factors~\cite{collins}.  The 
latter are to be interpreted in terms of hadronic structure; they are objects like TMD PDFs and/or 
TMD FFs.  (We will refer to them collectively as TMDs.)
For example, the TMD factorization 
theorems for semi-inclusive deep inelastic scattering (SIDIS), Drell-Yan scattering (DY), and inclusive $e^+e^-$ 
annihilation into back-to-back hadrons ($e^+e^- \to H_1 + H_2 + X$) are schematically:
\begin{align}
d \sigma_{\text{\tiny SIDIS}} & = \sum_f \mathcal{H}_{f, \text{\tiny SIDIS}}(\alpha_s(\mu),\mu/Q) \otimes F_{f/H_1}(x,k_{1T};\mu,\zeta_1) 
	\otimes D_{H_2/f}(z,k_{2T};\mu,\zeta_2) 
	 & + \; & Y_{\text{\tiny SIDIS}} \, , \label{eq:SIDISschem0}  \\
d \sigma_{\rm \text{\tiny DY}} & = \sum_f \mathcal{H}_{f, \rm DY}(\alpha_s(\mu),\mu/Q) \otimes F_{f/H_1}(x_1,k_{1T};\mu,\zeta_1) \otimes 
F_{\bar{f}/H_2}(x_2,k_{2T};\mu,\zeta_2) 
	 & + \; & \; Y_{\rm \text{\tiny Drell-Yan}} \, , \label{eq:DYschem0} \\
d \sigma_{\text{\tiny $e^+e^-$}} & = \sum_f \mathcal{H}_{f, \text{\tiny $e^+e^-$}}(\alpha_s(\mu),\mu/Q) \otimes D_{H_1/\bar{f}}(z_1,k_{1T};\mu,\zeta_1)
	\otimes D_{H_2/f}(z_2,k_{2T};\mu,\zeta_2)   	 & + \; &  \; Y_{\text{\tiny $e^+e^-$}} \, . \label{eq:epemschem0} 
\end{align} 
The first term in each equation is a generalized product of three factors.  These closely resemble a literal
TMD parton model description, and we will call them the ``TMD terms."  The first factor of each TMD term is a hard part, 
$\mathcal{H}(\alpha_s(\mu),\mu/Q)$, specific to the process, with the other two factors being the universal TMD PDFs 
and/or FFs, $F_{f/H_{1,2}}$ and $D_{H_{1,2}/f}$.  
The kinematical arguments, $Q$, $x_{1,2}$ and $z_{1,2}$ have standard definitions 
which can be found, for example, in Ref.~\cite{collins}, chapter 12.14 for SIDIS, chapter 14.5 for Drell-Yan scattering, and 
chapter 13.2 for $e^+ e^-$ annihilation into back-to-back hadrons.

The TMDs may in general contain a mixture of both perturbative and nonperturbative contributions.  
But regardless of whether or 
not they are predominantly described by perturbative or nonperturbative behavior, they are universal, 
and so may be regarded as being associated with individual specific 
hadrons.
\footnote{
In the case of ``naive time reversal odd''  TMDs, like the Sivers function~\cite{Sivers:1989cc}, universality is predicted to be generalized.  In particular
the Sivers function should appear with an opposite sign in Drell-Yan and
SIDIS.   The Sivers function is a TMD PDF describing the 
probability to find a quark of particular transverse momentum inside a transversely polarized hadron.  Though originally 
thought to vanish at leading power due to time-reversal and parity (TP) invariance~\cite{Collins:1992kk}, it was later shown through explicit 
calculations to be a leading power effect~\cite{Brodsky:2002cx,Brodsky:2002rv}.   In Ref.~\cite{Collins:2002kn}, it was shown that the 
TP invariance argument, in the context of a detailed consideration of TMD factorization and the role of Wilson lines, actually gives a  
leading power Sivers function that flips sign in the Drell-Yan process as compared to SIDIS.}
The last terms, $Y_{\rm SIDIS/Drell-Yan/e^+e^-}$, in Eqs.~(\ref{eq:SIDISschem0})-(\ref{eq:epemschem0}), are corrections 
for the region of large transverse momentum of order $Q$ where a description in terms of factorized TMD functions is no 
longer appropriate. These are called the ``$Y$-terms." Throughout this paper, we will assume that 
we are working with cross sections that are 
unpolarized and integrated over azimuthal angles.

The individual factors in Eqs.~(\ref{eq:SIDISschem0})-(\ref{eq:epemschem0}) contain dependence on auxiliary parameters
$\mu$, $\zeta_1$ and $\zeta_2$, 
though $\zeta_1$ and $\zeta_2$ are not independent and are related to 
the physical hard scale $Q$ via $\sqrt{\zeta_1 \zeta_2} = Q^2$.  In full QCD, the auxiliary parameters are exactly 
arbitrary, though to optimize the convergence properties of perturbatively calculable parts, a choice of $\mu \sim \sqrt{\zeta_1} \sim \sqrt{\zeta_2} \sim Q$ 
should generally be made.  From here forward
we will assume that the 
auxiliary parameters have already been fixed at order $Q$  
so that we may rewrite Eqs.~(\ref{eq:SIDISschem0})-(\ref{eq:epemschem0}) in 
the more compact form:
\begin{align}
d \sigma_{\text{\tiny SIDIS}} & = \sum_f \mathcal{H}_{f, \text{\tiny SIDIS}}(\alpha_s(Q)) \otimes F_{f/H_1}(x,k_{1T};Q) 
	\otimes D_{H_2/f}(z,k_{2T};Q) 
	 & + \; & Y_{\text{\tiny SIDIS}} \, ,
	&& \qquad {\rm SIDIS} \label{eq:SIDISschem}  \\
d \sigma_{\rm \text{\tiny DY}} & = \sum_f \mathcal{H}_{f, \rm \text{\tiny DY}}(\alpha_s(Q)) \otimes F_{f/H_1}(x_1,k_{1T};Q) \otimes 
F_{\bar{f}/H_2}(x_2,k_{2T};Q) 
	 & + \; & \; Y_{\rm \text{\tiny Drell-Yan}} \, ,
	&& \qquad {\rm Drell-Yan} \label{eq:DYschem} \\
d \sigma_{\text{\tiny $e^+e^-$}} & = \sum_f \mathcal{H}_{f, \text{\tiny $e^+e^-$}}(\alpha_s(Q)) \otimes D_{H_1/\bar{f}}(z_1,k_{1T};Q)
	\otimes D_{H_2/f}(z_2,k_{2T};Q)   	 & + \; &  \; Y_{\text{\tiny $e^+e^-$}} \, .
	&& \qquad {\text{\tiny ${\rm e^+e^- \to H_1 + H_2 + X}$}} \label{eq:epemschem} 
\end{align} 
A principal goal of the TMD factorization theorem is to unify the description of 
all TMD-factorizable processes like Eqs.~(\ref{eq:SIDISschem})-(\ref{eq:epemschem}) (and potentially others), across all scales 
where perturbation theory is valid, and including the treatment of 
nonperturbative input.
Thus, for instance, constraints on $F_{f/H_1}(x_1,k_{1T};Q)$ obtained from measurements using Eq.~\eqref{eq:DYschem} may be reused 
in Eq.~\eqref{eq:SIDISschem} to constrain $D_{H_2/f}(z,k_{2T};Q)$. Likewise, constraints on $D_{H_2/f}(z,k_{2T};Q)$ 
obtained from Eq.~\eqref{eq:epemschem} may be used in Eq.~\eqref{eq:SIDISschem} to constrain $F_{f/H_1}(x_1,k_{1T};Q)$. 
The internal consistency of such measurements tests the 
TMD factorization theorem and its associated universality properties and, moreover, validates  
the interpretation of TMD FFs and TMD PDFs as objects that can be consistently 
associated with intrinsic hadronic structure.  
Similarly, an observed dependence on the species of hadrons $H_1$ and $H_2$ reveals information about the 
intrinsic structure of the specific hadrons in terms of their elementary quark and gluon degrees of freedom.

We will work within the recent TMD-factorization theorem of Collins~\cite{collins}, especially chapters 10, 13 and 14,
which applies at least to the classic electroweak processes in Eqs.~(\ref{eq:SIDISschem})-(\ref{eq:epemschem}). 
This formalism is similar to, and originates in, the earlier Collins-Soper-Sterman (CSS) formalism of Refs.~\cite{Collins:1981uk,Collins:1981uw,Collins:1984kg}. 
The CSS formalism has been applied in particular to the construction of numerical calculations of transverse momentum 
distributions of Drell-Yan pairs and heavy electroweak vector bosons in high energy hadron-hadron collisions. (See 
Refs~\cite{Ladinsky:1993zn,Balazs:1995nz,Balazs:1997sk,Balazs:1997xd} and other references provided at the website~\cite{resbos}.)
Extensions of the CSS formalism were given for SIDIS in Refs.~\cite{Meng:1995yn,Nadolsky:1999kb,Nadolsky:2000ky}. 
Though very similar in structure and implementation to the CSS formalism, 
there are important differences between the original CSS formalism and the TMD-factorization 
formalism of Ref.~\cite{collins}.  One complication is in how to define TMD PDFs in a way that is 
consistent with factorization formulas like Eqs.~(\ref{eq:SIDISschem})-(\ref{eq:epemschem}), and to identify them 
with the functions that are parameterized in phenomenology.
Reference~\cite{Collins:2003fm} contains a useful overview of some of the problems as they appeared 
approximately a decade ago. The complications outlined in Ref.~\cite{Collins:2003fm} were mostly
resolved in Ref.~\cite{collins}, and it is to this formulation of TMD factorization we are referring  
in Eqs.~(\ref{eq:SIDISschem})-(\ref{eq:epemschem}). 
A thorough overview of the differences between 
the standard CSS formalism and the formulation of Ref.~\cite{collins} is beyond 
the scope of the current article, however, and we leave it for future work. 

In addition, there are by now many other CSS-like treatments of TMD PDFs and FFs, 
some with important differences in details.  One notable example is that of Refs.~\cite{Ji:2004wu,Ji:2004xq}.
Also, a TMD formalism starting from soft-collinear effective theory (SCET) considerations was derived in 
Ref.~\cite{GarciaEchevarria:2011rb} and was 
shown to be equivalent, up to details, to the TMD factorization formalism of~\cite{collins} in Refs.~\cite{Collins:2012uy,Echevarria:2012js}.
A thorough comparison of different CSS-like formalisms is also beyond the scope of this article, though we 
expect the general conclusions to be independent of the specific formalism.  
We again leave a comparison of the details of different resummation and/or CSS-like approaches to future work.

For the sake of clarity, we reserve the term ``TMD-factorization" to apply not just
to the individual equations in Eqs.~(\ref{eq:SIDISschem0})-(\ref{eq:epemschem0}) and Eqs.~(\ref{eq:SIDISschem})-(\ref{eq:epemschem}), 
but rather to the full set of equations together with the essential properties of each factor that follow from a factorization derivation, 
including the infrared safety of the hard part and other perturbatively calculable quantities, the universality 
of the separate TMD functions, the evolution equations, and the applicability of an operator product expansion for each separate
TMD function in the limit of small $b_T$ to match to the collinear formalism.  
For the purpose of this article, the most important aspect of the full TMD-factorization formalism of Ref.~\cite{collins} 
is that it is tailored to the treatment of the individual, well-defined operator definitions for the TMDs, 
and it maps directly onto the partonic 
picture displayed in the TMD terms in Eqs.~(\ref{eq:SIDISschem})-(\ref{eq:epemschem}).

In a strict parton model description, 
PDFs and FFs are treated literally as process-independent number densities. In real QCD,
however, they acquire anomalous dependence on the hard scale $Q$.  
In both collinear and TMD factorization, scaling violations can be computed from evolution equations. 
Moreover, the behavior of TMD functions under evolution is closely related to the details of the derivation 
of factorization, provided that a factorization theorem exists. 
Evolution ultimately 
culminates in the $Q$ dependence of the individual universal factors 
in Eqs.~(\ref{eq:SIDISschem})-(\ref{eq:epemschem}).
Calculating and
observing scaling violations is, therefore, important for testing QCD factorization in 
both the collinear and TMD cases. 

In the collinear case, evolution is described by 
the well-known Dokshitzer-Gribov-Lipatov-Alterelli-Parisi (DGLAP) equations~\cite{Gribov:1972ri,Altarelli:1977zs,Dokshitzer:1977sg}. 
The TMD factorization formalism provides a 
different, but analogous, set of evolution equations for the TMD functions  
in Eqs.~(\ref{eq:SIDISschem})-(\ref{eq:epemschem}).  As in the collinear case, the TMD evolution equations emerge from  
steps of factorization.  
(See Ref.~\cite{Collins:2012ss} for a recent overview of the 
relationship between TMD factorization and evolution.)

An increasing number of phenomenological applications of TMD factorization now focus on 
separating and identifying TMDs in experimental measurements. (See, again, Refs.~\cite{Boer:2011fh,Balitsky:2011jrs}.) 
One of the primary goals of such studies is to extract detailed information about the nonperturbative quark/gluon structure of specific hadrons. 
In the treatment of TMD evolution,
it is therefore becoming increasingly important
to incorporate, within the evolution formalism, the interpretation of TMD PDFs and FFs 
as actual descriptions of the quark and gluon structure of specific 
hadrons in formulas like Eqs.~(\ref{eq:SIDISschem})-(\ref{eq:epemschem}); that is, 
to use a complete TMD factorization formulation 
such as that in~\cite{collins}.
For example, $F_{f/H_1}(x_1,{\bm k}_{1T};Q)$, should be regarded as a function specific to hadron $H_1$ and quarks of flavor $f$, 
similar to its collinear counterpart, $f_{f/H_1}(x_1; Q)$.
While PDFs and FFs are universal with respect to the processes under consideration, they do depend on the 
types of hadrons they describe.  
And by observing this type of dependence, one may hope to 
acquire information about the quark-gluon structure of 
specific hadronic bound states.

\begin{figure}
  \centering
  \includegraphics[scale=.4]{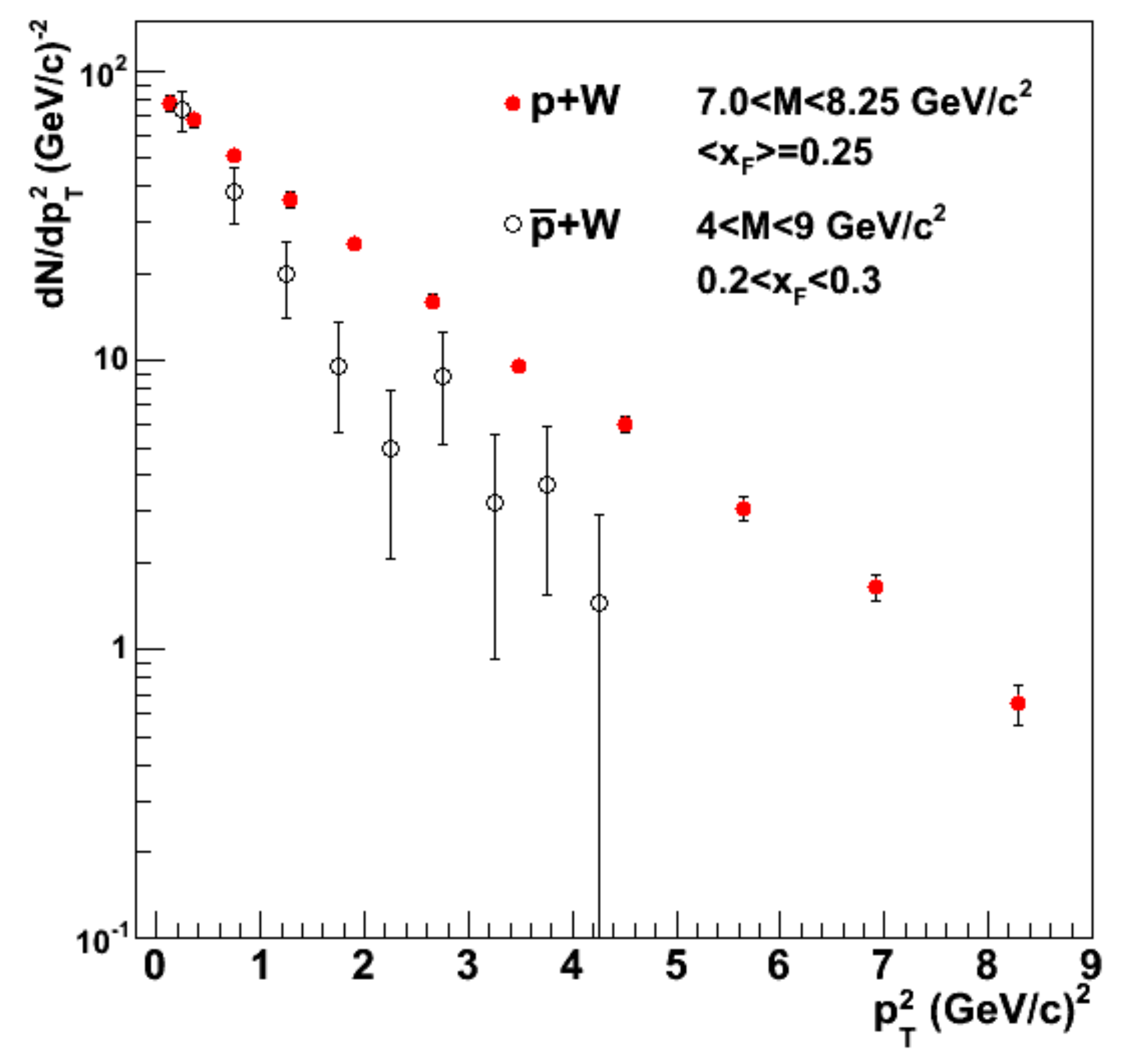}
  \caption[(color online). Drell-Yan transverse momentum distributions in proton-tungstenversus antiproton-tungstenscattering.
  The data are from, respectively and have been normalized to the same value at the lowest $P_T^2$.  
The data from use a beam energy of $400$~GeV (27.4 GeV center-of-mass energy) while the data from have a 
  beam energy of $125$~GeV (15.4 GeV center-of-mass energy). ]{Drell-Yan transverse momentum distributions in proton-tungstenversus antiproton-tungstenscattering.
  The data are from~\cite{Oliver:1978av} and~\cite{Anassontzis:1987hk}, respectively and have been normalized to the same value at the lowest $P_T^2$.  
The data from~\cite{Oliver:1978av}
use a beam energy of $400$~GeV (27.4 GeV center-of-mass energy) while the data from~\cite{Anassontzis:1987hk} have a 
  beam energy of $125$~GeV (15.4 GeV center-of-mass energy). }
  \label{ppbarTungsten}
\end{figure}

These points can be best highlighted with specific examples of the types of questions 
one hopes to address in studies that rely on TMD factorization.  One may consider, for instance, the difference 
between valence and sea quark TMD PDFs. 
Chiral quark 
soliton models~\cite{Reinhardt:1988fz,Diakonov:1987ty,Christov:1995vm,Weigel:2008zz}
suggest that the transverse momentum width of sea
quarks in a proton may be as much as three times broader than that of 
the valance distribution~\cite{Wakamatsu:2008ki,Schweitzer:2012hh,Schweitzer:2012dd}.\footnote{In interpreting the broader sea 
quark distribution in Ref.~\cite{Schweitzer:2012hh}, it is important to note that the functional form of the nonperturbative behavior for sea quarks is non-Gaussian.}
A very direct way to test this would 
be to compare transverse momentum distributions for $pA$ and $\bar{p}A$ Drell-Yan collisions in experiments 
done with exactly the same kinematics, where $A$ is 
a nucleus target.  In the $pA$ case, the 
(quark-in-proton) $\times$ (antiquark-in-A) TMD PDF combination 
appears in the factorization theorem in Eq.~\eqref{eq:DYschem} whereas in the $\bar{p}A$ case it is the 
(antiquark-in-antiproton) $\times$ (quark-in-A) 
TMD PDF combination that enters.\footnote{Of course, the (antiquark-in-proton) $\times$ (quark-in-A) TMD combination also appears 
in the $pA$ case, but this tends to be suppressed in the kinematics of fixed-target nuclear experiments.}  
As such, the difference between transverse momentum distributions concerns the difference between sea quark versus 
valence quark TMD PDFs.
Ideally, to see an effect, a 
comparison should be done with both experiments performed at exactly the same values of $x_1$, $x_2$ and $Q$ to 
avoid 
mixing dependence on the species of the TMD PDF with variations in kinematics. 
At present, the closest we can find to such a 
comparison in existing data is for proton-Tungsten~\cite{Oliver:1978av} and 
antiproton-Tungsten~\cite{Anassontzis:1987hk} production of Drell-Yan pairs with overlapping bins in $Q$ and $x_F$.  
The transverse momentum distributions, normalized to the same values in the lowest $P_T^2$ bins, are shown in 
Fig.~\ref{ppbarTungsten}.  At first sight, the trend appears to be consistent with the behavior described 
in~\cite{Schweitzer:2012dd}.  
However, the range of $Q$ for the proton-tungstendata 
is cut off at a significantly larger value of $Q$ ($7$~GeV) as compared with the antiproton-tungstendata ($4$~GeV).  
Also, most of the data in Fig.~\ref{ppbarTungsten} are at rather large $P_T$, while the TMD terms 
only account for the region of a few GeV.  Therefore, the $Y$ term is certainly needed to reliably 
establish any difference in the intrinsic, nonperturbative $P_T$ dependence.
Furthermore, the proton-Tungsten
data are at $1.8$ times the center-of-mass energy of the antiproton-tungstendata;  the proton-Tungsten
data have a beam energy of $400$~GeV (27.4 GeV center-of-mass energy) whereas the antiproton-tungstendata have a beam energy of only
$125$~GeV (15.4 GeV center-of-mass energy).   
Since higher energies also lead to a broadening of the distribution, it is not possible 
yet to determine whether the trend actually indicates a genuine difference between intrinsic sea and valence distributions or if it is
merely an artifact of kinematics or effects from large $P_T$.  
In addition, the recent study of SIDIS data from Ref.~\cite{Airapetian:2012ki} in Ref.~\cite{Signori:2013mda} does find sensitivity
to the sea quark versus valence quark distributions using a collection of Gaussian fits for the low $P_T$ TMD PDF, 
but finds a slightly broader distribution for sea quark TMD PDFs only when fitting data with $Q^2 > 1.4$~GeV$^2$ -- 
see Fig.~5(a) of Ref.~\cite{Signori:2013mda}. Making a slightly larger $Q^2$ cut of $Q^2 > 1.6$~GeV$^2$ for the fit, 
the trend appears in the opposite direction -- see Fig.~6(a) of Ref.~\cite{Signori:2013mda}.
Reference~\cite{Signori:2013mda} also finds important sensitivity to quark flavor.  
Stronger constraints on $x$, $z$ and $Q$ dependence of the separate (anti)proton and tungstenTMD PDFs
are needed to establish consistency across different experiments and with nonperturbative theoretical 
approaches.

Our hope is that future improved constraints on 
nonperturbative input, along with more data in similar experiments, will help to clarify
the possible difference between valence quark versus sea quark transverse distributions in global fits. 
The purpose of the present paper is not to attempt to address or answer this and 
similar nonperturbative questions; we mention
it here only to motivate the need to formulate TMD factorization as in Eqs.~(\ref{eq:SIDISschem})-(\ref{eq:epemschem}) by  
including, for instance, the dependence of the TMD functions like $F_{f/H_1}(x_1,{\bm k}_{1T};Q)$ on hadron species.

In order to converge toward the types of TMD factorization theorem analyses discussed 
in the previous paragraphs, it will 
be necessary to incorporate all the important elements of the TMD factorization theorem, including 
predictions about the nonperturbative, large-distance behavior at large $b_T$.
A unique aspect of TMD evolution is that the kernel for the evolution itself 
becomes nonperturbative in the region of large transverse distances.  
However,
one of the important predictions of the TMD factorization
theorem in~\cite{collins}, and a central component to the analysis of TMD evolution, 
is that the nonperturbative contribution to evolution is totally universal, not only with respect to different 
processes, but also with respect to the species of hadrons involved 
and kinematical variables like $x$ and $z$.
Furthermore, the soft evolution is independent of whether the TMDs are PDFs or FFs, and is independent of 
whether the hadrons and/or partons 
are polarized.\footnote{The only dependence is on whether the target partons 
are quarks or gluons.} Therefore, a parametrization of the nonperturbative
 evolution from the $Q$ dependence in one observable 
strongly constrains the evolution of many other observables 
across a wide and diverse variety of different kinds of experiments 
and for both TMD PDFs and TMD FFs. This strong form 
of universality is, therefore, an important 
basic test of the TMD factorization theorem.  
It is related to the soft factors -- the vacuum expectation 
values of Wilson loops -- that are needed in the TMD definitions for
consistent factorization with a minimal number of arbitrary cutoffs.   
As such, constraining the nonperturbative component of the 
evolution probes fundamental aspects of soft QCD. 

In practical fitting, this means that the $Q$ dependence for small transverse momentum must be fitted in some set 
of processes before it is used in other calculations. In principle, this 
would ideally be done for fixed 
values of kinematical parameters like $x$ and $z$ and for fixed species of external particles to avoid confounding
dependence on $Q$ with dependence on these other variables.  In studies of hadron structure, it becomes especially 
critical to keep the species of hadron fixed in the extraction of nonperturbative evolution since the difference 
in structure of separate hadrons is often the main objective.
An example is the comparison of proton-tungstenand 
antiproton-tungstendata sets shown in Fig.~\ref{ppbarTungsten} and discussed above.  

Data that allow for this ideal ``apples-to-apples" comparison as a way to extract the nonperturbative 
evolution are sparse, and data for $Q$ are usually  
correlated with $x$ and $z$, making these studies difficult to implement in practice.  
Moreover, ideal  implementations involve global fitting with a large quantity of 
diverse data making the fitting process complicated and labor intensive.
However, it is important that such studies be performed
if the Collins TMD factorization formalism is to be 
used reliably to probe universal nonperturbative aspects of quark and gluon degrees of freedom within 
a pQCD TMD factorization framework. 

Important work in this direction includes the recent analysis in 
Ref.~\cite{Anselmino:2013lza} which analyzes 
an impressively large collection of unpolarized SIDIS data, including 
data from HERMES~\cite{Airapetian:2012ki,Airapetian:2009jy} and 
COMPASS~\cite{SIDISdata}.  
Moreover, Ref.~\cite{Anselmino:2013lza}  does find some improvement in fitting when 
allowing for $Q$ evolution (see discussion section).  
Very recently, a study in Ref.~\cite{Echevarria:2014xaa} has analyzed a very large quantity of data and determined
that nonperturbative evolution effects consistent with those of Ref.~\cite{Landry:2002ix,Konychev:2005iy,Nadolsky:1999kb,Nadolsky:2000ky} 
are needed to explain the Sivers effect.
In Ref.~\cite{Anselmino:2013lza}, 
an example of an apples-to-apples comparison of the type described above would be the
extraction of the $x$-dependence of the TMD PDF for a quark or anti-quark in a \lid target
from the red curves in the last row of plots in Fig.~10, such that $Q$ and $z$ are approximately 
fixed and $x$ varies from $9.90 \times 10^{-3}$ to $4.42 \times 10^{-2}$. 
Likewise, an extraction of $Q$ dependence that avoids confounding 
TMD-evolution effects with $x$, $z$ and hadron species dependence, would be
a fit of the red curves in the last column of Fig.~10 of Ref.~\cite{Anselmino:2013lza} (although the two bins are immediately adjacent 
so that the lever arm for determining $Q$ dependence is rather small).  

We will perform such an analysis in 
Sec.~\ref{sec:numerics} of this article.
Studies of SIDIS data are especially valuable 
for isolating $x$ and $z$ dependence and flavor dependence - a strategy also used 
in Ref.~\cite{Signori:2013mda} which does find important flavor dependence.  
Drell-Yan and heavy boson production measurements 
with fixed energy and targets, but widely varying $Q$, are ideal for extracting nonperturbative 
components of TMD-evolution.  
Also, $e^+ e^-$-annihilation experiments such as Belle~\cite{Seidl:2008xc} will 
also be useful for constraining the $Q$ dependence in a true apples-to-apples analysis - see the analysis  
in Ref.~\cite{Boer:2008fr}.

Detailed global fits in the past, within the CSS formalism, have also found 
substantial effects from the nonperturbative
component of the Collins-Soper (CS) evolution kernel~\cite{Landry:2002ix,Konychev:2005iy,Nadolsky:1999kb,Nadolsky:2000ky}, 
even at relatively large $Q$ ($Q \gtrsim 10$~GeV). 
See, especially, the recent discussion in Ref.~\cite{Guzzi:2013aja}.  
Even at large $Q$,  measurements of standard model 
parameters have been found to be sensitive to the nonperturbative component of evolution.  
For example, nonperturbative effects 
have been known for some time to be a 
relevant issue in
measurements of the mass of the $W$ boson~\cite{Nadolsky:2004vt}, and 
recent measurements of the $W$ boson mass in Ref.~\cite{LopesdeSa:2013zga} 
find particular sensitivity to the nonperturbative component of the CS evolution kernel.
Very recently, Ref.~\cite{Echevarria:2014xaa} has found that a nonperturbative component 
to evolution is important for describing the Sivers effect.

By contrast, there 
are recent 
claims~\cite{Echevarria:2012pw,Sun:2013dya,Sun:2013hua}, following 
alternative formalisms, that a nonperturbative component to evolution is 
fundamentally unnecessary for studying evolution, even in regions of $Q$ as low as $\sim 2.0$~GeV and 
at almost zero transverse momentum. 

The current phenomenological situation is further complicated by the observation 
that parametrizations obtained by extrapolating large $Q$ fits to 
small $Q$ implies suspiciously rapid evolution in the region of a few GeV, a result 
very clearly demonstrated in the 
recent work of Sun and Yuan~\cite{Sun:2013dya,Sun:2013hua} -- 
see especially Fig.~2 of Ref.~\cite{Sun:2013hua}. 

To summarize, with studies of hadronic structure becoming increasingly relevant in implementations of evolution, 
it is important to begin to obtain a more detailed phenomenological understanding 
of the behavior of the nonperturbative contribution to evolution, particularly if results are 
to be understood in terms of a unified pQCD TMD-factorization picture. As we will frequently emphasize in this paper, 
this will require a careful account 
of the dependence on hadron species, dependence on whether the processes involve PDFs or FFs, and 
dependence on kinematic variables like $x$ and $z$. 

Semi-inclusive deep inelastic scattering experiments are particularly suited to an examination of 
the variation in transverse momentum distributions with relatively low $Q$ and
approximately fixed $x$, $z$ and $Q$~\cite{Bacchetta:2008xw}. 
The COMPASS experiment has recently released data~\cite{SIDISdata} for charged hadron SIDIS measurements 
that are differential in all kinematical parameters and cover a range of moderately low values of $Q$.  
We will use this to study the $Q$ dependence in the region of small $Q$ within the same experiment and 
for approximately fixed $x$, $z$.  The species of hadrons in Ref.~\cite{SIDISdata} is not fixed; the target is 
\lid while the 
measured final state particles 
include all positively (negatively)
 charged hadrons.  Furthermore, as recently illustrated in Ref.~\cite{Signori:2013mda}, there can 
be significant flavor dependence in the TMD functions. However, the specific mixtures of hadrons in Ref.~\cite{SIDISdata} are not varied 
within the experiment, 
so for the purpose of extracting universal $Q$ dependence, we will 
take this to be a reasonable 
proxy for fixed hadron species. 
While the range in $Q$ in Ref.~\cite{SIDISdata} is too small to allow for 
reasonably accurate 
fits to the nonperturbative evolution, it is significant enough that we can use it to rule 
out any dramatic variations with $Q$ that might be suggested by direct direct extrapolations from large $Q$ fits, and to 
infer certain general aspects of the very large coordinate space $Q$ dependence.  

At the same time, we find it important to sound notes of caution regarding the extraction of nonperturbative information 
from the TMD factorization formalism in the region of such small $Q$.  The TMD factorization equations in 
Eqs.~(\ref{eq:SIDISschem})-(\ref{eq:epemschem}) are derived by systematically neglecting non-factorizing error terms of 
order $(M/Q)^a$ with $a > 0$ and $M$ a typical hadronic mass scale.  (The error terms are not shown explicitly in 
Refs.~(\ref{eq:SIDISschem})-(\ref{eq:epemschem}).)  As $Q$ begins to approach sizes of order the proton mass, it 
becomes reasonable to question the overall validity of the TMD factorization formalism in this region of kinematics.  
In a similar vein, the construction of the distinct TMD and (process dependent) $Y$ terms in Eq.~(\ref{eq:SIDISschem})-(\ref{eq:epemschem}) relies on 
having widely separate scales in transverse momentum, from $q_T \lesssim M$ to $q_T \sim \mathcal{O}(Q)$.  When $Q$ is
not so much larger than $M$, the construction of distinct  TMD and $Y$-terms becomes questionable.  Moreover, we will continue to 
make the now standard assumption that the $Y$-term can be neglected in studies that focus on intrinsic nonperturbative 
transverse momentum.  Given the complications discussed above with separating a $Y$-term and a 
TMD term at low $Q$, this is an assumption that can and should be questioned in future studies.  Indeed, our observations in Sec.~\ref{sec:pertcons} 
suggest that the $Y$ term becomes important.  In the region of $Q \sim 1.0$~GeV refinements and extensions to the TMD factorization formalism 
may be necessary.

Nevertheless, our general results regarding the importance of the nonperturbative regions of $b_T$ are sufficiently robust
that they provide important guidance for future dedicated global fitting and Monte Carlo efforts that incorporate questions about 
intrinsic nonperturbative structure.  The discussion above of sea versus valence quark distributions is meant to provide one  
typical example of such questions.         

We emphasize that there is  
overlap between this paper and 
those of Sun and Yuan~\cite{Sun:2013dya,Sun:2013hua}, which clearly illustrate a much slower evolution in existing data 
from what might be expected from a direct extrapolation of large $Q$ fits to low $Q$. 
However, 
while both our analysis and that of Sun-Yuan find a very soft rate of evolution in the region of 
moderate $Q$, this observation is 
interpreted within the contexts of different formalisms.  The result is that we arrive at very 
different 
conclusions regarding the nature and relevance of the large-distance nonperturbative region and the relationship to larger $Q$ fits.  
Namely, Sun-Yuan argue that the nonperturbative component of evolution is unnecessary to study of evolution in the 
$Q \sim 2.0$~GeV region whereas we find that detailed knowledge of nonperturbative evolution becomes especially 
important at this order of $Q$.
Since data from lower $Q$ are ideal for constraining the strongly universal nature of the 
nonperturbative evolution at large transverse sizes, we argue that it should be included in future global fits that 
utilize TMD factorization if a unified treatment of TMD factorization is the ultimate goal. 
This paper was also influenced by the recent workshop proceedings~\cite{Collins:2013zsa}.

In Sec.~\ref{sec:evolution}  we review the TMD factorization formalism of Ref.~\cite{collins} as it pertains to 
phenomenological extractions with SIDIS data.  In Secs.~\ref{sec:approxes} and~\ref{sec:numerics} we 
estimate the rate of variation in the $P_T$-width due to 
TMD evolution over the moderate $Q$ and small $P_T$ intervals that are 
observed in the SIDIS data of Ref.~\cite{SIDISdata}.  
In Sec.~\ref{sec:largebT}, we analyze the importance of nonperturbatively large $b_T$ behavior, and 
in Sec.~\ref{sec:pertcons} we interpret the empirically observed low $Q$ dependence of the $P_T$ distribution 
in terms of the perturbative QCD TMD factorization theorem.
In Sec.~\ref{sec:sunyuan} we remark on the difference in the treatment of perturbative QCD 
evolution in the treatment by Collins in Ref.~\cite{collins} and Sun-Yuan in Ref.~\cite{Sun:2013dya,Sun:2013hua}.
We discuss the overall
interpretation and consequences of our observations in Sec.~\ref{sec:discussion}, and 
offer recommendations for a way forward, with an emphasis on testing the universal and intrinsic nature of 
any nonperturbative hadronic structures. 


\section{Brief Review of Evolution Formulas}
\label{sec:evolution}
Here we briefly summarize the basic formulas of TMD factorization.  For a complete derivation, we refer the 
reader to Ref.~\cite{collins}, particularly chapters 10 and 13.

The transversely differential cross section (Eq.~(13.116) of Ref.~\cite{collins}) for SIDIS 
in transverse coordinate space, corresponding to Eq.~\eqref{eq:SIDISschem}, takes the form:
\begin{align}
\frac{d \sigma}{d P_T^2}  & \propto
\mathcal{H}(\alpha_s(Q)) \int d^2\bm{b}_T e^{i\bm{b}_T\cdot\bm{P}_T}
\; \tilde{F}_{H_1}(x,b_T;Q,Q^2) \, \tilde{D}_{H_2}(z,b_T;Q,Q^2)  \; \; + \; \; Y_{\rm SIDIS} \,  \nonumber \\
\nonumber \\
 & \propto {\rm F.T.} \;\; \tilde{F}_{H_1}(x,b_T;Q,Q^2) \, \tilde{D}_{H_2}(z,b_T;Q,Q^2)  \; \; + \; \; Y_{\rm SIDIS} \, . \label{eq:evolution0}
\end{align}
The two-dimensional Fourier transform is needed to convert the transverse coordinate space 
expression into momentum space, and we drop overall factors since, for our purposes, we are only 
interested in the $P_T$-dependence. 
On the second line we introduce the convenient ``${\rm F.T.}$" notation, where the ``${\rm F.T.}$" represents 
the Fourier transform and all factors not related specifically to the $b_T$-dependence.  
We will mainly be working with the coordinate space integrand in the first term in Eq.~\eqref{eq:evolution0}.
We will also make the simplifying assumption of quark 
flavor independence, so we have dropped flavor indices; flavor dependence is easily
restored in later formulae.\footnote{See, however, Ref.~\cite{Bacchetta:2008xw}.} 
The scales in Eq.~\eqref{eq:evolution0} are chosen 
to be $Q$ so that the hard part, $\mathcal{H}(\alpha_s(Q))$, has good convergence properties. 
The kinematical variables $x$, $z$ and $Q$ for SIDIS are defined in the usual way, and correspond to 
those of Ref.~\cite{SIDISdata}.  In our notation, $P$ is the 
four momentum of the produced hadron, $Q^2$ = $-q^2$ where $q$ is the virtual photon momentum, $x = Q^2/(2 P_{H_1} \cdot q)$ where
$P_{H_1}$ is the incoming hadron four-momentum, and $z  = P_{H_1} \cdot P / (P_{H_1} \cdot q)$.  
$P_T$ is the transverse momentum 
of the produced hadron in a frame where 
both the incoming hadron and the virtual photon have zero transverse momentum;
see Fig.~1 of Ref.~\cite{SIDISdata}.

The TMD functions in Eq.~\eqref{eq:SIDISschem0} obey a set of evolution equations which 
we will simply quote here for easy reference.  They are the Collins-Soper (CS) equations for each TMD:
\begin{equation}
\frac{\partial \ln \tilde{F}(x,b_T;\mu,\zeta_1)}{\partial \ln \sqrt{\zeta_1} } = \frac{\partial \ln \tilde{D}(z,b_T;\mu,\zeta_2)}{\partial \ln \sqrt{\zeta_2} } = \tilde{K}(b_T;\mu)\, , \label{eq:CS} 
\end{equation}
and the renormalization group (RG) equations
\begin{align}
\frac{d \tilde{K}(b_T;\mu)}{d \ln \mu}  = & - \gamma_K(\alpha_s(\mu)) \, ,  && \qquad {\rm RG \; CS \; kernel} \label{eq:RGCS} \\
\frac{d \ln \tilde{F}(x,b_T;\mu,\zeta_1)}{d \ln \mu} = & \; \gamma_{\rm PDF}(\alpha_s(\mu);\zeta_1 /\mu^2)\, ,  && \qquad {\rm RG \; TMD \; PDF } \label{eq:RGPDF} \\
\frac{d \ln \tilde{D}(z,b_T;\mu,\zeta_2)}{d \ln \mu} = & \; \gamma_{\rm FF}(\alpha_s(\mu);\zeta_2 /\mu^2)\, .   && \qquad {\rm RG \; TMD \; FF } \label{eq:RGFF}
\end{align}
Again, we refer the reader to Ref.~\cite{collins} for details -- see especially Eqs.~(13.47),~(13.49),~(13.50) and the discussion beginning with Sec.~13.15.4.
The anomalous dimensions $\gamma_K(\alpha_s(\mu))$ and $\gamma_F(\alpha_s(\mu);\zeta_F /\mu^2)$ are perturbatively calculable, and we will keep up to order $\alpha_s$ terms.
The CS kernel, $\tilde{K}(b_T;\mu)$, is also perturbatively calculable as long as $b_T \ll \sim 1 / \Lambda_{\rm QCD}$.

Over short transverse distance scales, $1/b_T$ becomes a legitimate hard scale, and the transverse coordinate  
dependence in the TMD PDFs can itself be calculated in perturbation theory.  With the choice of renormalization 
scale $\mu \sim 1 / b_T$,   
$\alpha_s(\sim 1 / b_T)$ approaches zero for small sizes due to asymptotic freedom, 
thus ensuring that 
the small size transverse coordinate dependence is optimally calculable in perturbation theory.  
For very large $b_T$, the transverse coordinate dependence corresponds to intrinsic nonperturbative behavior 
associated with the hadron wave function. 
(In momentum space, this corresponds to the onset of effects from intrinsic bound state transverse momentum in the hadron wavefunction)  There, a 
prescription is needed to tame the growth of $\alpha_s(1 / b_T)$ and match to a nonperturbative, large distance description of the $b_T$-dependence.  
The renormalization group scale is therefore chosen to be
\begin{equation}
\mu_b \equiv C_1/| {\bm b}_{\ast}(b_T)| \, , \label{eq:mub}
\end{equation}
where ${\bm b}_{\ast}(b)$ is a function of $b_T$ that equals $b_T$ at small $b_T$, but 
freezes in the limit  where $b_T$ becomes nonperturbatively large, i.e., when $b_T$ is larger 
than some fixed $b_{\rm max}$.  This function must obey
\begin{equation}
{\bm b}_{\ast}({\bm b}_T) = 
\begin{dcases}
{\bm b}_T & b_T \ll b_{\rm max} \\
{\bm b}_{\rm max} & b_T \gg b_{\rm max} \, . \label{eq:bdef}
\end{dcases}
\end{equation}
The most common taming prescription is
\begin{equation}
\label{bstar}
{\bm b}_{\ast}(\trans{b}) \equiv \frac{\trans{b}}{\sqrt{1+b_T^2/b_{\rm max}^2}}.
\end{equation}
Although any function obeying Eq.~\eqref{eq:bdef} is consistent with both 
TMD factorization and the standard CSS formalism, 
Eq.~\eqref{bstar} is one of the simplest choices and is the one that we will adopt in this paper.
The factor $C_1$ is an arbitrary numerical constant that can be chosen to minimize higher 
order corrections.  It is typically fixed at $C_1 = 2 e^{-\gamma_{\rm E}}$.

To put Eq.~\eqref{eq:evolution0} into a convenient form for perturbative calculations, we need 
to rewrite each TMD function evolved from the reference scale $\mu_b$ of Eq.~\eqref{eq:mub}.
Following Ref.~\cite{collins} Eq.~(13.70) (along with Eq.~(13.64)) we have for the TMD FF
\begin{align}
\tilde{D}_{H_2}(z,b_T;Q,Q^2) & =  \tilde{D}_{H_2}(z,b_{\ast};\mu_b,\mu_b^2) 
\exp \left\{ -g_2(z,b_T;\bma) - g_K(b_T;\bma) \ln \left( \frac{Q}{Q_0} \right) \right.  \nonumber \\
& \left. + \ln \left( \frac{Q}{\mu_b} \right) \tilde{K}(b_{\ast};\mu_b)  +  \int_{\mu_b}^{Q} \frac{d \mu^\prime}{\mu^\prime} 
\left[ \gamma_{\rm FF}(\alpha_s(\mu^\prime);1) - \ln \left( \frac{Q}{\mu^\prime} \right) \gamma_K(\alpha_s(\mu^\prime)) \right] \right\} \, .
\label{eq:FFatQ}
\end{align}
The mirror expression for the TMD PDF is
\begin{align}
\tilde{F}_{H_1}(x,b_T;Q,Q^2) & =  \tilde{F}_{H_1}(x,b_{\ast};\mu_b,\mu_b^2) 
\exp \left\{ -g_1(x,b_T;\bma) - g_K(b_T;\bma) \ln \left( \frac{Q}{Q_0} \right) \right.  \nonumber \\
& \left. + \ln \left( \frac{Q}{\mu_b} \right) \tilde{K}(b_{\ast};\mu_b)  +  \int_{\mu_b}^{Q} \frac{d \mu^\prime}{\mu^\prime} 
\left[ \gamma_{\rm PDF}(\alpha_s(\mu^\prime);1) - \ln \left( \frac{Q}{\mu^\prime} \right) \gamma_K(\alpha_s(\mu^\prime)) \right] \right\} \, .
\label{eq:PDFatQ}
\end{align}
The functions $\tilde{F}_{H_1}(x,b_T;\mu_b,\mu_b^2) $ and $\tilde{D}_{H_2}(z,b_T;\mu_b,\mu_b^2)$ now 
have optimal perturbative behavior at small $b_T$.  They are calculable, via an operator product expansion, in 
terms of collinear PDFs and FFs and Wilson coefficients with powers of small $\alpha_s(\mu_b)$ and perturbative coefficients 
that are well-behaved in the limit of $Q \gg \Lambda_{\rm QCD}$ (and contain no large logs of $b_T$).
The functions $g_1(x,b_T;\bma)$, $g_2(z,b_T;\bma)$ and $g_K(b_T;\bma)$ correspond to $g_{j/H_A}(x,b_T)$, 
$g_{H_A/f}(z_A,b_T)$, and $g_K(b_T)$ in 
Eqs.~(13.70) and (13.110) of Ref.~\cite{collins}. The definition of $g_K(b_T;\bma)$ is given in Eq.~(13.60) of 
Ref.~\cite{collins} and the definition of $g_2(z,b_T)$ ($g_{H_A/f}(z_A,b_T)$) is given in Eq.~(13.68), and there is  
an exactly similar definition for $g_1(x,b_T;\bma)$ ($g_{j/H_A}(x,b_T)$).  The functions $g_1(x,b_T;\bma)$ and $g_2(z,b_T;\bma)$ are specific to 
the type of hadron and the fragmentation function, respectively.  
The interpretation is that they describe the 
corrections needed to account for the higher orders and 
intrinsic nonperturbative transverse motion 
of the bound state partons 
in the limit of  
large $b_T$.\footnote{In our notation, we have included $\bma$ as an explicit auxiliary parameter 
in $g_1(x,b_T;\bma)$, $g_2(z,b_T;\bma)$ and $g_K(b_T;\bma)$ to emphasize that these functions depend
on the choice of $\bma$.} 

It is important to note that, although $g_K(b_T;\bma)$ is totally universal, 
$g_1(x,b_T;\bma)$ and $g_2(z,b_T;\bma)$ depend in general on the species of 
the incoming and outgoing hadrons respectively, as well as on the fact that one TMD is 
a PDF while the other is an FF, just as in the case of collinear PDFs and FFs.

Let us introduce two further definitions to simplify notation.  
The purpose of the present paper is not to implement a detailed perturbative treatment of 
the small $b_T$-dependence, but rather to investigate the large $b_T$ behavior at relatively small $Q$.
Therefore, let us define,
\begin{equation}
\label{eq:newgpdf}
- g_{\rm PDF} (x,b_T;\bma)  \equiv -g_1(x,b_T;\bma) + \ln \left( \tilde{F}_{H_1}(x,b_{\ast};\mu_b,\mu_b^2)  \right) \, ,
\end{equation}
and 
\begin{equation}
\label{eq:newgff}
- g_{\rm FF} (z,b_T;\bma)  \equiv -g_2(z,b_T;\bma) + \ln \left( \tilde{D}_{H_2}(z,b_{\ast};\mu_b,\mu_b^2) \right) \, .
\end{equation}
Then, Eqs.~(\ref{eq:FFatQ})-(\ref{eq:PDFatQ}) become
\begin{align}
\tilde{D}_{H_2}(z,b_T;Q,Q^2) & =  
\exp \left\{ -g_{\rm FF} (z,b_T;\bma)  - g_K(b_T;\bma) \ln \left( \frac{Q}{Q_0} \right) \right.  \nonumber \\
& \left. + \ln \left( \frac{Q}{\mu_b} \right) \tilde{K}(b_{\ast};\mu_b)  +  \int_{\mu_b}^{Q} \frac{d \mu^\prime}{\mu^\prime} 
\left[ \gamma_{\rm FF}(\alpha_s(\mu^\prime);1) - \ln \left( \frac{Q}{\mu^\prime} \right) \gamma_K(\alpha_s(\mu^\prime)) \right] \right\} \, ,
\label{eq:FFatQ2}
\end{align}
and
\begin{align}
\tilde{F}_{H_1}(x,b_T;Q,Q^2) & =  
\exp \left\{ -g_{\rm PDF} (x,b_T;\bma)  - g_K(b_T;\bma) \ln \left( \frac{Q}{Q_0} \right) \right.  \nonumber \\
& \left. + \ln \left( \frac{Q}{\mu_b} \right) \tilde{K}(b_{\ast};\mu_b)  +  \int_{\mu_b}^{Q} \frac{d \mu^\prime}{\mu^\prime} 
\left[ \gamma_{\rm PDF}(\alpha_s(\mu^\prime);1) - \ln \left( \frac{Q}{\mu^\prime} \right) \gamma_K(\alpha_s(\mu^\prime)) \right] \right\} \, .
\label{eq:PDFatQ2}
\end{align}
Using the TMD PDF and FF of Eqs.~\eqref{eq:FFatQ2} and~\eqref{eq:PDFatQ2} in 
Eq.~\eqref{eq:evolution0} gives the cross section in the compact form:
\begin{align}
\frac{d \sigma}{d P_T^2} \propto  
    {\rm F.T.} \exp \left\{  \vphantom{ \ln \left( \frac{Q}{Q_0}\right)} \right. & \left. 
         - g_{\rm PDF} (x,b_T;\bma) 
         - g_{\rm FF} (z,b_T;\bma) 
         - 2 g_K(b_T;\bma) \ln \left( \frac{Q}{Q_0}\right)  \right.  \nonumber \\
  & + \left. 2 \ln \left( \frac{Q}{\mu_b} \right) \tilde{K}(b_{\ast};\mu_b) 
         +  \int_{\mu_b}^Q \frac{d \mu^\prime}{\mu^\prime} \left[ \gamma_{\rm PDF}(\alpha_s(\mu^\prime);1) 
         + \gamma_{\rm FF}(\alpha_s(\mu^\prime);1)
                  - 2 \ln \left( \frac{Q}{\mu^\prime} \right) \gamma_K(\alpha_s(\mu^\prime)) \right]\right\} \nonumber \\ 
                  & + Y_{\rm SIDIS} \, .  \label{eq:evolution}
\end{align}
The functions $g_{\rm PDF} (x,{\bm b}_T;\bma)$ and $ g_{\rm FF} (z,{\bm b}_T;\bma)$ parametrize the 
intrinsic large $b_T$ behavior associated with the TMD PDF and the TMD fragmentation 
function respectively.  They are independent of $Q$.  In our notation, they also include, via the 
definitions in Eqs.~(\ref{eq:newgpdf})-(\ref{eq:newgff}), the 
matching to the small $b_T$ behavior that is calculable using collinear factorization.
The terms in the exponent on the second line of Eq.~\eqref{eq:evolution}
arise from solving the evolution equations in terms of the perturbatively calculable anomalous 
dimensions, $\gamma_{\rm PDF}$, $\gamma_{\rm FF}$, 
$\gamma_K$ and the perturbative Collins-Soper (CS) evolution kernel $\tilde{K}(b_T;\mu)$;  
the function $g_K(b_T;\bma)$ on the first line is the correction to the CS kernel at large $b_T$ 
which includes nonperturbative effects.
Note that there is an underlying simplicity in TMD evolution in that there is a single universal function $\tilde{K}(b_T;\mu)$ that
governs the evolution of the cross section at small $P_T$, though in Eq.~\eqref{eq:evolution} it has been split into three parts: 
the terms involving $\gamma_K(\alpha_s(\mu^\prime))$, $\tilde{K}(b_{\ast};\mu_b)$, and $g_K(b_T;\bma)$.

It should be noted that there are multiple ways of ultimately expressing solutions to the evolution equations. 
The most convenient choice depends on the goals at hand, and on which particular physical phenomena one wishes 
to probe.  We will use one example in Sec.~\ref{sec:pertcons}.

The TMD terms in Eqs.~(\ref{eq:SIDISschem})-(\ref{eq:epemschem}) are derived using the approximation that $P_T \ll Q$.
For an accurate calculation of the full cross section, a correction term, the $Y$-term, is need for the region $P_T \sim Q$, 
and this is symbolized by the last term in Eq.~\eqref{eq:evolution}.  From here forward, we will neglect the 
$Y$-term contribution and focus only on the TMD term, which remains common practice in phenomenological 
studies done at moderate $Q$. We will remark on how legitimate such an approximation is in 
Sec.~\ref{sec:discussion}.

With the $b_T$-dependence of the perturbatively calculable part 
of Eq.~\eqref{eq:evolution} frozen above a certain $b_{\rm max}$, 
the remaining 
evolution is described by the function $g_K(b_T;\bma)$, which is 
totally universal and independent of $Q$, $x$, or $z$.   
$g_K(b_T;\bma)$ generally contains both perturbatively 
calculable contributions and nonperturbative effects.
By its definition, Eq.~(13.60) of Ref.~\cite{collins}, it must vanish like a power 
at small $b_T$.  Detailed studies of power corrections in Refs.~\cite{Korchemsky:1994is,Tafat:2001in,Laenen:2000ij,Laenen:2000hs} 
suggest that $g_K(b_T;b_{\rm max})$ should vanish like $b_T^2$ (or an even power of $b_T$) as $b_T \to 0$. 
See, especially, Eq.~(6.2)
of Ref.~\cite{Tafat:2001in} and also  
the discussion around Eq.~(55) of Ref.~\cite{Laenen:2000ij}.

The value of $b_{\rm max}$, as well as the functional form for the matching in Eq.~\eqref{bstar}, 
is exactly arbitrary in full QCD.   
In practical applications, it is preferable to choose it to be large enough to maximize the perturbative 
content of the calculation, while small enough that only a solidly perturbative range of $b_T$ 
is included in the calculation of $\tilde{K}(b_{\ast};\mu_b)$.
If, on one hand, $\bma$ is chosen very large, then perturbative calculations are used 
at large $b_T$ where their validity is suspect. 
Large corrections from $g_K(b_T;\bma)$ would then be needed to recover the true cross section.   
On the other hand, if $b_{\rm max}$ is chosen too small, 
most of the work in fitting 
would go into reproducing results that might otherwise be accounted for by perturbation theory if $\bma$ were 
chosen larger.  

However, the formalism is 
set up to be neutral as to where the actual transition from perturbative to 
nonperturbative $b_T$-dependence actually
occurs, and at a given order there is no distinction made between neglected higher order corrections and 
unavoidably nonperturbative contributions. 
Therefore, assuming $Q \gg \Lambda_{\rm QCD}$, a desired degree of precision may be in principle achieved 
point-by-point in all $b_T$ with a suitable combination of higher order calculations and constrained 
nonperturbative but universal input, and this remains true for any choice of $b_{\rm max}$.  
Thus, both the 
traditional CSS formalism and the TMD formalism of~\cite{collins} are exactly model independent in the sense 
that they accommodate any parametrization 
of nonperturbative large-$b_T$ physics.\footnote{Of course, nonperturbative physics need 
not be regarded as a type of model input if first principles nonperturbative calculational methods are available.}

The perturbative part of the CS kernel, $\tilde{K}(b_{\ast};\mu_b)$, is defined in 
Eq.~\eqref{eq:evolution} partly by the choice of ${\bm b}_{\ast}({\bm b_T})$, including the value 
of $b_{\rm max}$.  Then all 
remaining information about the large $b_T$ behavior of 
the CS evolution kernel, including but not limited to nonperturbative effects, is contained completely in $g_K(b_T;\bma)$.  
(Note that $g_K(b_T;\bma)$ may also contain perturbatively calculable contributions.)
As mentioned above, the cross section is exactly independent of $\bma$.  In practical applications, however, dependence on $\bma$ typically 
does arise due to incomplete knowledge of the exact form of $g_K(b_T;\bma)$ at large $b_T$.

A frequently used ansatz for $g_K(b_T;\bma)$ is
\begin{equation}
g_K(b_T;\bma) =  g_2(\bma) \frac{1}{2} b_T^2 \, , \label{eq:gaussform}
\end{equation}
where $g_2(\bma)$ is a Gaussian fit parameter.  This choice for $g_K(b_T;\bma)$, if positive and reasonably large, imposes 
a very strong Gaussian suppression of the nonperturbative regions of $b_T$ in Eq.~\eqref{eq:evolution} 
whenever $Q$ becomes significantly larger than $Q_0$.
As implied by the notation, $g_2(\bma)$ should be expected 
to take on different values depending on the 
choice of $\bma$.  (Indeed, if $\bma$ is changed, then the
functional form of $g_K(b_T;\bma)$ may also change.) For more on
this point, see section~\ref{sec:pertcons}.

The first applications of CSS evolution in the context of hadronic structure studies in spin physics were performed by 
Boer~\cite{Boer:2001he} within the original form of the CSS formalism from 
Refs.~\cite{Collins:1981uk,Collins:1981uw,Collins:1984kg}.  In Ref.~\cite{Aybat:2011zv}, parametrizations of 
the TMD PDFs were constructed out of previous nonperturbative fits within the updated version of the 
CSS formalism of Ref.~\cite{collins}, and were presented in a form where the contributions to separate operator 
definitions of the TMD PDFs and fragmentation functions could be automatically identified.  These parametrizations 
were constructed from nonperturbative functions that were extracted in earlier work in the old version of the CSS formalism for 
Drell-Yan scattering~\cite{Landry:2002ix,Konychev:2005iy}, and were combined with fixed scale SIDIS fits at 
low $Q$ that arose in the context of hadronic structure studies~\cite{Schweitzer:2010tt}.  
A direct extrapolation of the Drell-Yan fits to low $Q$ gives evolution that is too rapid (see, again, Ref.~\cite{Sun:2013hua}, Fig.~2), and in 
Ref.~\cite{Aybat:2011zv} this was conjectured to be due 
to the role of larger $x$ in the small $Q$ fits, so an $x$-dependent function was inserted to 
obtain a fit that interpolated between all of the fits, within the TMD evolution formalism.  
(Note that Ref.~\cite{Qiu:2000hf} finds 
that the transverse momentum width depends significantly on $\sqrt{s}$ as well 
as $Q$.)
By adjusting the fit parameters between those of~\cite{Landry:2002ix} and~\cite{Konychev:2005iy}, 
a theoretical error of approximately a factor of two was estimated.   
Though rough, and limited by the scarcity of TMD-style fits that included evolution for nonperturbative parts, 
this provided a clear illustration of how actual TMD fits map to the TMD factors of the Collins TMD factorization formalism, 
with the TMD parametrizations themselves mapping to the operator matrix element definitions that emerge from the 
TMD factorization derivation.

Another direct application of the TMD evolution formalism was applied later to the Sivers function, a polarized TMD PDF 
important for studies of hadron structure, in Ref.~\cite{Aybat:2011ge}.  
Quantitative estimates of the amount of suppression in the evolution of the Sivers asymmetry  were  
presented in Refs.~\cite{Aybat:2011ta,Boer:2013zca}, again based on extrapolations of earlier extractions of nonperturbative 
parameters from Drell-Yan scattering.  
Boer~\cite{Boer:2013zca} provided a treatment in the more traditional language of applications of the CSS formalism.
That the nonperturbative input is based on prior extractions is 
crucial in this class of phenomenological studies wherein a central goal is to establish and/or 
test the universality of nonperturbative functions, particularly the strong universality of the nonperturbative evolution.  
Reference~\cite{Aybat:2011ta} found general consistency 
between HERMES and COMPASS 
data and the extrapolations from large $Q$ fits, lending general support 
for the applicability of the TMD factorization formalism in the low $Q$ region, 
but the data corresponded to different ranges of $x$ and so 
the analysis was not 
totally in line with the apples-to-apples treatment described in the introduction.

It was recently illustrated very clearly in Ref.~\cite{Sun:2013dya,Sun:2013hua} 
that the rapid evolution given by extrapolating the nonperturbative extractions from Drell-Yan cross sections at large $Q$ is 
too fast to adequately account very generally for data in the region of $Q$ of order a few GeV. 
Therefore, the details of the nonperturbative contribution to evolution in the region 
of small $Q$ need to be reinvestigated. 

To maintain consistency with the general aim 
of extracting properties intrinsic to specific hadrons as outlined in the introduction, 
we would ideally vary $Q$ while holding $x$, $z$, and hadron species fixed.  
In experiments, however, these variables are correlated, and practical fitting becomes
challenging.  We will appeal, in the next section, to the multi-differential 
COMPASS  data from Ref.~\cite{SIDISdata} to study the variation in the 
multiplicity distribution with small variations in $Q$ and roughly 
fixed $x$ and $z$ bins within the same experiment.

\section{Empirical Rate of Evolution in the Region of Moderate $Q$}
\label{sec:approxes}

Empirically, the SIDIS data in Ref.~\cite{SIDISdata} reveal that the differential cross section 
as a function of $P_T$ is reasonably well-described by a Gaussian functional form in the region of 
small $P_T$ (see, e.g., Fig. 4 of Ref.~\cite{SIDISdata}), with a width that broadens very slightly with increasing $Q$.  
In this section, we will quantify this rate of change within the language of TMD evolution.

In Ref.~\cite{SIDISdata}, the data for hadron multiplicities are fitted using a Gaussian form, 
\begin{equation}
\frac{d \sigma}{d P_T^2} \propto \exp \left\{ - \frac{P_T^2}{\langle P_T^2\rangle} \right\} ,  \label{eq:bspacedata}
\end{equation}
and the resulting $\langle P_T^2\rangle$ values are presented.
Expressed in terms of the two dimensional Fourier transform from $b_T$-space, Eq.~\eqref{eq:bspacedata} becomes
\begin{equation}
\frac{d \sigma}{d P_T^2} \propto  {\rm F.T.}
\exp \left\{ - \frac{b_T^2 \langle P_T^2 \rangle }{4} \right\} \, . \label{eq:bspacedatafit}
\end{equation}
The parameter $\langle P_T^2 \rangle$ 
is in general a function of $x$, $z$, and $Q$.

Therefore, to match to the evolved formula, Eq.~\eqref{eq:evolution}, we assume that all the 
terms in the exponent of  Eq.~\eqref{eq:evolution} can 
be approximated as quadratic.  In particular, we need a quadratic ansatz for the functions 
$g_{\rm PDF} (x,b_T;\bma)$ and  $g_{\rm FF} (z,b_T;\bma)$:
\begin{equation}
\label{eq:ganzi}
g_{\rm PDF} (x,b_T;\bma) \propto  g_{\rm FF} (z,b_T;\bma) \propto b_T^2 \, . 
\end{equation}
A note of caution is needed here because the actual behavior of $g_{\rm PDF} (x,b_T;\bma)$ and
$g_{\rm FF} (z,b_T;\bma)$ includes, via the definitions in Eqs.~\eqref{eq:newgpdf} and~\eqref{eq:newgff}, 
non-power law effects from collinear perturbation theory that are important for 
accurately describing the small $b_T$ region.
This corresponds to the behavior 
of the large $P_T$ tail, and accounting for it properly would involve a careful treatment of the $Y$ term as well.  For the
moderate $Q$ range of the COMPASS data that we consider in this article, 
where a Gaussian fit actually provides a good description of the data, 
we work within the \emph{conjecture} that
the small $b_T$ behavior from $g_{\rm PDF} (x,b_T;\bma)$ and $g_{\rm FF} (z,b_T;\bma)$ is negligible.   However, the 
details of the initial-scale treatment of $g_{\rm PDF} (x,b_T;\bma)$ and $g_{\rm FF} (z,b_T;\bma)$
may become important when extending to much larger $Q$.  
Also, we echo again the cautionary remarks in the introduction 
regarding the possible importance of power-law $(M/Q)^a$ corrections that are normally neglected as part of the 
TMD factorization derivation.

A result of CS evolution is that, for the TMD term, the $Q$-dependence of the logarithm of the $b_T$-dependence 
is linear in $\ln (Q)$ -- see, e.g., Eq.~(3.3) of
Ref.~\cite{Collins:1984kg}. 
Let us therefore define,
\begin{equation}
\label{eq:sigdef}
\tilde{\sigma}_{\rm TMD \; term} \equiv \mathcal{H}(\alpha_s(Q)) \tilde{F}_{H_1}(x,b_T;Q,Q^2) \, \tilde{D}_{H_2}(z,b_T;Q,Q^2) \; .
\end{equation}
That is, it is the Fourier transform of the TMD term in Eq.~\eqref{eq:evolution0}, corresponding to $\tilde{W}$ in Eq.~(3.3) of
Ref.~\cite{Collins:1984kg}.
Then, 
\begin{equation}
\label{eq:basicevol}
\left. \frac{d \ln \tilde{\sigma}_{\rm TMD \; term}}{d \ln Q^2}  \right|_{\rm b_T \; dep} = \left. \tilde{K}(b_T;\mu_0) \right|_{\rm b_T \; dep} \, .
\end{equation}
Importantly, the right side is independent of $Q$, $x$ and $z$.
Still assuming that the $Y$-term can be neglected, 
and using Eq.~\eqref{eq:bspacedatafit}, we then make the approximation that 
\begin{equation}
\tilde{\sigma}_{\rm TMD \; term}
\approx
\exp \left\{ - \frac{b_T^2 \langle P_T^2 \rangle }{4} \right\} \, . \label{eq:bspacedatafitsig}
\end{equation}

Another note of caution is needed 
here because the right side of Eq.~\eqref{eq:basicevol} includes only the TMD term's contribution to the cross section 
and not the $Y$ term, while in Eq.~\eqref{eq:bspacedatafitsig} we have approximated 
$\tilde{\sigma}_{\rm TMD \; term}$ by the actual fit to the cross section from Eq.~\eqref{eq:bspacedatafit}.  For now 
we assume this to be a reasonable starting approximation.  The actual $Q$-dependence of the cross section 
including the $Y$ term will have corrections relative to what is obtained from the combination 
of Eq.~\eqref{eq:basicevol} and Eq.~\eqref{eq:bspacedatafitsig}.

For small $P_T$, the $P_T$-shape of the data in Ref.~\cite{SIDISdata} 
is empirically 
observed to broaden slightly as $Q$ increases, but remains quite well described by a Gaussian parametrization.
(See, however, the later discussion of tail effects in Sec.~\ref{sec:largebT}.)
The evolved differential cross section obtained from Eq.~\eqref{eq:basicevol} remains 
Gaussian after evolution, within the approximation above, only if the right side of Eq.~\eqref{eq:basicevol} can be approximated 
as quadratic in $b_T$ with a negative coefficient.
Therefore, if the observed Gaussian 
shape is to be maintained as $Q$ varies, then Eq.~\eqref{eq:bspacedatafit} must take the form
\begin{equation}
\frac{d \sigma}{d P_T^2} \propto  {\rm F.T.}
\exp \left\{ - \frac{b_T^2}{4}\left( 
	\langle P_T^2 \rangle_0 + 4 C_{\rm evol} \ln \left( \frac{Q_2}{Q_1} \right) \right) \right\} \, . \label{eq:bspacedatafit2}
\end{equation}
Here, $\langle P_T^2 \rangle_0$ may depend only on $x$ and $z$ (it is independent of $Q$)
and $C_{\rm evol}$ is a numerical parameter that is, in principle, independent of $x$ and $z$.
$Q_1$ and $Q_2$ are initial and final hard scales.

If $x$ and $z$ are held fixed, then the variation of $\langle P_T^2\rangle$ with $Q$ can be found
directly from the $b_T$-space integrand in Eq.~\eqref{eq:bspacedatafit2}:
\begin{equation}
\Delta \langle P_T^2\rangle(Q_1,Q_2) \approx  4 C_{\rm evol} \ln \left( \frac{Q_2}{Q_1} \right) \, , 
\label{eq:g2bound}
\end{equation}
where we define 
\begin{equation}
\Delta \langle P_T^2\rangle(Q_1,Q_2) = \langle P_T^2\rangle_{Q = Q_2} - \langle P_T^2\rangle_{Q = Q_1} \, . 
\label{eq:deltapt2}
\end{equation}
We will next use Eq.~\eqref{eq:g2bound} to extract approximate bounds on $C_{\rm evol}$ from 
experimental results for $\Delta \langle P_T^2\rangle(Q_1,Q_2)$.

The only aspect of TMD factorization that we have used so far is 
Eq.~\eqref{eq:basicevol}. 
Specifically, we have applied it to the case of the COMPASS data for the small 
range of $Q$ where the $P_T$ distribution appears to remain approximately Gaussian even after evolution to 
obtain Eq.~\eqref{eq:bspacedatafit2}.
We do not 
address at this stage the question of whether $ \tilde{K}(b_T;\mu_0)$ is governed primarily by perturbative 
or nonperturbative $b_T$-dependence.  
While $C_{\rm evol}$ resembles $g_2$ in a quadratic approximation to 
$g_K(b_T;\bma)$, here it should be emphasized that it is meant merely to approximate the collective effect of 
all the $Q$-dependent terms in the exponent of Eq.~\eqref{eq:evolution}, in a way consistent with 
Eq.~\eqref{eq:basicevol}, and it should not be 
identified at this stage with any specific perturbative or nonperturbative terms.  Of course, perturbative contributions are not quadratic, 
so the quadratic ansatz for the right side of Eq.~\eqref{eq:basicevol} is a poor one for small $b_T$.  
We will nevertheless attempt to use it to capture 
the general $Q$-dependence of the $P_T$-width in the 
vicinity of small $Q$ variations where the 
data appear from~\cite{SIDISdata} to be reasonably well-described 
by Gaussian fits.  We will 
further analyze the reliability  
of such an approximation in the next few sections.  
Since the right side of Eq.~\eqref{eq:basicevol} is universal and 
$x$, $Q$, and $z$ independent, then a test of the universal value for $C_{\rm evol}$ probes the 
assumptions that led to the use of Eq.~\eqref{eq:bspacedatafit2} as a model, such as the Gaussian functional 
form and the neglect of the $Y$-term.

In a full treatment of evolution, there is also a $Q$ dependence that affects 
only the normalization of the cross section.  Since we are mainly interested in the variation in the 
width, we ignore any such contributions and focus only on the broadening of the Gaussian 
shape.

\section{Estimates of $C_{\rm evol}$ from Unpolarized SIDIS}
\label{sec:numerics}
\begin{table*}
\caption{
Estimated upper bounds on the evolution parameter $C_{\rm evol}$ in units of GeV$^2$,
with positively charged produced hadrons.  All values of $Q$ are in units of GeV.  See the text
for an explanation of the difference between $\langle C_{\rm evol} \rangle$, $C_{\rm evol}^{\rm min}$,
and $C_{\rm evol}^{\rm max}$.}
\vspace{5mm}
\begin{tabular}{|l|l|l|l|c|l|l|l|}
	\hline \hline
\multicolumn{8}{|c|}
	{\rule[-3mm]{0mm}{8mm} $C_{\rm evol} =
	\Delta \langle P_T^2\rangle(Q_1,Q_2) / \left(4 \ln \left( \frac{Q_2}{Q_1} \right) \right)$,
	{\rm Positively Charged Hadrons}}   \\
\multicolumn{8}{|c|}
	{\rule[-3mm]{0mm}{8mm}  $0.2 < z < 0.25$ }   \\
\textbf{\rm $\sqrt{\langle Q_1^2 \rangle},Q_1^{\rm max}, Q_1^{\rm min}$}
         & \textbf{$\langle P_T^2 \rangle$} & \textbf{\rm $\sqrt{\langle Q_2^2 \rangle},Q_2^{\rm max}, Q_2^{\rm min}$} 
         & \textbf{$\langle P_T^2 \rangle$} & \textbf{\rm $\langle x_{\rm bj}\rangle$}
	&  $\langle C_{\rm evol} \rangle $ & $C_{\rm evol}^{\rm min}$ & $C_{\rm evol}^{\rm max}$
\\  \hline \hline
1.109,1.225,1.0 & 0.212 & 2.017,2.236,1.871 & 0.229 & 0.0213--0.0216  & 0.0070 & 0.0052	 &	  0.0099 \\
1.049,1.095,1.0 & 0.177 & 2.114,2.449,1.871 & 0.226 & 0.0295--0.0323  & 0.0175 &  0.0137  &  0.0229  \\
\hline
\end{tabular}
\begin{tabular}{|l|l|l|l|c|l|l|l|}
	\hline \hline
\multicolumn{8}{|c|}
	{\rule[-3mm]{0mm}{8mm} $0.25 < z < 0.3$}  \\
\textbf{\rm $\sqrt{\langle Q_1^2 \rangle},Q_1^{\rm max}, Q_1^{\rm min}$}
         & \textbf{$\langle P_T^2 \rangle$} & \textbf{\rm $\sqrt{\langle Q_2^2 \rangle},Q_2^{\rm max}, Q_2^{\rm min}$} & \textbf{$\langle P_T^2 \rangle$} & \textbf{\rm $\langle x_{\rm bj}\rangle$}
	&  $\langle C_{\rm evol} \rangle $ & $C_{\rm evol}^{\rm min}$ & $C_{\rm evol}^{\rm max}$
\\  \hline \hline
1.109,1.225,1.0 & 0.241 & 2.017,2.236,1.871 & 0.253 & 0.0213--0.0216  & 0.0051 & 0.0040	 &  0.0071 \\
1.049,1.095,1.0 & 0.202 & 2.114,2.449,1.871 & 0.249 & 0.0295--0.0323  & 0.0170 & 0.0132  &  0.0221   \\
\hline
\end{tabular}
\begin{tabular}{|l|l|l|l|c|l|l|l|}
	\hline \hline
\multicolumn{8}{|c|}
	{\rule[-3mm]{0mm}{8mm} $0.3 < z < 0.35$}  \\
\textbf{\rm $\sqrt{\langle Q_1^2 \rangle},Q_1^{\rm max}, Q_1^{\rm min}$}
         & \textbf{$\langle P_T^2 \rangle$} & \textbf{\rm $\sqrt{\langle Q_2^2 \rangle},Q_2^{\rm max}, Q_2^{\rm min}$} & \textbf{$\langle P_T^2 \rangle$} & \textbf{\rm $\langle x_{\rm bj}\rangle$}
	&  $\langle C_{\rm evol} \rangle $ & $C_{\rm evol}^{\rm min}$ & $C_{\rm evol}^{\rm max}$
\\  \hline \hline
1.109,1.225,1.0 & 0.263 & 2.017,2.236,1.871 & 0.283 & 0.0213--0.0216  & 0.0083 & 0.0062  & 0.0117 \\
1.049,1.095,1.0 & 0.230 & 2.114,2.449,1.871 & 0.276 & 0.0295--0.0323  & 0.0165 & 0.0130  & 0.0216    \\
\hline
\end{tabular}
\label{table:g2valuespos}
\caption{Estimated upper bounds on the evolution parameter $C_{\rm evol}$ in units of GeV$^2$,
with negatively charged produced hadrons.}
\vspace{5mm}
\begin{tabular}{|l|l|l|l|c|l|l|l|}
	\hline \hline
\multicolumn{8}{|c|}
	{\rule[-3mm]{0mm}{8mm} $C_{\rm evol} =
	\Delta \langle P_T^2\rangle(Q_1,Q_2) / \left(4 \ln \left( \frac{Q_2}{Q_1} \right) \right)$,
	{\rm Negatively Charged Hadrons}}   \\
\multicolumn{8}{|c|}
	{\rule[-3mm]{0mm}{8mm}  $0.2 < z < 0.25$ }   \\
\textbf{\rm $\sqrt{\langle Q_1^2 \rangle},Q_1^{\rm max}, Q_1^{\rm min}$}
          & \textbf{$\langle P_T^2 \rangle$} & \textbf{\rm $\sqrt{\langle Q_2^2 \rangle},Q_2^{\rm max}, Q_2^{\rm min}$} & \textbf{$\langle P_T^2 \rangle$} & \textbf{\rm $\langle x_{\rm bj}\rangle$}
	&  $\langle C_{\rm evol} \rangle $ & $C_{\rm evol}^{\rm min}$ & $C_{\rm evol}^{\rm max}$
\\  \hline \hline
1.109,1.225,1.0 & 0.207 & 2.017,2.236,1.871 & 0.233 & 0.0213--0.0216  & 0.0109 & 0.0081	 & 0.0155  \\
1.049,1.095,1.0 & 0.167 & 2.114,2.449,1.871 & 0.233 & 0.0295--0.0323  & 0.0234 & 0.0183  &  0.0306   \\
\hline
\end{tabular}
\begin{tabular}{|l|l|l|l|c|l|l|l|}
	\hline \hline
\multicolumn{8}{|c|}
	{\rule[-3mm]{0mm}{8mm} $0.25 < z < 0.3$}  \\
\textbf{\rm $\sqrt{\langle Q_1^2 \rangle},Q_1^{\rm max}, Q_1^{\rm min}$}
          & \textbf{$\langle P_T^2 \rangle$} & \textbf{\rm $\sqrt{\langle Q_2^2 \rangle},Q_2^{\rm max}, Q_2^{\rm min}$} & \textbf{$\langle P_T^2 \rangle$} & \textbf{\rm $\langle x_{\rm bj}\rangle$}
	&  $\langle C_{\rm evol} \rangle $ & $C_{\rm evol}^{\rm min}$ & $C_{\rm evol}^{\rm max}$
\\  \hline \hline
1.109,1.225,1.0 & 0.233 & 2.017,2.236,1.871 & 0.264 & 0.0213--0.0216  & 0.0133 & 0.0100 & 0.0188 \\
1.049,1.095,1.0 & 0.193 & 2.114,2.449,1.871 & 0.256 & 0.0295--0.0323  & 0.0222 & 0.0174 & 0.0291    \\
\hline
\end{tabular}
\begin{tabular}{|l|l|l|l|c|l|l|l|}
	\hline \hline
\multicolumn{8}{|c|}
	{\rule[-3mm]{0mm}{8mm} $0.3 < z < 0.35$}  \\
\textbf{\rm $\sqrt{\langle Q_1^2 \rangle},Q_1^{\rm max}, Q_1^{\rm min}$}
         & \textbf{$\langle P_T^2 \rangle$} & \textbf{\rm $\sqrt{\langle Q_2^2 \rangle},Q_2^{\rm max}, Q_2^{\rm min}$} & \textbf{$\langle P_T^2 \rangle$} & \textbf{\rm $\langle x_{\rm bj}\rangle$}
	&  $\langle C_{\rm evol} \rangle $ & $C_{\rm evol}^{\rm min}$ & $C_{\rm evol}^{\rm max}$
\\  \hline \hline
1.109,1.225,1.0 & 0.254 & 2.017,2.236,1.871 & 0.291 & 0.0213--0.0216  & 0.0154 & 0.0114	& 0.0217  \\
1.049,1.095,1.0 & 0.220 & 2.114,2.449,1.871 & 0.284 & 0.0295--0.0323  & 0.0229 & 0.0179 & 0.0300  \\
\hline
\end{tabular}
\label{table:g2valuesneg}
\end{table*}
\begin{figure*}
\centering
  \begin{tabular}{c@{\hspace*{10mm}}c}
    \includegraphics[scale=0.4]{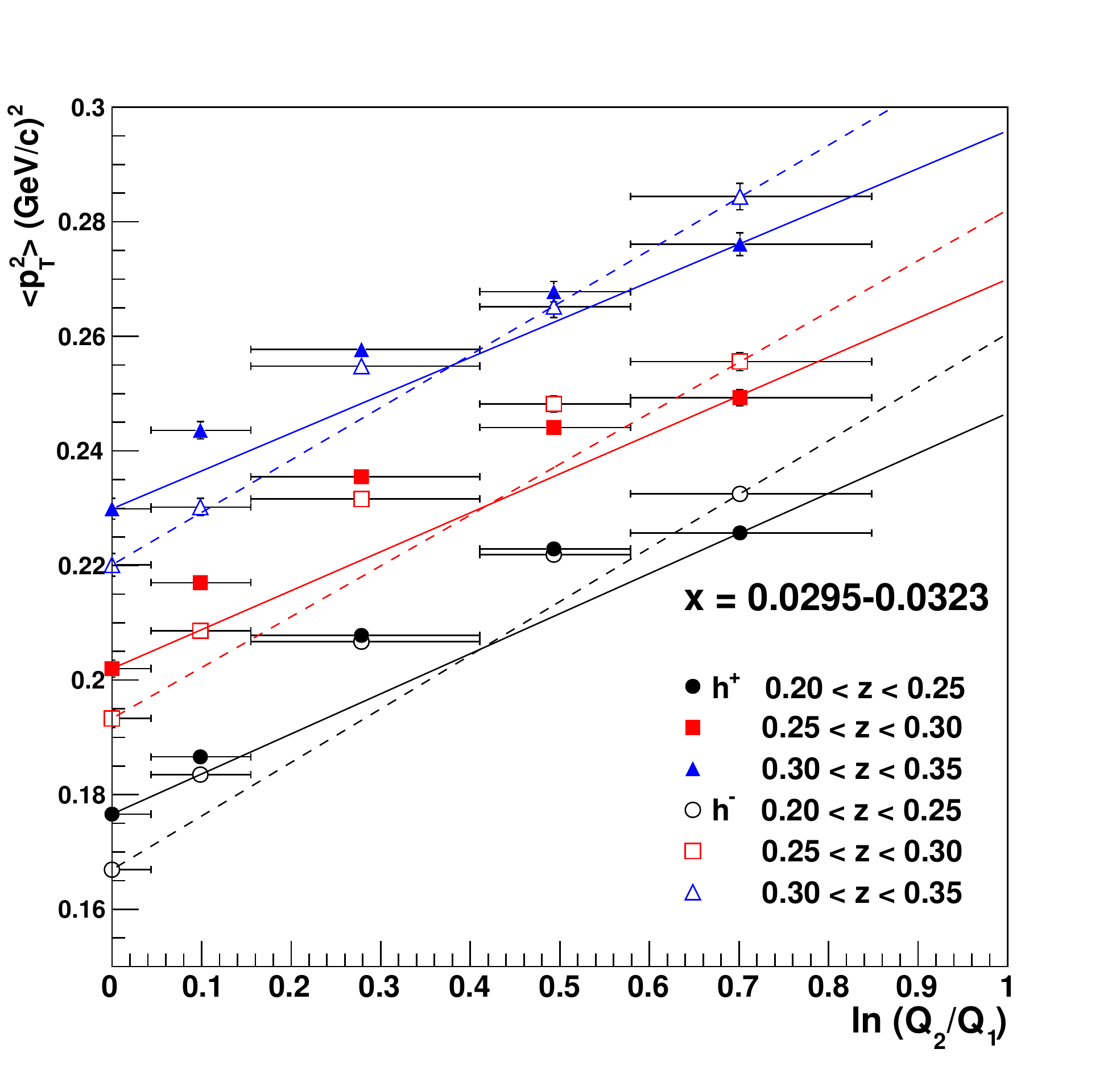}
    &
    \includegraphics[scale=0.4]{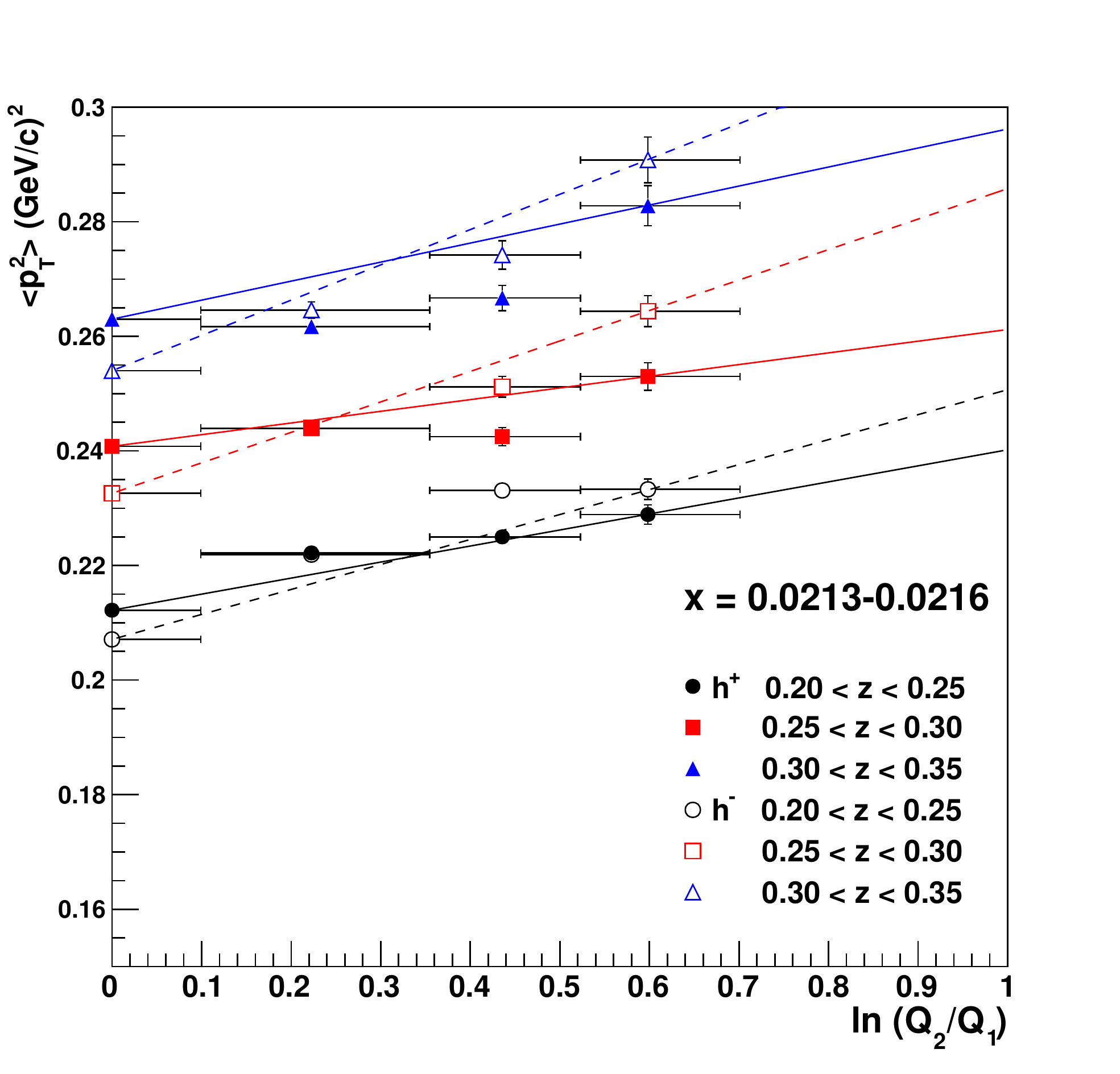}
    \\
    (a) & (b)
    \\[5mm]
   \end{tabular}
\caption{(color online).  Linear fits, calculated using Eq.~\eqref{eq:g2bound}, connecting 
low to high $Q$ using $C_{\rm evol}$.  The horizontal bars 
show the bin widths in $Q$.  The vertical bars are the 
errors of the Gaussian fits reported in Ref.~\cite{SIDISdata}. 
Plot (a) is for $x_{\rm bj} = 0.0295-0.0323$ and plot (b) is for $x_{\rm bj} = 0.0213-0.0216$.  The solid 
and open points are for positive and negative produced hadrons respectively.  The linear slopes are calculated using the 
largest and smallest
$Q_2$, $Q_1$ values. (See text for details.)}
\label{fig:linearplots}
\end{figure*}

Evolution leads to a well-known broadening of the $P_T$ width with $Q$ at fixed $x$ and $z$.  For a 
significant effect to be clearly observable, one must examine fixed $x$ and $z$ bins over 
sufficiently broad ranges of $Q$.  In Ref.~\cite{SIDISdata}, Figs.~5 and~6 allow 
$Q$ intervals of order $\sim 1.0$~GeV for 
fixed $x$  and $z$ bins to be identified across 
several bins in $Q$.  
In each panel, the fifth and sixth columns of vertical blocks correspond 
to fixed $x_{\rm bj}$ and $z$ bins with four and five $Q^2$-bins, respectively.  Since these give the 
maximum variation in $Q$, they are the data we will use in our analysis to obtain conservative 
limits on the amount of evolution at moderately small $Q$.  In addition, we exclude data with $z > .35$ to avoid
complications with the large $z$ region, and to be certain that we are away from any possible
significant resonance effects in the remnant.  
We stress, however, that these reservations about the large $z$ region 
apply only to the approach to quite small $Q$ region where contamination from 
resonance effects might become a serious concern.  At larger $Q$ such exclusions
would be excessively conservative.

The incoming target is always \lid
and the final state is inclusive in all species of charged hadrons.

Tables~\ref{table:g2valuespos},~\ref{table:g2valuesneg} show the results for $C_{\rm evol}$ 
from Eq.~\eqref{eq:deltapt2} for each 
$x_{\rm bj}$ and $z$ bin.   (Spreadsheets will be made available at Ref.~\cite{webpage}.)
A limitation of this analysis is the unavoidably large $Q$ bin sizes relative to $Q$ itself in 
the moderate $Q$ region.  
To estimate the error from large $Q$ bin sizes, we have therefore calculated $C_{\rm evol}$ using 
the following three methods:
First, for $Q_2$ and $Q_1$ we use the average $\langle Q^2 \rangle$ for the top and bottom $Q^2$ bins, 
respectively, in Figs.~5 and~6 of Ref.~\cite{SIDISdata}.  
The result is called $\langle C_{\rm evol} \rangle$ in the Tables~\ref{table:g2valuespos},~\ref{table:g2valuesneg}.
Next, in order to obtain an estimated 
upper bound on the evolution we use the value of $Q$ for the top edge of the lowest bin, called 
$Q_1^{\rm max}$ in the tables, for $Q_1$, and the bottom edge of the largest $Q^2$-bin, called $Q_2^{\rm min}$ 
in the tables, for $Q_2$.  This will tend to underestimate $\ln(Q_2/Q_1)$ and thus give 
a value for $C_{\rm evol}$ that is too large.  
The result is called $C_{\rm evol}^{\rm max}$ in 
the tables.  Similarly, to get an estimated lower bound on $C_{\rm evol}$, we use the value of $Q$ for the
bottom edge of the lowest bin, called $Q_1^{\rm min}$ in the tables, for $Q_1$, and the upper edge 
of the largest bin, called $Q_2^{\rm max}$ in the tables, for $Q_2$.  This will tend to overestimate
$\ln(Q_2/Q_1)$ and thus will tend to give a value for $C_{\rm evol}$ that is too small.  The result is called 
$C_{\rm evol}^{\rm min}$ in Tables~\ref{table:g2valuespos},~\ref{table:g2valuesneg}.
Plots showing the extraction of $C_{\rm evol}$ are presented in Fig.~\ref{fig:linearplots}.

Another source of error is the cutoff at $P_T = 0.85$~GeV in the fits of Ref.~\cite{SIDISdata}, where the 
Gaussian description starts to break down.  Variations in the precise cutoff, as well as variations in the 
precise functional form of fit, may affect the variation in the overall width of the distribution with $Q$.  We will 
address this further in Sec.~\ref{sec:largebT}.  

The trend in Tables~\ref{table:g2valuespos},~\ref{table:g2valuesneg} and Fig.~\ref{fig:linearplots}
suggests a small yet non-vanishing $Q$-dependence in the $P_T$ width; the lowest value of 
$C_{\rm evol}$ is $0.0040$~GeV$^2$ and the largest value is $0.0306$~GeV$^2$.  
Since $C_{\rm evol}$ would be 
expected to be universal if TMD factorization with a neglected $Y$ term is valid, then observable correlations 
in the value of $C_{\rm evol}$ in Fig.~\ref{fig:linearplots} with $x$, $z$ and hadron charge suggests that a truly accurate treatment 
requires the $Y$-term.  Still, Fig.~\ref{fig:linearplots} also suggests reasonable upper limits on the size of the evolution. 
In the next few sections, we 
will interpret this in the context of an analysis of the importance of contributions from different regions of $b_T$.
We will comment further on the size of $C_{\rm evol}$ and its relevance to $g_2$ in section~\ref{sec:pertcons}. 

\section{Relevance of Large $b_T$}
\label{sec:largebT}

\begin{figure}[t]
\centering
\includegraphics[scale=0.45]{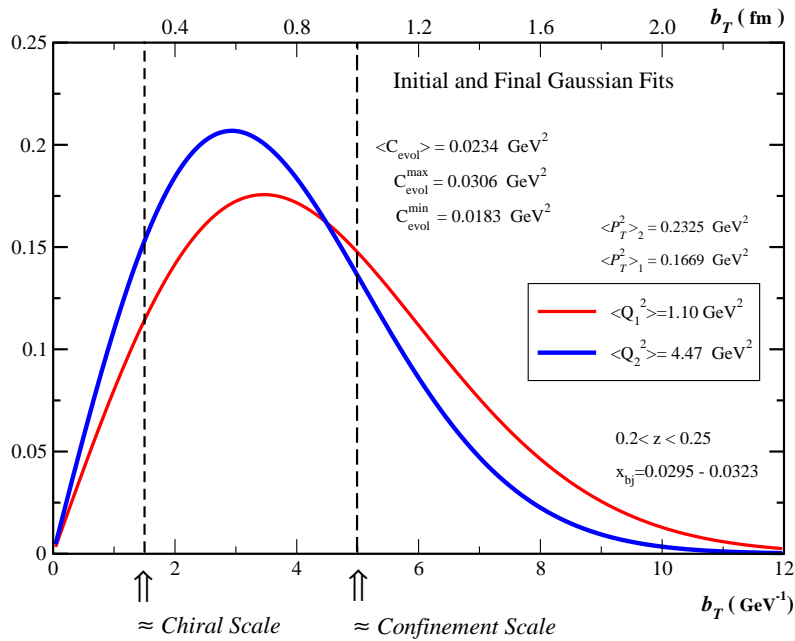}
\caption{
Coordinate space Gaussian fits showing the largest variation in the width found in Tables~\ref{table:g2valuespos},~\ref{table:g2valuesneg} with 
a change from $\sqrt{\langle Q_1^2 \rangle} = 1.049$~GeV to $\sqrt{\langle Q_2^2 \rangle} = 2.114$~GeV. 
The precise function being plotted is Eq.~\eqref{eq:maxfits} with the initial (red) and final (thick blue) $\langle P_T^2 \rangle$ COMPASS values in Eq.~\eqref{eq:compassfits}.
(See online for color.) The peak moves toward smaller values with increasing $Q$.
These curves  correspond to the first entry (smallest $z$ bin) of Table~\ref{table:g2valuesneg} and the second row (largest $x_{\rm bj}$ bin).  We have 
marked the approximate chiral symmetry breaking scale from Ref.~\cite{Schweitzer:2012dd} at $b_T \approx 1.5$~GeV$^{-1}$ and 
the approximate confinement scale at $b_T \approx 5.0$~GeV$^{-1}$.  Note that at the top of the graph we have also shown the horizontal axis in fm to provide a more intuitive sense of 
relevant size scales.
Compare the dominant regions of $b_T$ here with larger $Q$ curves of Fig.~4 in Ref.~\cite{Konychev:2005iy}.}
\label{fig:bdists}
\end{figure}

In the context of applications like those outlined in the introduction, it is important to 
recognize that, although the $b_T$-dependence has
both perturbative and nonperturbative contributions, the TMD factorization 
theorem is valid for all $b_T$, including $b_T \gg 1 / \Lambda_{\rm QCD}$, so long as $Q$ is large enough that the 
expansion of $\mathcal{H}_{f, \text{\tiny process}}(\alpha_s(Q))$ in each of 
Eqs.~(\ref{eq:SIDISschem})-(\ref{eq:epemschem}) is perturbatively well-behaved.  
TMD factorization, therefore, retains important predictive 
power for all $b_T$, regardless of how much of the $b_T$-dependence itself is  
perturbatively describable. Part of that predictive power comes from 
the universality of the TMD functions, analogously to the collinear PDFs of collinear factorization, 
and from the very strong universality of the CS kernel, including the nonperturbative parts contained in $g_K(b_T;\bma)$.

As $Q$ is increased, the dominant contribution to the cross section  
becomes localized in coordinate space around small $b_T$ so that the
nonperturbative $b_T$ contribution becomes less important~\cite{Parisi:1979se}. 
For extremely large $Q$, it is 
expected that the nonperturbative contribution can be ignored altogether.  
Alternatively, at moderate values of $Q$, $\alpha_s(Q)$ might be small 
enough that TMD factorization is completely valid, 
yet the $b_T$-dependence may still contain a large, or even dominant, nonperturbative large-$b_T$ contribution.  
The latter situations are ideal for extracting information about non-pertubative
hadron structure in terms of elementary quark and gluon degrees of freedom within a valid pQCD TMD factorization formalism.
Moreover, measurements at relatively small $Q$ are ideal for measuring and testing the strongly universal 
nature of the nonperturbative scaling violations contained within $g_K(b_T;\bma)$.  

Within the CSS formalism, estimates of the importance of nonperturbative $b_T$-dependence vary widely in the existing
literature. For example, Ref.~\cite{Collins:1984kg} estimates that the cross section can be 
reliably assumed to be totally insensitive to the nonperturbative region for $Q \sim 10^{8}$~GeV.  
Global fits to large $Q$ behavior,  such as that discussed in the recent analysis of Ref.~\cite{Guzzi:2013aja}, find a small but still 
important contribution from the nonperturbative 
component of the evolution factor for values $Q$ of order heavy vector boson masses.  
Another method for estimating the nonperturbative content of the $b_T$-dependence within the CSS formalism 
was given in Refs.~\cite{Qiu:2002mu,Qiu:2000hf} and similarly finds   
that nonperturbative input remains important for $Q$ of order heavy vector boson masses. Refs.~\cite{Qiu:2002mu,Qiu:2000hf}
further note that the relative contribution from the nonperturbative regime also has significant dependence on $\sqrt{s}$.
By contrast, it has been suggested in Refs.~\cite{Echevarria:2012pw,Sun:2013dya,Sun:2013hua}, within the context of 
similar but alternative evolution formalisms, that accounting for nonperturbative evolution can be avoided entirely even at scales of 
order $Q \sim 1.0$ to $2.0$~GeV.  

The question of the relevance of the nonperturbative region in the Collins TMD-factorization theorem
may be addressed directly in the context of the COMPASS measurements by using the fits to estimate the 
important range of $b_T$.\footnote{Despite the notation, the $b_T$ in the TMD-factorization formula  
is not an impact parameter like that appearing in generalized parton distributions for exclusive processes.  Therefore, it should not be taken 
to represent the total size of either the target or final state hadron.}
We have plotted the fits obtained by the COMPASS
collaboration~\cite{SIDISdata} in coordinate space as the solid lines in Fig.~\ref{fig:bdists}. 
Since the transverse momentum space distribution is obtained from a two dimensional 
Fourier transform from the coordinate space expression, we have also included a factor of $b_T$.  Also, since 
we are primarily interested in the width of the distribution, we normalize 
to unity in the integration over $b_T$.  That is, instead of Eq.~\eqref{eq:bspacedata}
the curves in Fig.~\ref{fig:bdists} are for 
\begin{equation}
\frac{b_T \langle P_T^2 \rangle}{2} \exp \left\{ - \frac{b_T^2 \langle P_T^2 \rangle }{4} \right\} \, . \label{eq:maxfits}
\end{equation}
Applying the integration $\int_0^{\infty} \, d b_T$ gives unity.
Thus, up to a normalization, Eq.~\eqref{eq:maxfits} is the integrand of the Fourier transform to coordinate space 
for the region of small $P_T$.

The initial and final Gaussian slope parameters $\langle P_T^2 \rangle$ that we have used in Fig.~\ref{fig:bdists}
correspond to the \emph{largest} parameter $C_{\rm evol}$ that is found in 
Tables~\ref{table:g2valuespos},~\ref{table:g2valuesneg}.  This gives an estimate of the \emph{maximum} reasonable rate  
of variation in the width 
with changes in $Q$ of order $\sim 1.0$~GeV and so is consistent with a strategy of placing rough
upper limits on the rate of evolution that can reasonably be expected at low $Q$. The largest value for $C_{\rm evol}$
corresponds to the second row of the first entry in Table~\ref{table:g2valuesneg}, and the corresponding slope parameters 
from Ref.~\cite{SIDISdata} are:
\begin{equation}
\label{eq:compassfits}
\langle P_T^2 \rangle_{Q_1 = 1.049 \, {\rm GeV}} = 0.1669 \pm 0.0012 \, {\rm GeV}^2\,; 
\qquad \langle P_T^2 \rangle_{Q_2 = 2.114 \, {\rm GeV}} = 0.2325 \pm 0.0011  \,  {\rm GeV}^2 \, ,
\end{equation}
where the uncertainties are the quoted statistical uncertainties from the fit only.

The resulting curves shown in Fig.~\ref{fig:bdists} are peaked around 
$b_T \sim 3.0$~GeV$^{-1}$ with tails extending out to nearly 
$b_T \sim 10.0$~GeV$^{-1}$, i.e. up to transverse sizes about twice that of the 
proton charge radius, suggesting that
the effect of nonperturbative input is substantial, at least in this region of moderate $Q$.  
For comparison, typical values for $b_{\rm max}$ used in the CSS formalism 
are between about $\sim 0.3$~GeV$^{-1}$~\cite{Qiu:2000hf} and 
$\sim 1.0$~GeV$^{-1}$~\cite{Konychev:2005iy}. More relevant are estimates of the physical
transverse distance scales over which nonperturbative physics is expected to become important.
Using a chiral quark soliton model~\cite{Reinhardt:1988fz,Diakonov:1987ty,Christov:1995vm,Weigel:2008zz}, 
Ref.~\cite{Schweitzer:2012hh} 
estimates a chiral symmetry breaking scale of about $\sim 0.3 \, {\rm fm} \sim 1.5$~GeV$^{-1}$ and a confinement
scale of about $5$~GeV$^{-1}$.  These estimates are built on earlier instanton 
models~\cite{Shuryak:1981ff,Diakonov:1985eg,Schafer:1996wv,Polyakov:1996kh} which likewise find 
a typical instanton size of $\sim 0.3$~fm. 
The $5$~GeV$^{-1}$ confinement scale is also consistent with a proton charge radius of $\sim 0.88$~fm~\cite{Beringer:1900zz} and a bag model  
radius of roughly $\sim 1.2\, {\rm fm}$ (See Ref.~\cite{Bhaduri:1988gc} and references therein).\footnote{We mention the bag model here
since it continues to be used in nonperturbative model treatments of special TMD functions.  See, for example, Refs.~\cite{Yuan:2003wk,Avakian:2008dz,Courtoy:2008dn,Avakian:2010br}.}  
Both are of order $1.0\, {\rm fm} \sim 5.0 \, {\rm GeV}^{-1}$.  The points where various categories of nonperturbative 
physics are estimated to become relevant have been marked 
by arrows in Fig.~\ref{fig:bdists}.

From the general features of Fig.~\ref{fig:bdists}, we conclude that, for the differential cross section in the limit of $P_T \to 0$, 
the relevant range of $b_T$ is likely to be nearly dominated by the nonperturbative region of $b_T$ for 
$Q \sim 1.0\, {\rm GeV}$ to $\sim 2.0\, {\rm GeV}$.
\begin{figure*}
  \centering
  \begin{tabular}{c@{\hspace*{4mm}}c}
    \includegraphics[scale=0.67]{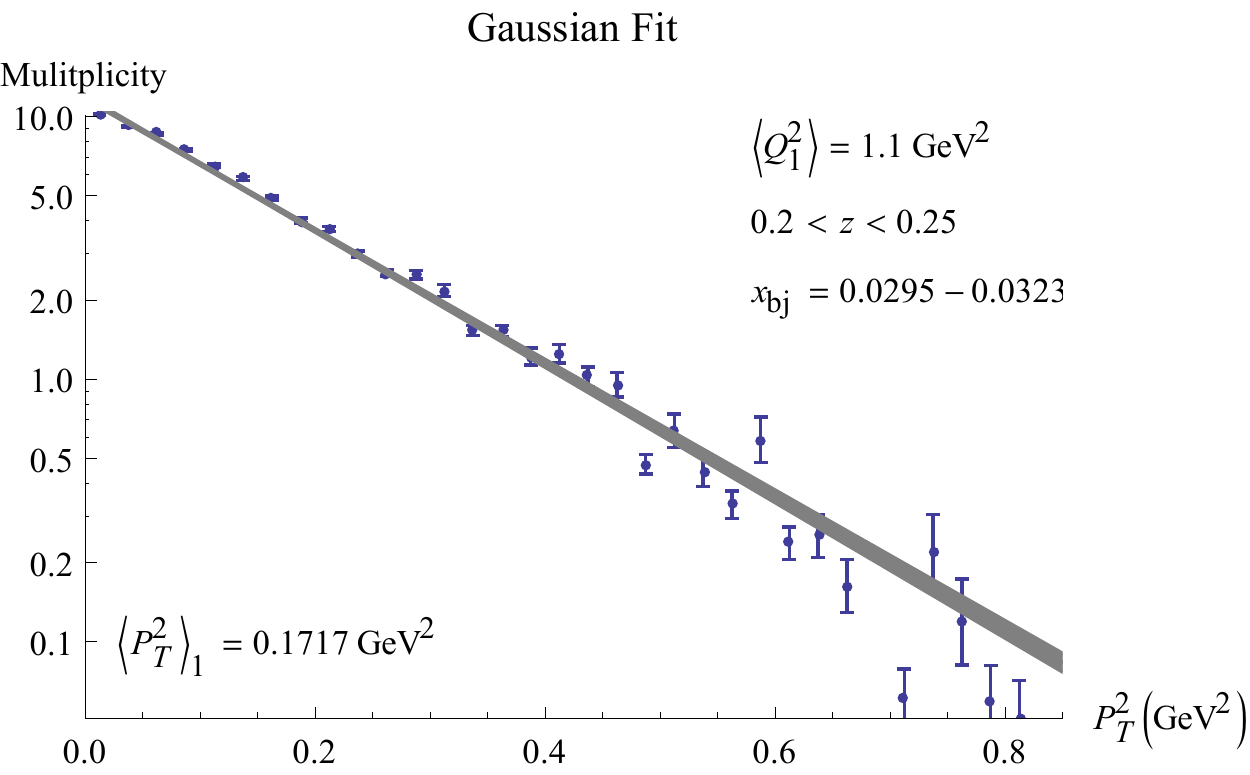} &     \includegraphics[scale=0.67]{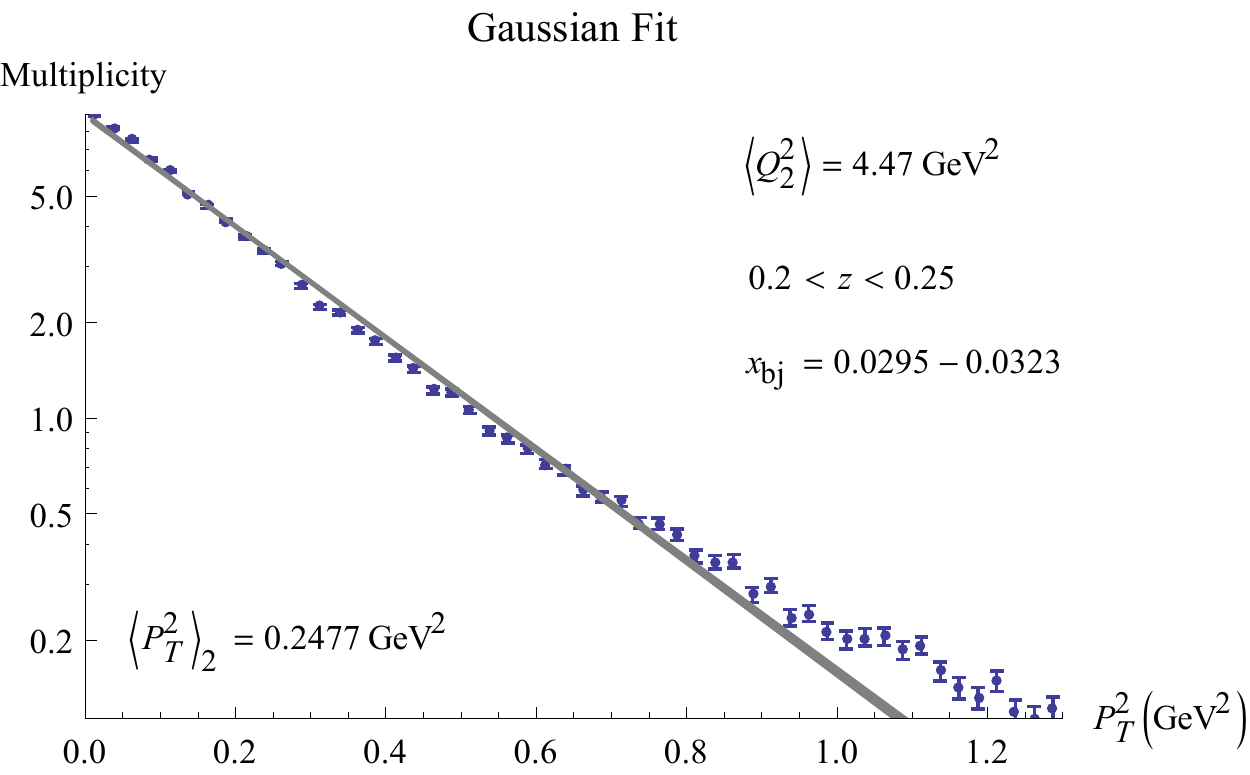}
    \\
    (a) &     (b)
\\[7mm]
    \includegraphics[scale=0.67]{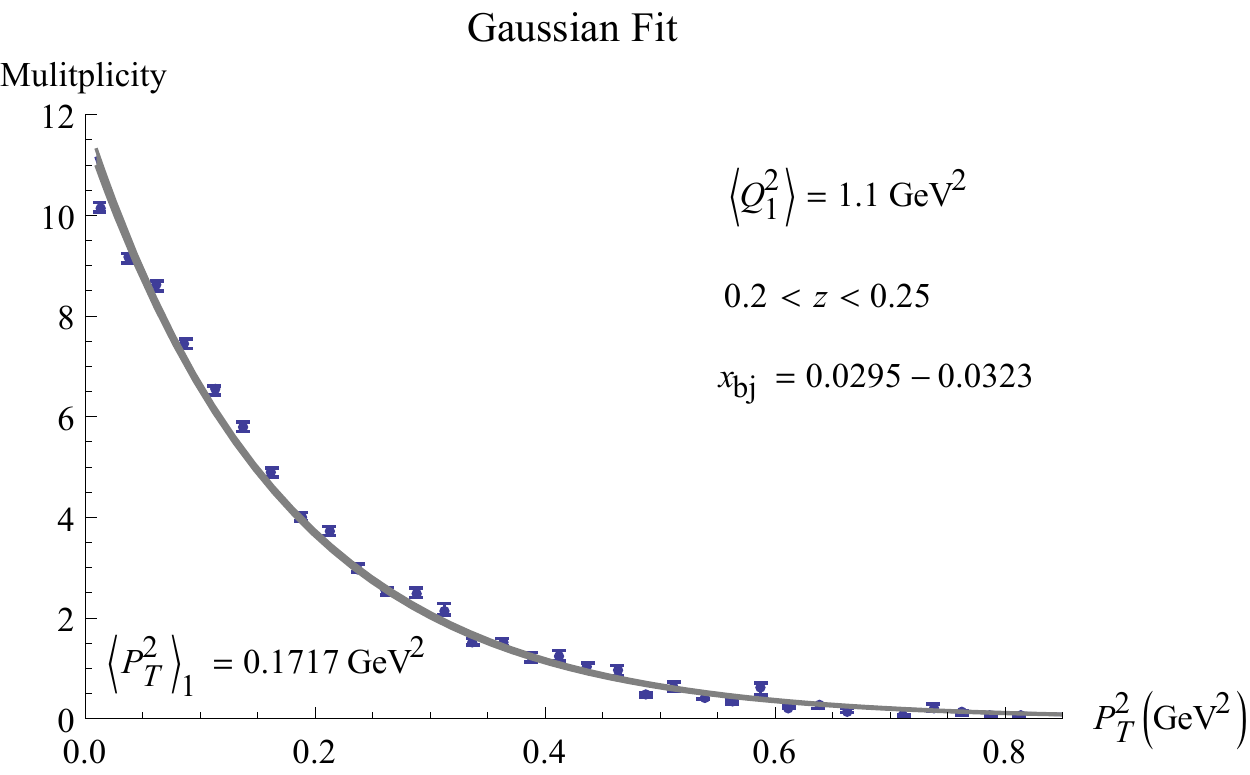} &     \includegraphics[scale=0.67]{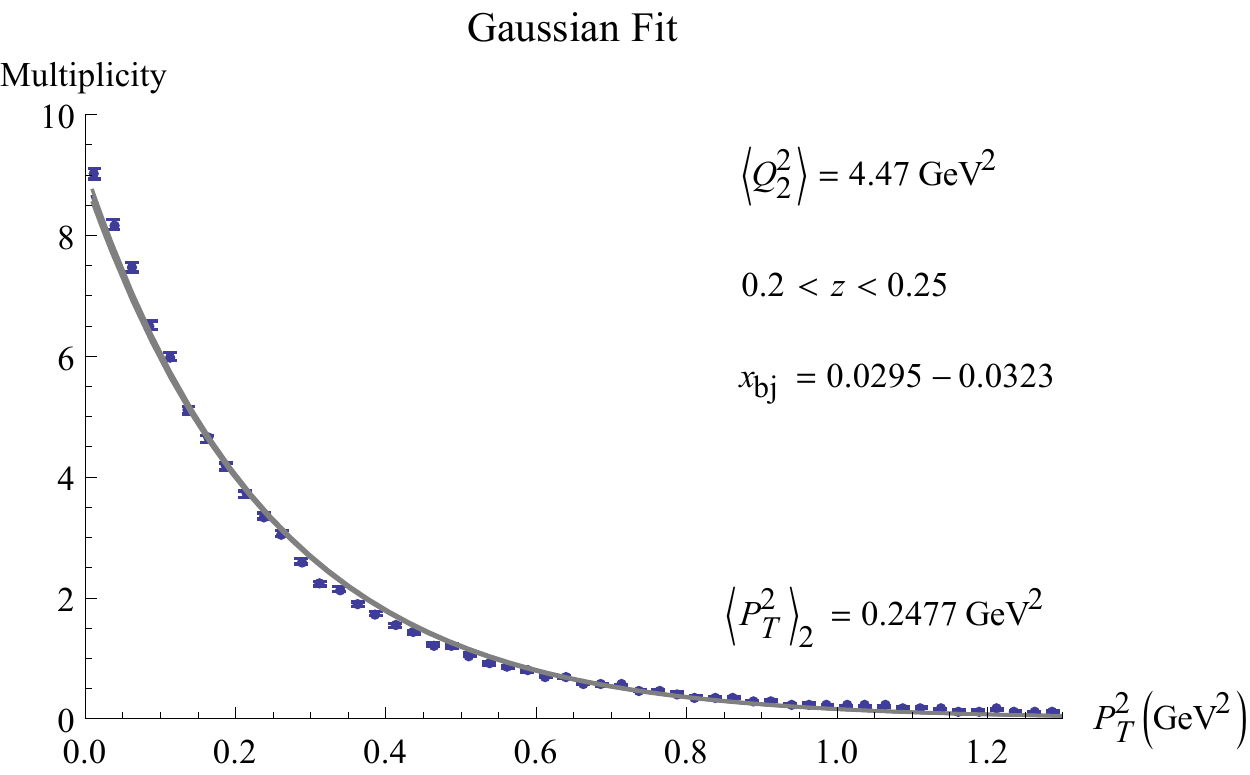}
    \\
    (c)  &  (d)
    \\[7mm]
  \end{tabular}
  \caption{(a) Gaussian fit for $Q = 1.049$~GeV, all $P_T$. (b) Gaussian fit for $Q = 2.114$~GeV, all $P_T$. (c, d) Same as (a, b) but on a linear axis. The gray band represents a 99\% confidence band for the fit parameters, where only the reported statistical errors have been included. (See online for color.)}
  \label{fig:bdists2}
\end{figure*}
\begin{figure}[t]
\centering
\includegraphics[scale=0.45]{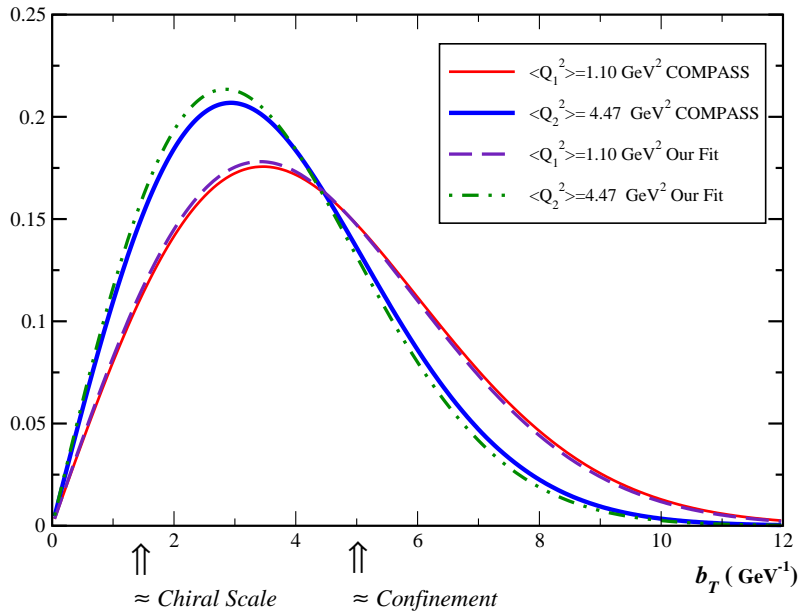}
\caption{
Gaussian fits again showing the largest variation in the width found in Tables~\ref{table:g2valuespos},~\ref{table:g2valuesneg}.  
The solid red and thick blue curves are the same 
as those in Fig.~\ref{fig:bdists}, in which the fit is restricted to the region of 
$P_T \leq 0.85$~GeV.  
The purple dashed and green dot-dashed curves are from the refit Gaussian curves in Fig.~\ref{fig:bdists2} that use all $P_T$ and correspond to Eq.~\eqref{eq:maxfits}
with the initial and final $\langle P_T^2 \rangle$ from Eq.~\eqref{eq:ourfits}. (See online for color.)}
\label{fig:theorybdistsa}
\end{figure}
\begin{figure*}
\centering
  \begin{tabular}{c}
    \includegraphics[scale=0.85]{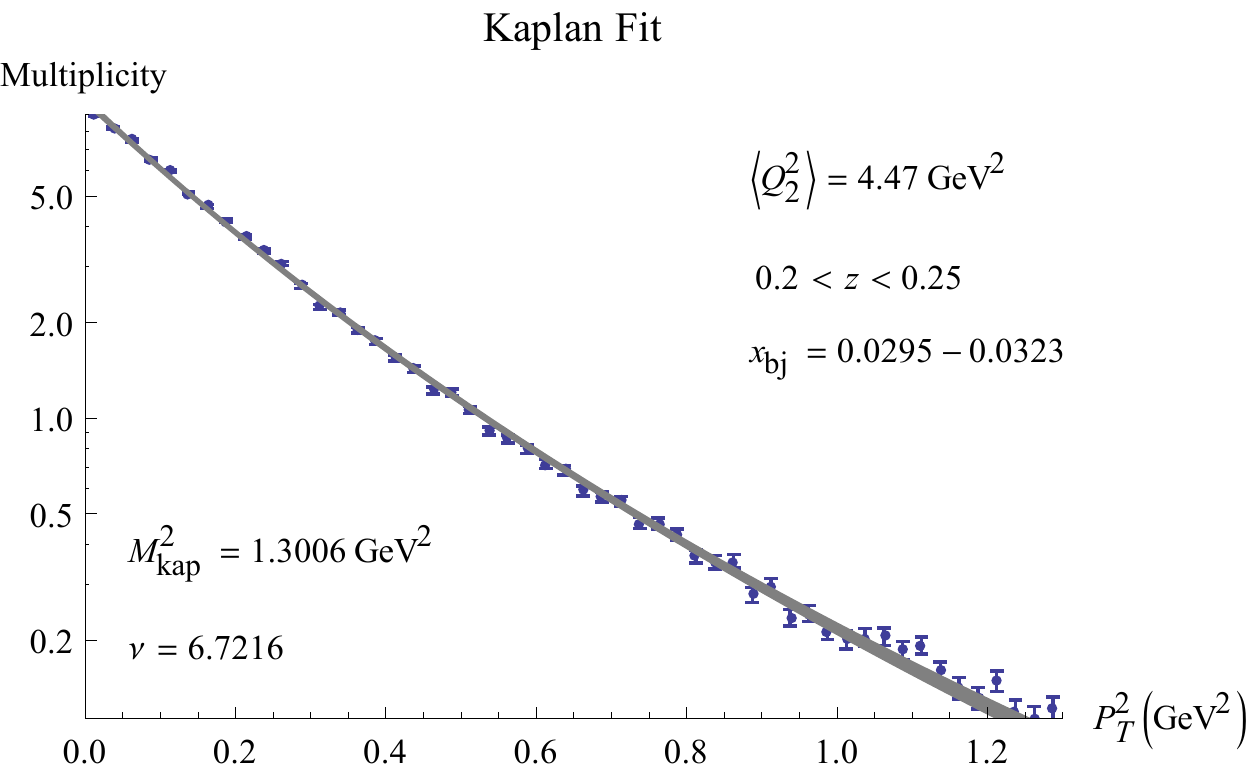}
    \\
    (a)
    \\
    \includegraphics[scale=0.85]{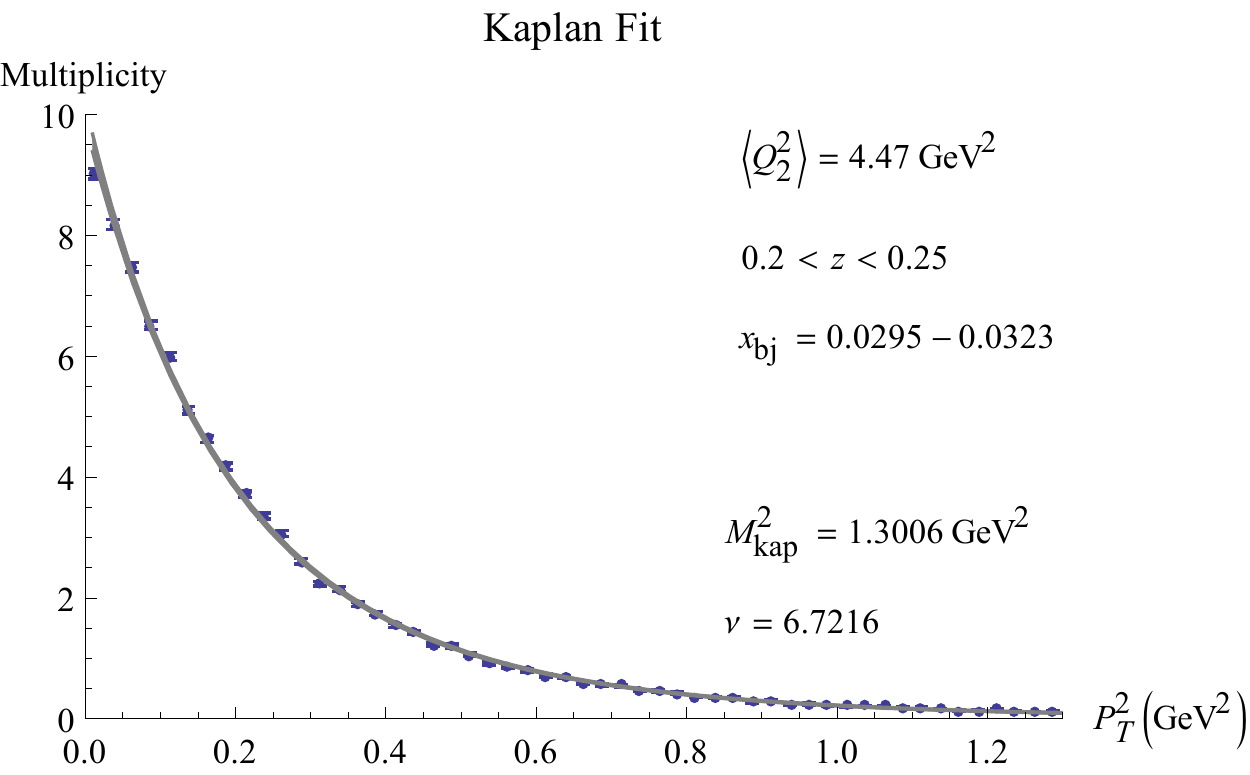}
    \\
    (b)
    \\[5mm]
   \end{tabular}
\caption{Fits of the Kaplan function, Eq.~\eqref{eq:kaplan}, for  $Q = 2.114$~GeV and for all $P_T$ with (a) a logarithmic plot and (b) 
a linear plot. The gray band represents a 99\% confidence band for the fit parameters, where only the reported statistical errors have been included. (Color online.)}
\label{fig:kaplanplots}
\end{figure*}

The robustness of this conclusion might be questioned on 
the grounds that the fits from~\cite{SIDISdata} apply to a restricted range, $P_T < 0.85$~GeV.  One could
speculate that including more of the large $P_T$ tail might result in an enhanced 
relative contribution from small $b_T$.
To address this, we have performed our own fit of the Gaussian form using the 
same data from Ref.~\cite{SIDISdata} that gave the two curves 
for $Q = 1.049$~GeV and $Q = 2.114$~GeV in Fig.~\ref{fig:bdists}, 
but now for the entire range of $P_T$ (up to $P_T \gtrsim 1.0$~GeV).\footnote{An accurate description of this large $P_T$ region requires the $Y$-term 
rather than a fit based entirely on the TMD terms.  However, fitting the TMD functions using the full range of $P_T$ is a useful test 
of the sensitivity of our 
general conclusions about relevant ranges of $b_T$ to the treatment of the $P_T$ tail within fits.}  
We perform the fitting in Wolfram Mathematica.
The new Gaussian fits are shown in Fig.~\ref{fig:bdists2}.  From the plot, 
it is clear that the values we find for the Gaussian slopes, $\langle P_T^2 \rangle_{Q_1 = 1.049 \, {\rm GeV}}$ and 
$\langle P_T^2 \rangle_{Q_2 = 2.114 \, {\rm GeV}}$, are so close 
to the COMPASS values that the curves in Fig.~\ref{fig:bdists} are nearly unchanged, despite the 
inclusion of larger $P_T$.  Instead of Eq.~\eqref{eq:compassfits}, we find: 
\begin{equation}
\label{eq:ourfits}
\langle P_T^2 \rangle_{Q_1 = 1.049 \, {\rm GeV}}^{\rm New \; Fits} = 0.1717 \pm 0.0011 \, {\rm GeV}^2\,; 
\qquad \langle P_T^2 \rangle_{Q_2 = 2.114 \, {\rm GeV}}^{\rm New \; Fits} = 0.2477 \pm 0.0008 \,  {\rm GeV}^2 \, ,
\end{equation}
where again the uncertainties are statistical uncertainties from the fit only.  The difference between 
the COMPASS fits in Eq.~\eqref{eq:compassfits} and our fits in Eq.~\eqref{eq:ourfits}
for $Q_1 = 1.049$~GeV is $0.0048$~GeV$^2$ and for $Q_2 = 2.114$~GeV it is $0.0152$~GeV$^2$.  This difference 
gives a sense of the 
systematic uncertainty due to the upper cutoff on $P_T$. 
Note that this uncertainty is of order the values of $C_{\rm evol}$ found in Tables~\ref{table:g2valuespos},~\ref{table:g2valuesneg} and Fig.~\ref{fig:linearplots}, 
suggesting that the precise value of $C_{\rm evol}$ 
has significant sensitivity to the way the large $P_T$ region is cutoff.  

To see how the new fits affect the coordinate space distribution, Eq.~\eqref{eq:maxfits}, we
have replotted  in Fig.~\ref{fig:theorybdistsa} the original curves from Fig.~\ref{fig:bdists} along with the curves using the new parameters
in Eq.~\eqref{eq:ourfits}.
It is clear that neglecting the large $P_T$ values has little influence on the general 
features of the fits discussed above; namely, that there is a large contribution from intervals of $b_T$ 
deep in the nonperturbative region.

A further critique could be made regarding the use of a Gaussian form on the grounds that
analyticity considerations~\cite{Schweitzer:2012hh} imply a power law fall-off for the large $P_T$ behavior 
of TMD correlation functions.  
Moreover, a power law behavior
$1/P_T^2$ (up to logarithmic corrections and the effects of evolution
of collinear PDFs) is a prediction of pQCD (see, for example, Ref.~\cite{Bacchetta:2008xw}). This power law
behavior is tied to singular behavior in the transverse position at small $b_T$.\footnote{The true large $P_T$ behavior of the TMD functions is not
directly meaningful at very large $P_T$, since TMD factorization (without the $Y$ term) is
inapplicable once the $P_T$ is comparable with $Q$. Clearly, the $Y$-term will be need be incorporated in the future to deal with these issues.}
Figure~\ref{fig:bdists2}(b) shows that the Gaussian form does have
some slight difficulty accounting for the full range of $P_T$ for the larger $Q_2 = 2.114 \, {\rm GeV}$ value.  
To address this, we have again refitted the $Q_2 = 2.114 \, {\rm GeV}$ data
but instead of Eq.~\eqref{eq:bspacedata}, we have used a Kaplan functional form: 
\begin{equation}
\label{eq:kaplan}
\frac{d \sigma}{d P_T^2} \propto \frac{1}{\left(1 + \frac{P_T^2}{M_{\rm kap}^2} \right)^{\nu}} \, .
\end{equation}
The result, shown in Fig.~\ref{fig:kaplanplots}, gives a slightly more successful fit than the Gaussian 
fit of Fig.~\ref{fig:bdists2}(b). When switching from the Gaussian fit to the Kaplan fit it is possible to quantify the goodness of the two fits. 
We use a straightforward coefficient of determination, $R^2$, which is defined in the usual 
way~\cite{probability} as $1-\text{SS}_\text{res}/\text{SS}_\text{T}$, where $\text{SS}_\text{res}$ is the residual sum of squares of each data point 
and the fit and $\text{SS}_\text{T}$ is the total sum of squares. This coefficient is a simple measure of the goodness of the fit that approaches unity 
for a perfect fit. In this case, the $R^2$ fit parameter rises modestly from $0.9918$ to $0.9988$ when moving from the Gaussian form to the Kaplan fit. 
The final Kaplan fit parameters are $M_{\rm kap}^2 = 1.3006$~GeV$^{2}$ and $\nu = 6.7216$.

For the lower value of $Q$, $Q = 1.049$~GeV, the Gaussian form actually gives a better fit than the Kaplan form.
Indeed, from Fig.~\ref{fig:bdists2}(a) it can be seen that even the Gaussian fit 
tends to overshoot the data slightly at large $P_T$.  
This could be due to the role of resonances at very small $Q$.

As with the Gaussian form, we may examine the Kaplan fit in coordinate space.  Instead of 
Eq.~\eqref{eq:maxfits} we have
\begin{equation}
\label{eq:kaplanbdist}
\frac{2 b_T^{\nu}  
M_{\rm kap}}{\Gamma(\nu)} \left( \frac{M_{\rm kap}}{2} \right)^{\nu} \; K_{1 - \nu}\left( b_T M_{\rm kap} \right) \, ,
\end{equation}
where $K_{1 - \nu}$ is the order $1 - \nu$ modified Bessel function of the second kind.
Again, we have imposed in Eq.~\eqref{eq:kaplanbdist} the normalization condition that the 
integration $\int_0^{\infty} \, d b_T$ is unity.
\begin{figure}[t]
\centering
\includegraphics[scale=0.4]{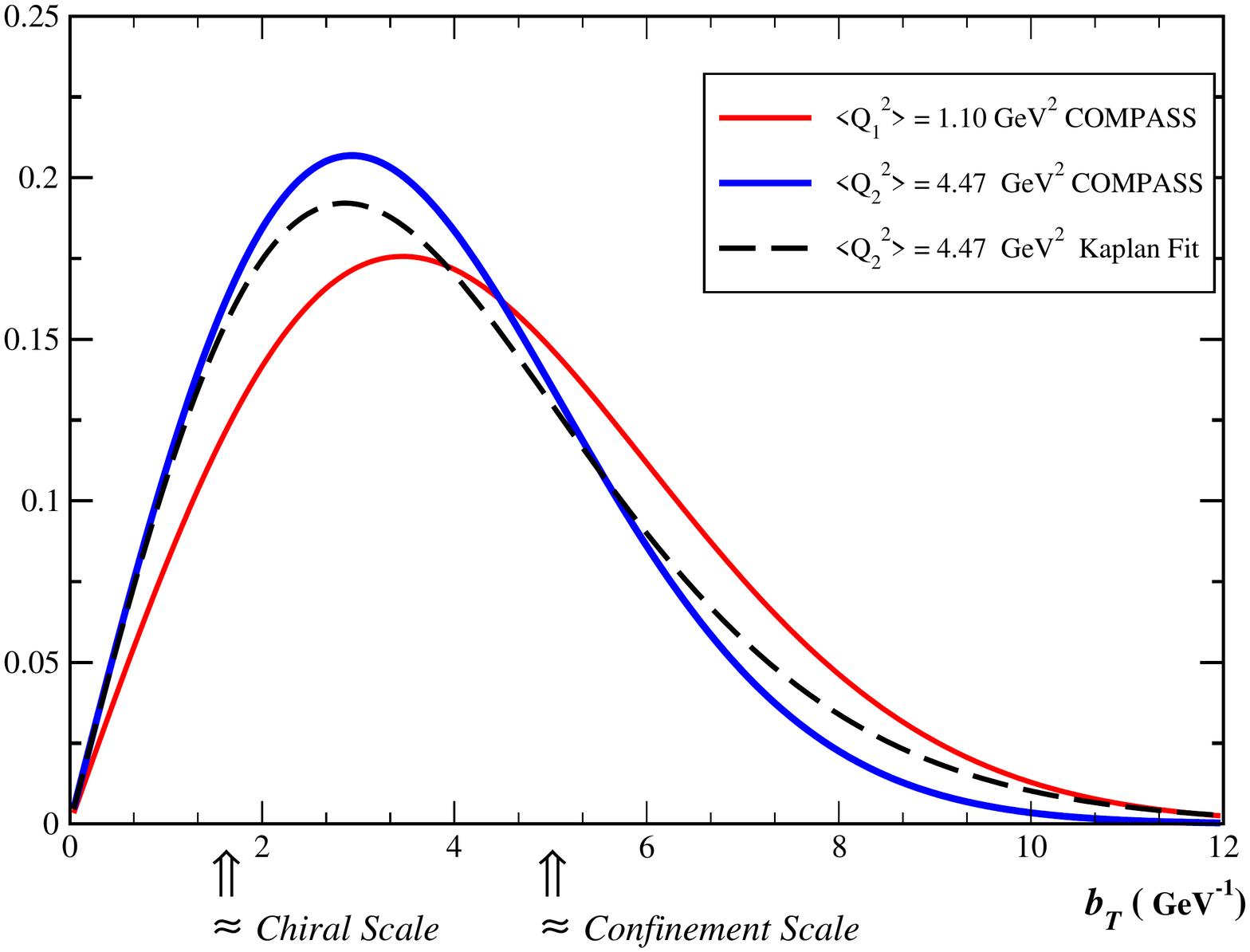}
\caption{
The black dashed curve shows the $b_T$ space function in Eq.~\eqref{eq:kaplanbdist} for $Q_2 = 2.114 \, {\rm GeV}$. 
This corresponds to the fit obtained in transverse momentum space using the Kaplan function in Eq.~\eqref{eq:kaplan}.
The fits themselves are shown in Figs.~\ref{fig:kaplanplots} and yield parameters $M_{\rm kap}^2 = 1.3006$~GeV$^{2}$ and $\nu = 6.7216$.
(See text for discussion.) For easy comparison, we have again included the solid red and thick blue curves from Fig.~\ref{fig:bdists}, 
corresponding to the original fits obtained by the COMPASS collaboration at $\sqrt{\langle Q_1^2 \rangle} = 1.049$~GeV 
and $\sqrt{\langle Q_2^2 \rangle} = 2.114$~GeV, respectively. (Color online.)}
\label{fig:theorybdistsb}
\end{figure}

In coordinate space, the
difference between the Gaussian and the Kaplan fits can be examined by comparing Eq.~\eqref{eq:kaplanbdist} and Eq.~\eqref{eq:maxfits} with 
the fit parameters corresponding to $Q_2 = 2.114 \, {\rm GeV}$.   The result is  
shown in Fig.~\ref{fig:theorybdistsb}.  Again, the original COMPASS fits from Fig.~\ref{fig:bdists} are shown as 
the solid red ($Q = 1.049$~GeV) and blue ($Q_2 = 2.114$~GeV) curves. 
From  Fig.~\ref{fig:theorybdistsb}, it can be seen that an analysis of the 
important regions of $b_T$ leads to roughly the same conclusions as in the case of the Gaussian fit. 
We conclude that the general observation of this section -- that  regions of $b_T$ deep into 
the nonperturbative regime are significant --
is robust for $P_T \to 0$ and for $Q \sim 1$~GeV to $\sim 2$~GeV, regardless of which 
functional form is used. 

\section{Comparison with TMD Evolution}
\label{sec:pertcons}

\subsection{Standard Evolution}

Next, we examine 
the evolved formula in Eq.~\eqref{eq:evolution} to estimate how well
it matches the change in widths of the Gaussian fits observed in Fig.~\ref{fig:linearplots} under 
different assumptions for $g_K(b_T;\bma)$.  
Let us consider the coordinate space factor in Eq.~\eqref{eq:evolution} of the TMD term, including 
an overall factor of $b_T$ in analogy with Eq.~\eqref{eq:maxfits}:
\begin{align}
    \frac{b_T}{N(Q)} \exp \left\{  \vphantom{ \ln \left( \frac{Q}{Q_0}\right)} \right. & \left. 
         - g_{\rm PDF} (x,b_T;\bma) 
         - g_{\rm FF} (z,b_T;\bma) 
         - 2 g_K(b_T;\bma) \ln \left( \frac{Q}{Q_0}\right)  \right.  \nonumber \\
  & + \left. 2 \ln \left( \frac{Q}{\mu_b} \right) \tilde{K}(b_{\ast};\mu_b) 
         +  \int_{\mu_b}^Q \frac{d \mu^\prime}{\mu^\prime} \left[ \gamma_{\rm PDF}(\alpha_s(\mu^\prime);1) 
         + \gamma_{\rm FF}(\alpha_s(\mu^\prime);1)
                  - 2 \ln \left( \frac{Q}{\mu^\prime} \right) \gamma_K(\alpha_s(\mu^\prime)) \right]\right\} \, .  \label{eq:evolution2}
\end{align}
$N(Q)$ is defined to be the integral $\int_0^{\infty} \, d b_T$ of the numerator, so that the full quantity is 
normalized to unity when integrating over $b_T$. We will require that for $Q = Q_0 = 1.049$~GeV, Eq.~\eqref{eq:evolution2} reduces to 
the $Q = 1.049$~GeV COMPASS Gaussian fit shown in Fig.~\ref{fig:bdists}.  That is, the input distributions are
\begin{align}
& - g_{\rm PDF} (x,b_T;\bma)  -  g_{\rm FF} (z,b_T;\bma) \nonumber \\
\nonumber \\
& \; = - \frac{b_T^2 \langle P_T^2 \rangle_{Q_0} }{4} - 2 \ln \left( \frac{Q_0}{\mu_b} \right) \tilde{K}(b_{\ast};\mu_b) 
         -  \int_{\mu_b}^{Q_0} \frac{d \mu^\prime}{\mu^\prime} \left[ \gamma_{\rm PDF}(\alpha_s(\mu^\prime);1) 
         + \gamma_{\rm FF}(\alpha_s(\mu^\prime);1)
                  - 2 \ln \left( \frac{Q_0}{\mu^\prime} \right) \gamma_K(\alpha_s(\mu^\prime)) \right] \, . \label{eq:inputdists}
\end{align}
With this choice for $- g_{\rm PDF} (x,b_T;\bma)  -  g_{\rm FF} (z,b_T;\bma)$, Eq.~\eqref{eq:evolution2} reduces exactly to
Eq.~\eqref{eq:maxfits} at $Q = Q_0$.\footnote{Recall, however, the note of caution immediately following Eq.~\eqref{eq:ganzi}.}

We use the one-loop \MSbar{} expressions for 
the anomalous dimensions with $C_1 = 2 e^{-\gamma_E}$, which are included in App.~\ref{sec:anondim}
 for reference. 
We use the approximation $\alpha_s(\mu) = 1 / 2 \beta_0 \ln(\mu/\Lambda_{\rm QCD})$ for the running coupling 
with 3 flavors and $\Lambda_{\rm QCD} = 0.2123$~GeV. (See App.~\ref{sec:runningcoupling} for more discussion 
of $\alpha_s(\mu)$ and the choice of $\Lambda_{\rm QCD}$.)
Then, the integrals in the one loop anomalous dimensions may be 
straightforwardly evaluated to obtain analytic expressions for all perturbative 
parts of the exponent in Eq.~\eqref{eq:evolution2}.  The 
explicit expression is given in App.~\ref{sec:evofacts}. 

\begin{figure*}[t]
\centering
  \begin{tabular}{c@{\hspace*{-10mm}}c}
\includegraphics[scale=0.35]{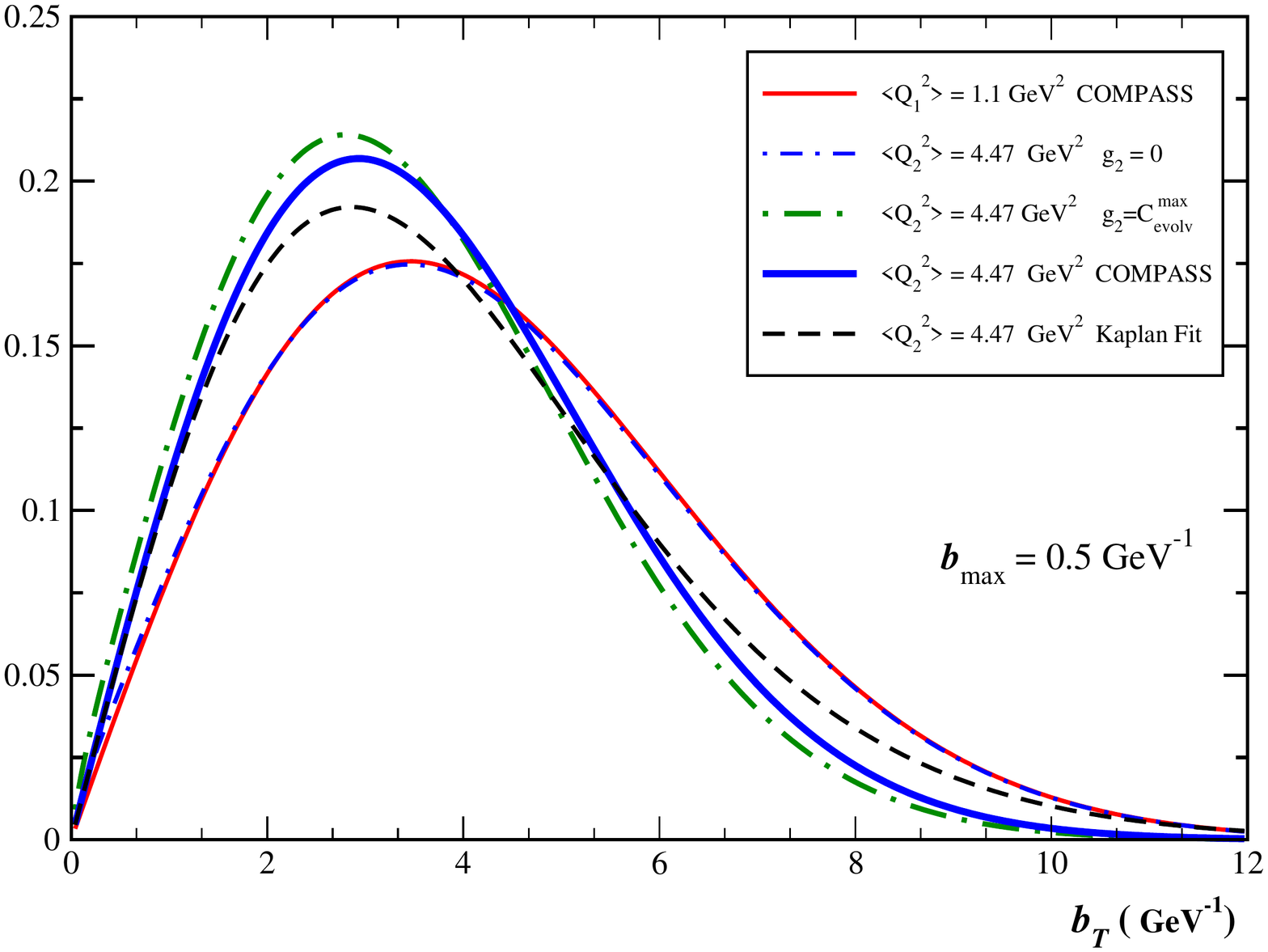}  & \includegraphics[scale=0.35]{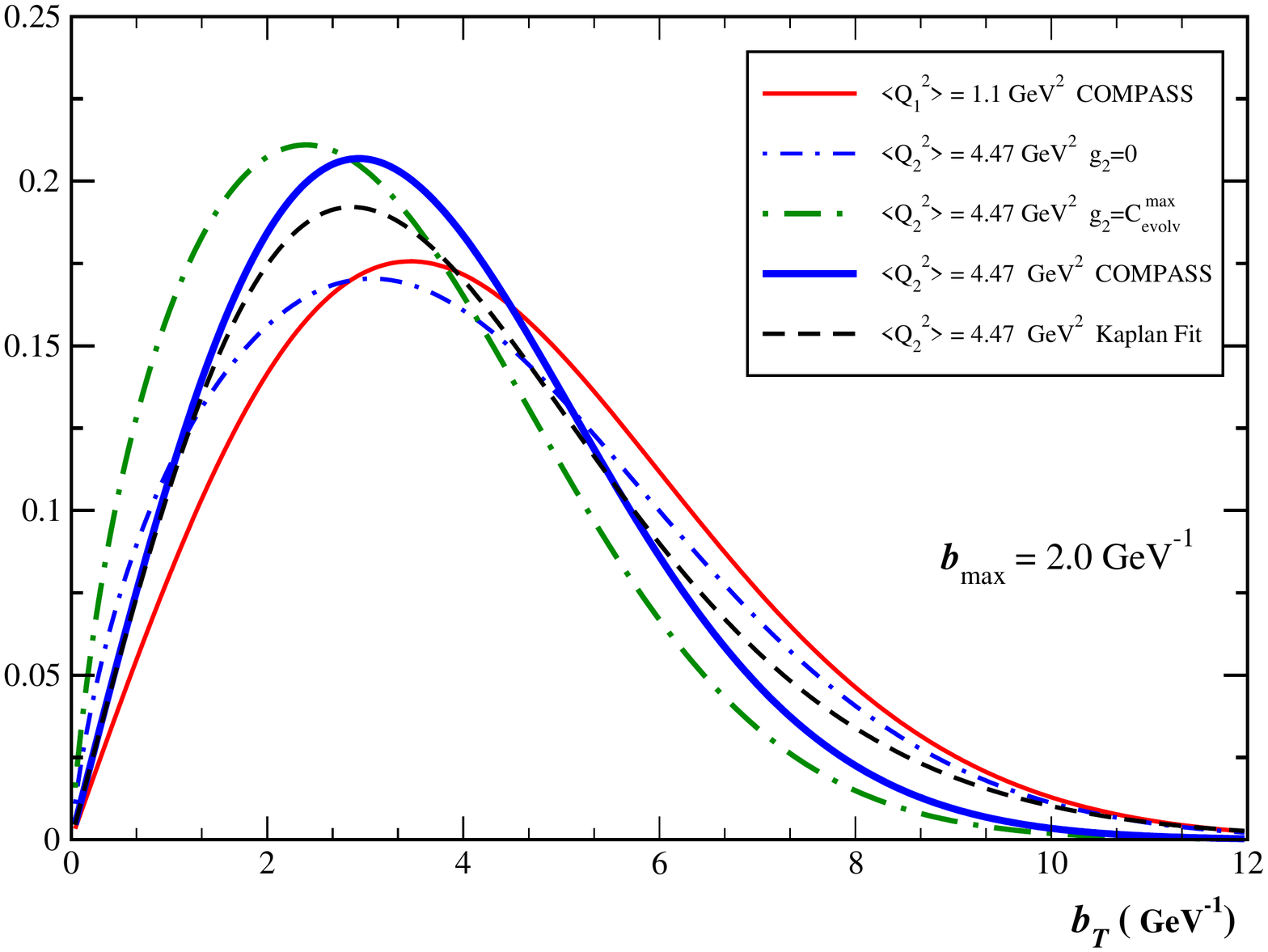}
\\
    (a) &     (b)
\\[7mm]
\includegraphics[scale=0.35]{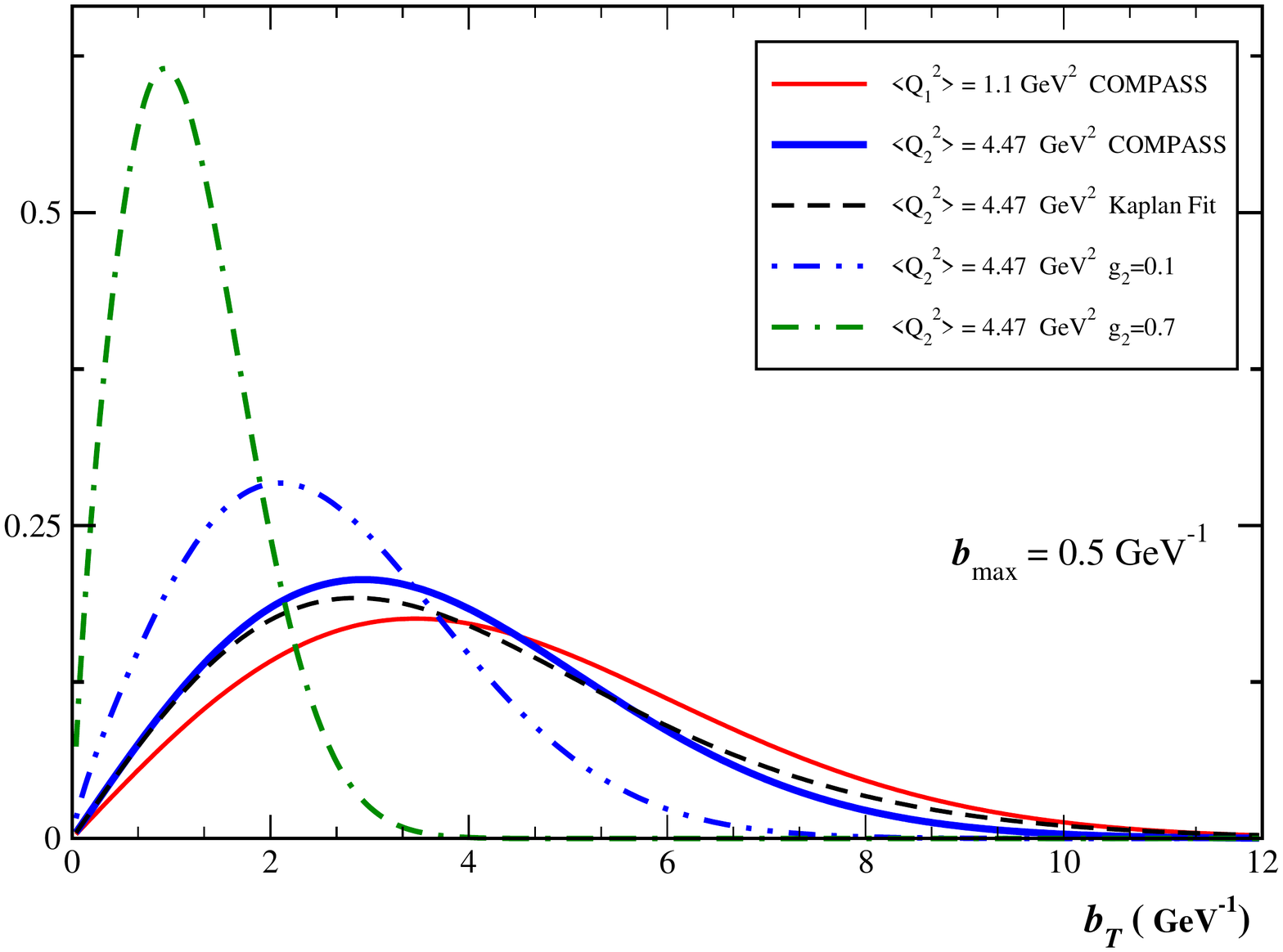} &  \includegraphics[scale=0.35]{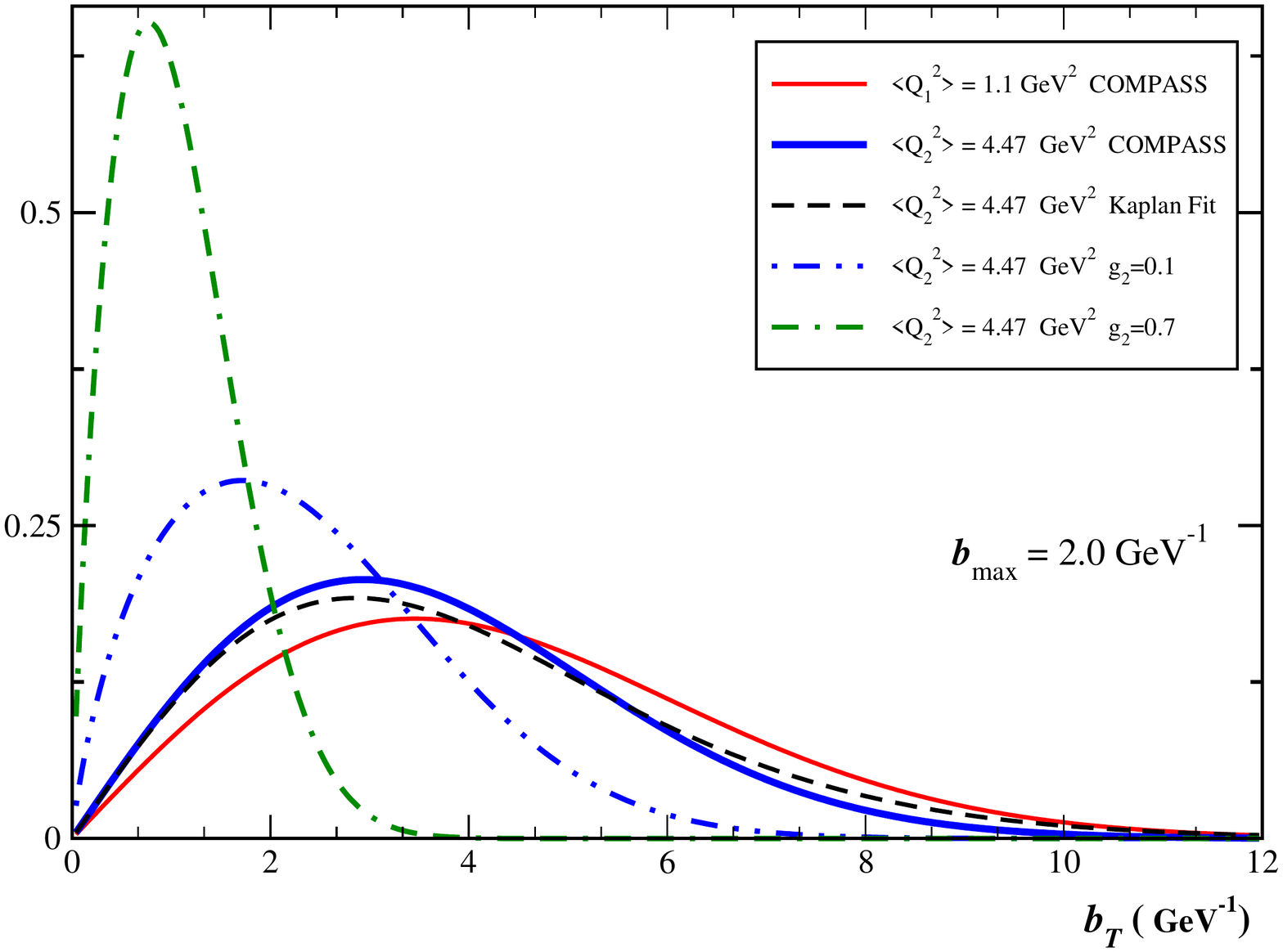}
\\
  (c)  &  (d)
    \\[7mm]
  \end{tabular}
\caption{
\underline{ Left Panels} (a) and (c): 
The solid red and thick blue lines (see online for color) are the same initial 
and final Gaussian fits obtained by COMPASS 
as in Fig.~\ref{fig:bdists} 
for $Q_1^2 = 1.1$~GeV$^2$ and $Q_2^2 = 4.47$~GeV$^2$ respectively.  
The black dashed curve is the Kaplan fit for $Q^2 = 4.47$~GeV$^2$, already shown in 
Fig.~\ref{fig:theorybdistsb}.
The dot-dashed lines are the TMD factorization expression in 
Eq.~\eqref{eq:evolution2} for 
the evolution to $Q_2^2 = 4.47$~GeV$^2$ with the 
Gaussian ansatz from Eq.~\eqref{eq:gaussform} for $g_K(b_T;\bma)$ 
with $b_{\rm max} = 0.5$~GeV$^{-1}$.  
The positions of the peaks of the 
evolved distributions decrease with increasing $g_2$:  Figure (a) shows
the results for $g_2 = 0$ (blue dot-dashed) 
and $C_{\rm evol}^{\rm max} = 0.0306 \, {\rm GeV}^{2}$ (green  dot-dashed); 
Figure  (c) shows the result for  $g_2 = 0.1 \, {\rm GeV}^{2}$ 
(blue dot-dashed) and $g_2 = 0.7 \, {\rm GeV}^{2}$ 
(green dot-dashed).  
All curves are normalized to one in the integration over $b_T$.
\underline{Right Panels} (b) and (d):  Same as the left panels, 
but for $b_{\rm max} = 2.0$~GeV$^{-1}$.}
\label{fig:theorybdists3}
\end{figure*}

For $g_K(b_T;\bma)$, we start by using Eq.~\eqref{eq:gaussform}, with a conservative 
$\bma = 0.5$~GeV$^{-1}$ and several sample values of $g_2(\bma)$.  
We compare with the 
maximum observed rate of evolution seen in the COMPASS data --  the curves 
already shown in Fig.~\ref{fig:bdists}.  
The results are shown in 
Fig.~\ref{fig:theorybdists3}(a) and (c), where the dot-dashed curves show the evolution to  $Q^2 = 4.47$~GeV$^2$ for a 
range of sample values for $g_2$.  
There is ambiguity as to which values of $Q_0$ and $Q$ should 
be used in the evolution, given the differences between $Q^{\rm min}$, $Q^{\rm max}$ and $\sqrt{\langle Q^2 \rangle}$ 
in Tables~\ref{table:g2valuespos},~\ref{table:g2valuesneg}.  
In order to estimate roughly the approximate 
size of evolution effects, we will continue to use $\sqrt{\langle Q^2 \rangle}$ for the the initial and final values of $Q$.

We begin with $g_2 = 0$ and see essentially no effect on the $b_T$ distribution when $Q$ is varied; the integrand is  
small in the region of small $b_T$ where perturbative evolution would be substantial, and setting $g_2 = 0$ suppresses 
any nonperturbative contribution to evolution.   
Next, we consider $g_2 = C_{\rm evol}$, 
with the maximum value of $C_{\rm evol} = 0.0306$~GeV$^2$ 
found in Tables~\ref{table:g2valuespos},~\ref{table:g2valuesneg}.  Finally, we consider $g_2 = 0.1$~GeV$^{2}$ and 
$g_2 = 0.7$~GeV$^{2}$ which are values more typical of fits 
obtained at large $Q$, as well as the renormalon analysis value 
of $g_2 = 0.19$~GeV$^2$ in Ref.~\cite{Tafat:2001in}. (See, also, Fig.~1 of Ref.~\cite{Konychev:2005iy}.)

We have repeated this exercise for the much more liberal value of $\bma = 2.0$~GeV$^{-1}$, and 
the result is shown in Figs.~\ref{fig:theorybdists3}  (b) and (d).  In Figs.~\ref{fig:theorybdists3}(a)-(d), a value of $g_2(\bma) \lesssim C_{\rm evol}^{\rm max}$ is
clearly preferred over values of $g_2(\bma) \geq 0.1$~GeV$^2$. 
Note that with $g_2 = 0$, there is very weak evolution in the $b_T$ shape relative to the variations in the width suggested 
by the COMPASS data in the small range of $Q$ values.
A choice of $g_2 = C_{\rm evol}^{\rm max} = 0.0306$~GeV$^2$ is roughly consistent with the 
upper limit on the rate of evolution observed in Tables~\ref{table:g2valuespos},~\ref{table:g2valuesneg} and Fig.~\ref{fig:linearplots}.
Thus, if we demand the ansatz in Eq.~\eqref{eq:gaussform} for the form of $g_K(b_T;\bma)$ for all $b_T$, then we estimate that the true value of $g_2$, at least 
for the kinematics of Tables~\ref{table:g2valuespos},~\ref{table:g2valuesneg}, 
must lie roughly in the range of $0 < g_2 \lesssim 0.03$~GeV$^2$.  

\subsection{Modified Large $b_T$ Behavior}
\label{sec:modlargebT}

Because of the strong universality of $g_K(b_T;\bma)$, the results of the last section 
seem on the surface  to indicate a discrepancy between 
the low $Q$ data and detailed and successful fits of the past that focus on larger $Q$, which tend to find 
$g_2 \gtrsim 0.1$~GeV$^2$~\cite{Landry:2002ix,Konychev:2005iy,Nadolsky:1999kb,Nadolsky:2000ky}.
For instance, values of $g_2$ have been found to be as large as $0.68$~GeV$^2$~\cite{Landry:2002ix}, and a 
value of $g_2 = 0.19$~GeV$^2$ is used in Ref.~\cite{Nadolsky:2000ky} for SIDIS in the CSS formalism, both using a 
value of $\bma = 0.5$~GeV$^{-1}$.
Moreover, the renormalon 
analysis of Ref.~\cite{Tafat:2001in} also suggests a $g_2$ of similar size for small $b_T$. (See, also, Fig.~1 of Ref.~\cite{Konychev:2005iy}.)
However, the quadratic ansatz in Eq.~\eqref{eq:gaussform}  
(which gives a Gaussian 
ansatz when it appears in the exponent of Eq.~\eqref{eq:evolution}) seems to impose excessive suppression 
of the very large nonperturbative $b_T$ region whenever $g_2 \gtrsim 0.1$~GeV$^2$. 
A critique of the purely Gaussian nonperturbative form was 
also given in Ref.~\cite{Qiu:2000hf}, where it was argued 
that the Gaussian form gives excessive sensitivity to nonperturbative input at large transverse momentum, 
and a power law, $\sim b_T^{0.3}$, with a $b_{\rm max} = 0.3$~GeV$^{-1}$ is suggested, though this is possibly 
an overly conservative choice, given our earlier discussion of $b_T$ regions in Fig.~\ref{fig:bdists}, and given 
that scales $\geq 3.0$~GeV are generally considered to be well within the perturbative 
region.  See related discussions of this in Ref.~\cite{Konychev:2005iy}. 
\begin{figure}[b]
\centering
\includegraphics[scale=0.4]{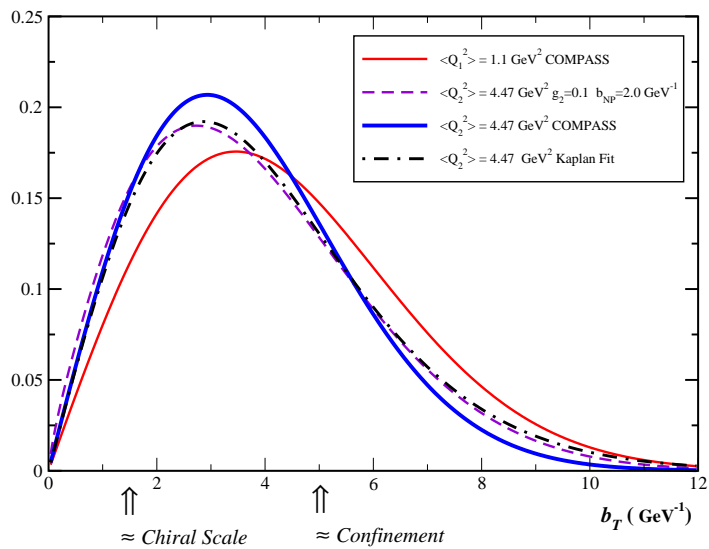}
\caption{
The solid red and thick blue curves are again the same initial and final Gaussian fits obtained by COMPASS 
for $Q^2 = 1.1$~GeV$^2$ and $Q^2 = 4.47$~GeV$^2$ respectively --
the same as in Fig.~\ref{fig:bdists} . (See online for color.) 
The black dot-dashed curve is again the Kaplan fit for $Q^2 = 4.47$~GeV$^2$ already shown in Fig.~\ref{fig:theorybdistsb}.
For comparison, the purple short-dashed curve is the TMD factorization expression in Eq.~\eqref{eq:evolution2}, but now
using Eq.~\eqref{eq:model} for $g_K(b_T;\bma = 0.5\, {\rm GeV}^{-1})$ with $b_{\rm NP} = 2.0$~GeV$^{-1}$ and $g_2 = 0.1$~GeV$^2$. 
This should be compared with the $g_2 \geq 0.1$~GeV$^{2}$ curves in Fig.~\ref{fig:theorybdists3} where the quadratic ansatz for $g_K(b_T;\bma)$ -- Eq.~\eqref{eq:gaussform} --  is used.
}
\label{fig:theorybdistlogmod}
\end{figure}

To resolve the apparent discrepancy discussed above,  we recall that large $Q$ fits, e.g. for $Q \gtrsim 10$~GeV, 
are sensitive mainly to the region of $b_T \lesssim 2.0$~GeV$^{-1}$.  See, for example, Fig.~4 of Ref.~\cite{Konychev:2005iy} and 
compare this with Fig.~\ref{fig:bdists}, where contributions from $b_T \gtrsim 2.0$~GeV$^{-1}$ dominate.
Now let us assume that nonperturbative effects become totally dominant at some large size scale $b_{\rm NP}$, 
where $g_K(b_T;\bma)$ acquires a more complicated and as-yet unknown precise form. Recall also that $g_K(b_T;\bma)$ 
is predicted to vanish as a power of $b_T^2$ at small $b_T$~\cite{Korchemsky:1994is,Tafat:2001in,Laenen:2000ij,Laenen:2000hs}.  
Thus, for $b_T \ll b_{\rm NP}$ the 
following expansion applies:
\begin{equation}
\label{eq:gexp}
g_K(b_T;\bma) = a_1 \left( \frac{b_T^2}{b_{\rm NP}^2} \right) + a_2 \left( \frac{b_T^4}{b_{\rm NP}^4} \right) + \cdots \, .
\end{equation} 
See also Eq.~(6.1) of Ref.~\cite{Tafat:2001in}.\footnote{Note, however, that Ref.~\cite{Tafat:2001in} predicts a linear rather than constant dependence at very large $b_T$.}
We conjecture that large $Q$ fits typically obtain a large $g_2$ because they 
are sensitive only to the first power-law correction in Eq.~\eqref{eq:gexp}.   
By contrast, at smaller $Q$ higher powers, and eventually the complete functional form, become important. 

We propose that the optimal way to proceed is to use a functional form for $g_K(b_T;\bma)$ 
that: a.) respects its strong universality set forth in TMD factorization by 
matching to earlier large $Q$ fits that use a Gaussian form but b.) avoids strong disagreement
with the results of the empirical analysis of SIDIS data from Sec.~\ref{sec:largebT}.  Thus, we impose the following conditions:
\begin{enumerate}
\renewcommand{\labelenumi}{(\roman{enumi})}
\renewcommand{\labelenumii}{\roman{enumi}.~\alph{enumii}}
\item At small $b_T^2$, the lowest order coefficient in Eq.~\eqref{eq:gexp}, i.e. $a_1 / b_{\rm NP}^2$, must be roughly $\gtrsim 0.1$~GeV$^2$ in order to 
be consistent with the values of $g_2 / 2$ found in Ref.~\cite{Landry:2002ix,Konychev:2005iy,Nadolsky:1999kb,Nadolsky:2000ky,Tafat:2001in}, thereby 
respecting the strong universality of $g_K(b_T;\bma)$. 
\item At $b_T \gg b_{\rm NP}$, $g_K(b_T;\bma)$ should become nearly constant, or at most logarithmic in $b_T$.
\end{enumerate}
As a simple example, we propose
\begin{equation}
\label{eq:model}
g_K(b_T;\bma) = \frac{g_2(\bma) b_{\rm NP}^2}{2} \ln \left( 1 + \frac{b_T^2}{b_{\rm NP}^2} \right) \, .
\end{equation}   
(See, also, Eq.~(6.14) of Ref.~\cite{Tafat:2001in}.)
Expanding around $b_T \ll b_{\rm NP}$ gives the first two terms,
\begin{equation}
\label{eq:expandedmod}
g_2(\bma) \frac{1}{2} b_T^2 - g_2(\bma) \frac{1}{4 b_{\rm NP}^2} b_T^4 + \cdots \, .
\end{equation}
In Fig.~\ref{fig:theorybdistlogmod} we illustrate 
how the 
low $Q$ dependence of the COMPASS data may be accommodated into earlier larger $Q$ fits by using the modified $g_K(b_T;\bma)$ 
from Eq.~\eqref{eq:model} with $\bma = 0.5$~GeV$^{-1}$, $g_2 = 0.1$~GeV$^2$ and $b_{\rm NP} = 2.0$~GeV$^{-1}$.\footnote{In general, $b_{\rm NP}$ may 
also be a function of $\bma$ but to simplify notation we do not show it explicitly in Eq.~\eqref{eq:model}.}  
Since the lowest order term in the expansion 
in Eq.~\eqref{eq:expandedmod} matches Eq.~\eqref{eq:gaussform} with $g_2 = \mathcal{O}(0.1~{\rm GeV}^2)$ and thus is generally consistent 
with earlier fits such as Ref.~\cite{Nadolsky:1999kb,Nadolsky:2000ky}.  In this way, moderate $Q$ data may be accommodated without introducing disagreement 
with important and universal nonperturbative contributions obtained in earlier fits, while simultaneously giving access to 
further universal nonperturbative information.

For now we propose Eq.~\eqref{eq:model} only as a simple example of how $g_K(b_T;\bma)$ might possibly 
be modified at very large $b_T$.  In practice, better and more detailed parametrizations may be needed, possibly obtainable from 
nonperturbative studies.

\section{Comparison Between Collins and Sun-Yuan Formalism}
\label{sec:sunyuan}

\begin{figure*}
  \centering
  \begin{tabular}{c@{\hspace*{1mm}}c}
    \includegraphics[scale=.33]{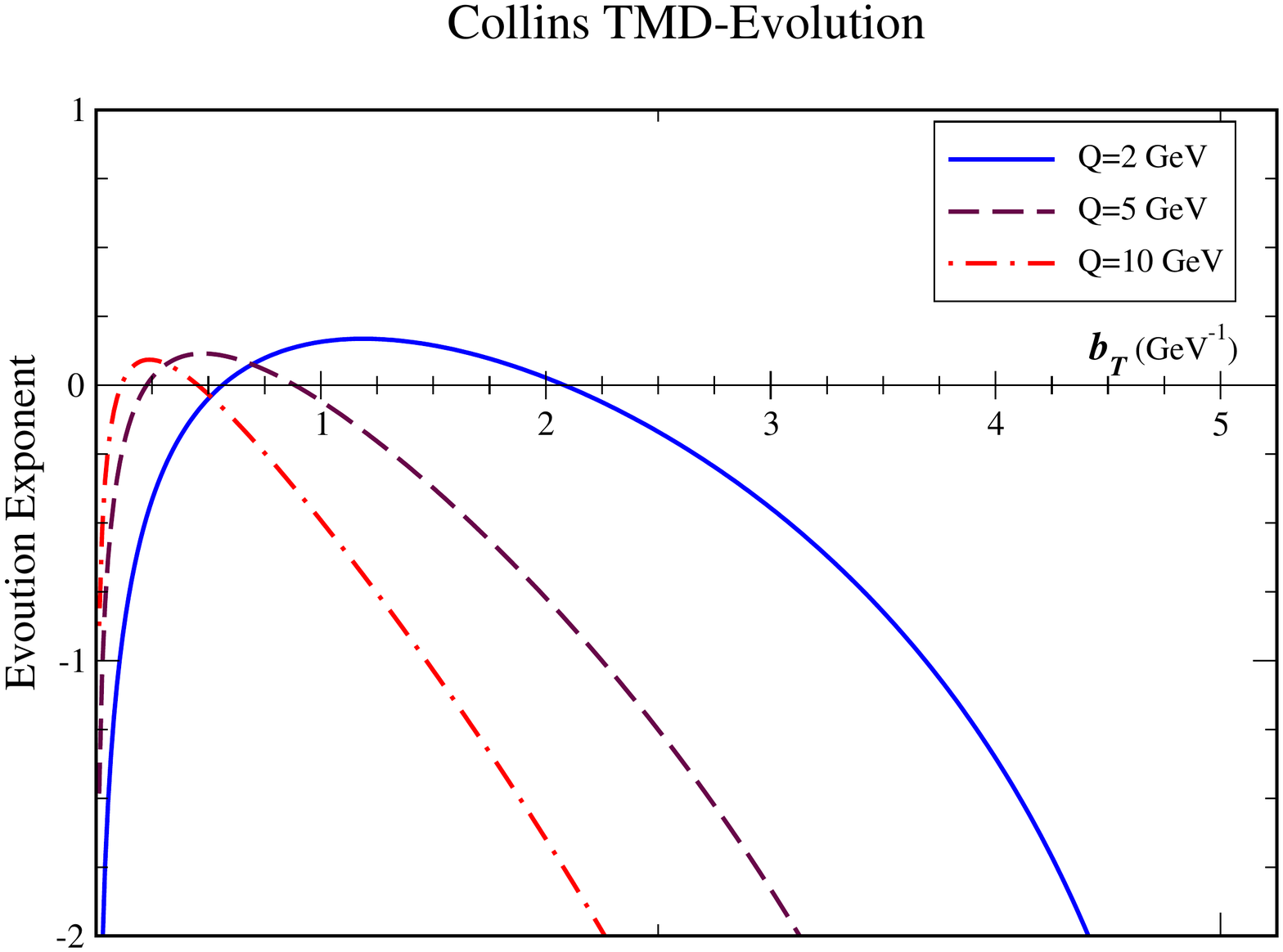}  &     \includegraphics[scale=.33]{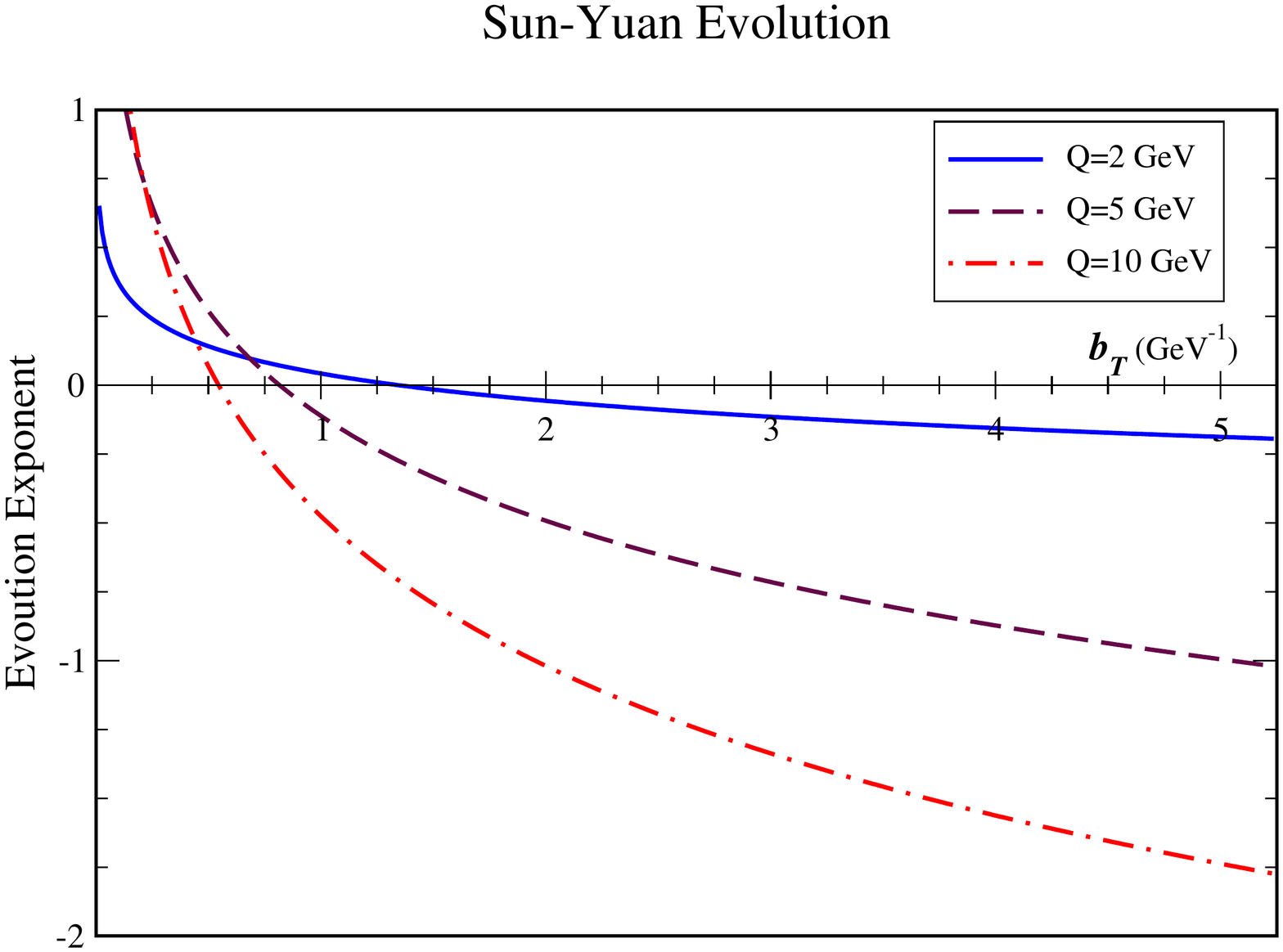}
 \\
 (a) & (b)
  \end{tabular}
  \caption{(color online). The $Q$ dependent terms in the perturbative parts of the exponents in (a) Eq.~\eqref{eq:evolution2} for the TMD factorization formalism 
  and (b) Eq.~\eqref{eq:evolution3} for the Sun-Yuan formalism.}
  \label{fig:bdists3}
\end{figure*}
\begin{figure}[t]
  \centering
    \includegraphics[scale=.35]{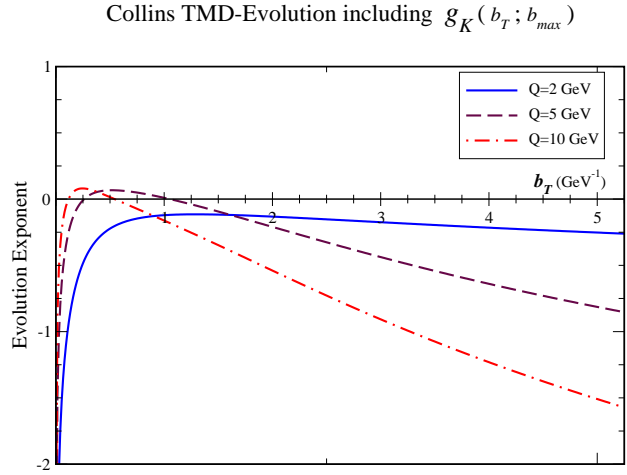}
  \caption{
  (color online). The $Q$ dependent terms in the Collins TMD factorization formalism  exponents from  Eq.~\eqref{eq:evolution2} using 
the large-$b_T$ function $g_K(b_T;\bma)$ in the form of Eq.~\eqref{eq:model} with $g_2 = 0.1$~GeV$^2$, $b_{\rm NP} = 2.0$~GeV$^{-1}$
and with $\bma=0.5\, {\rm GeV}^{-1}$.}
  \label{fig:bdists4}
\end{figure}

In Ref.~\cite{Sun:2013dya,Sun:2013hua}, Sun and Yuan argue that an alternative evolution factor 
should be adopted for the region of $Q \lesssim 10$~GeV.  
In their approach, the TMD PDFs are evolved relative to an arbitrary 
scale $Q_0 \sim 1.0$~GeV rather than the intrinsic hard scale $\sim 1/b_T$ of the TMD PDF.  Thus the Sun-Yuan 
formalism contains unresummed logarithms of $b_T$.
The Sun-Yuan form of evolution replaces the exponential factor in Eq.~\eqref{eq:evolution2} with, 
\begin{align}
    b_T \exp \left\{ - \frac{b_T^2 \langle P_T^2 \rangle_0 }{4} \right\} 
    \exp \left\{  -2 C_F \int_{Q_0}^Q \frac{d \mu^\prime}{\mu^\prime} \frac{\alpha_s(\mu^\prime)}{\pi} \left[ 
    \ln \left( \frac{Q^2}{{\mu^\prime}^2} \right) + \ln \left( \frac{Q_0^2 b_T^2}{C_1^2} \right) - \frac{3}{2}  \right]\right\} \, .
    \label{eq:evolution3}
\end{align}
See Eq.~(3) of Ref.~\cite{Sun:2013dya} and Eq.~(77) of Ref.~\cite{Sun:2013hua}. It is arrived at by 
extending a low order calculation of the $b_T$-dependence into the region of very large $b_T$.
From the point of view of doing practical calculations, there appears to be an advantage in that
there is no explicit Landau pole encountered in the evaluation of $\alpha_s(\mu)$ and thus, on the surface, no need
to include a nonperturbative component to the evolution. 
(See the discussion immediately before and after Eq.~(3) in Ref.~\cite{Sun:2013dya} for the rationale and motivation given
to use this form rather than the Collins TMD factorization or CSS form.)
By contrast, in the Collins TMD-factorization approach, the region
of smaller $Q$ is where the genuine nonperturbative $b_T$-dependence is understood
to become increasingly important, including in the evolution.  
The Collins TMD-factorization formalism includes a strategy of isolating and 
testing the strong universality of nonperturbative behavior at large $b_T$ while matching to an optimal 
perturbative treatment at $b_T \ll 1 / \Lambda_{\rm QCD}$.
Both the standard CSS formalism and the Collins TMD factorization formalism predict a greater input 
from nonperturbative evolution over low regions of $Q$.  Conversely, Sun-Yuan argue that the nonperturbative 
component is needed for evolution at large $Q$ but is negligible in the vicinity of small $Q$.  

While it is beyond the scope of this article to make a full comparison between these two approaches (see, however, 
Ref.~\cite{CRlargeb}), it is worthwhile to examine briefly 
whether the two formalisms are essentially equivalent ways of implementing the same evolution  
or whether they are in contradiction with one another.
To this end, we note that the purely perturbative contributions give rather different evolution exponents in the
region of $Q \lesssim 10$~GeV. This can be seen 
by directly computing the exponents in both Eq.~\eqref{eq:evolution2} and Eq.~\eqref{eq:evolution3} using the one-loop 
expressions for the anomalous dimensions and the running coupling.  The resulting 
analytic expressions for both the Collins TMD-factorization 
case and the Sun-Yuan case are given in App.~\ref{sec:evofacts}.
In Fig.~\ref{fig:bdists3}(a) we have plotted the exponent of the evolution factor in 
Eq.~\eqref{eq:evolution2}, keeping only the perturbative parts and dropping the $g_K(b_T;\bma)$.  
We have set $b_{\rm max}$ to infinity so that it is a purely perturbative expression with 
no explicit cutoff functions like Eq.~\eqref{eq:bdef}.
In Fig.~\ref{fig:bdists3}(b) we have plotted, for comparison, the Refs.~\cite{Sun:2013dya,Sun:2013hua} evolution 
exponent from Eq.~\eqref{eq:evolution3}. For $Q_0$ we use $Q_0 = \sqrt{2.4}$~GeV which is the value used in Refs.~\cite{Sun:2013dya,Sun:2013hua}.
As in this article, Refs.~\cite{Sun:2013dya,Sun:2013hua} neglect the $Y$ term and only account for the role of the TMD term.

Even in the region of $1.0 \; {\rm GeV} \lesssim Q \lesssim 10.0 \; {\rm GeV}$, it can be seen from these graphs that there are 
significant differences, both qualitatively and quantitatively, between the Collins and Sun-Yuan treatment of the perturbative parts of the evolution exponents.  
In addition, it is worth noting that Eq.~\eqref{eq:evolution3} does not obey Eq.~\eqref{eq:basicevol} for the TMD term, as can be checked explicitly using Eq.~\eqref{eq:sunyuanform}.  
The main difference is in the region of very small $b_T$ where the two expressions diverge with opposite signs.  
In the Collins TMD factorization treatment, $\alpha_s$ is allowed to run with $\sim 1 / b_T$ in both the 
CS kernel and in the TMD functions in such a way as to optimize the perturbative treatment in the small regions of $b_T$, whereas in the Sun-Yuan case the evolution is 
relative to a fixed scale $\mu_0$. As is clearly acknowledged in Refs.~\cite{Sun:2013dya,Sun:2013hua}, the Sun-Yuan treatment of 
evolution must break down
at large $Q$ (in fact it diverges above some $Q$) as the integrand of the Fourier transform becomes increasingly concentrated around perturbatively small $b_T$. 
In the Collins TMD factorization treatment, the evolution involves true nonperturbative physics at large $b_T$ whereas the Sun-Yuan 
formalism retains a perturbative treatment at large $b_T$.  Thus, the treatments differ at both large and small $b_T$.

It is worth further investigating the origin of the difference between Fig.~\ref{fig:bdists3}(a) and (b).
Because $\alpha_s$ does not run with $\sim 1/b_T$ in the Sun-Yuan kernel, it has very weak $b_T$-dependence in the region 
of very large $b_T$, so that on the surface there does not appear to be the same sensitivity to nonperturbative large $b_T$ physics
as in the Collins TMD factorization treatment.  However, our analysis from Sec.~\ref{sec:largebT} suggests that much of the relevant 
$b_T$ range in Figs.~\ref{fig:bdists3}(a,b) is well into the region characterized by the type of genuine nonperturbative physics one hopes 
to extract in a TMD analysis.

Taking into account the details of the nonperturbative part of the large $b_T$ behavior, such as is described in Sec.~\ref{sec:modlargebT}, 
allows one to obtain a reasonable description of the large $b_T$ behavior in a way that 
agrees with the qualitative behavior of the COMPASS
data while still matching to the optimized perturbative TMD factorization form of evolution from Ref.~\cite{collins} in the limit of small $b_T < b_{\rm max}$.  
Thus the Collins TMD-factorization formalism unifies the large and small $Q$ behavior in a single evolution formalism.
To illustrate this, we have replotted Fig.~\ref{fig:bdists3}(a) in Fig.~\ref{fig:bdists4}, but now with the $g_K(b_T;\bma)$ term included and using the sample
functional form in Eq.~\eqref{eq:model} with $\bma = 0.5$~GeV$^{-1}$, $g_2 = 0.1$~GeV$^2$, and $b_{\rm NP} = 2.0$~GeV$^{-1}$.  

Also, compare the categorization of relevant regions of $b_T$ at $Q \sim 2.0$~GeV 
in Figs.~\ref{fig:bdists},~\ref{fig:theorybdistsa}, and~\ref{fig:theorybdistsb} with Figs.~\ref{fig:bdists3} and~\ref{fig:bdists4}. From this we 
can see that the main difference between the Collins and Sun-Yuan treatments at $Q$ of $\sim 1.0$~GeV to $2.0$~GeV 
is in how nonperturbative, large size regions of $b_T$ 
behavior are taken into account.  The reliably perturbative, small $b_T$-behavior gives only a rather small contribution for such kinematics. 

As $Q$ increases above $\sim 2.0$~GeV, sensitivity to the treatment of perturbation theory at small $b_T$ becomes increasingly 
important. Future improvements to the global implementation of evolution, combined with increasingly sensitive experiments, 
may possibly be able to distinguish between the two ways of dealing with the perturbative component of evolution at small 
$b_T$ in future data even in the range of $Q \sim 2.0$~GeV to $10.0$~GeV. 

\section{Discussion}
\label{sec:discussion}

Since evolution gives the variation with respect to $Q$ (with all other variables fixed) then in principle 
a much larger range of $Q$ values than in 
Tables~\ref{table:g2valuespos},~\ref{table:g2valuesneg} needs to be 
taken into consideration in order to obtain reasonable constraints on the nonperturbative input to $g_K(b_T;\bma)$. 
We stress, therefore, that what we have presented in this paper should not be regarded as 
a self-contained fitting project, but rather an investigation of general features of moderate $Q$ behavior that need to 
be accounted for in treatments that operate within a complete TMD factorization 
formalism and with the goals and strategies outlined in the introduction.
Even with the small variations in $Q$ discussed in this paper, however, one is able to 
constrain general properties of $g_K(b_T;\bma)$. 
That the data are at relatively low $Q$ helps especially to 
constrain the form of the nonperturbative evolution function $g_K(b_T;\bma)$ in regions of 
very large $b_T$ that are inaccessible in fits that focus on large $Q$.
An important cautionary note, however, is that while the nonperturbative input in TMD factorization becomes increasingly important at smaller 
$Q$, the power suppressed terms of order $\mathcal{O}((M/Q)^a)$, with $a > 0$ and $M$ of order 
a hadronic mass, also become increasingly 
important.  It is notable, then, that when $Q \sim 1.1$~GeV, the hard scale is only slightly larger than 
the mass of hadron.  (See additional discussions below regarding the approach to the border of the region of 
validity of TMD factorization.)
Our method of extracting of the weak $Q$ dependence 
considers variations in $Q$ within a single experiment at fixed $x$ and $z$ bins.
To analyze several bins in $Q^2$, and thus place meaningful constraints on evolution, we considered
smaller $x$ values, relative to the valence region, in Tables~\ref{table:g2valuespos},~\ref{table:g2valuesneg} and Figs.~\ref{fig:linearplots} (a) and (b).

Our analysis is framed within the context of the Collins TMD factorization theorem, and by allowing 
a more general treatment of the non-perturtubative component of the CS kernel than the usual power law, we find that 
we may extend TMD factorization to lower $Q$ SIDIS measurements with no need to 
distort the perturbative part of evolution that is necessary to unify low $Q$ cross section measurements with large $Q$ ones. 
Moreover, we are able to maintain consistency with prior successful fits of nonperturbative parts done at larger $Q$.   
Finally, by maintaining the basic framework of Ref.~\cite{collins} we are able to focus on what can be learned 
about the nonperturbative yet totally universal component of evolution, thereby accessing fundamental nonperturbative
information.

Figures~\ref{fig:linearplots} (a) and (b) demonstrate that, although the variation in 
the $P_T$-shape with $Q$ is small, there is evidence of non-zero broadening 
due to Collins-Soper evolution.  The size of 
the evolution can be estimated from the range of values for $C_{\rm evol}$ found 
 in Tables~\ref{table:g2valuespos},~\ref{table:g2valuesneg}.

Tables~\ref{table:g2valuespos},~\ref{table:g2valuesneg} and Figs.~\ref{fig:linearplots} (a) and (b) show some variation 
of $C_{\rm evol}$ with $x$ and $z$.
Also, comparing Tables~\ref{table:g2valuespos},~\ref{table:g2valuesneg}, 
one sees a trend of larger $C_{\rm evol}$ for the
production of negative hadrons than for positive hadrons.
This suggests that a description in terms of the TMD term alone is a poor approximation 
at these relatively low values of $Q$, and that the $Y$-term is in certainly needed.  
Details of the calculation of the entries in Tables~\ref{table:g2valuespos},~\ref{table:g2valuesneg} 
and of our plots will be made available at~\cite{webpage}.

In addition to the recalling the important role of the $Y$-term, some other words of caution are necessary.  
A possible limitation of TMD studies such as this one, done at such small $Q$, is that 
they may begin to approach the boundary of the 
region of applicability for the TMD factorization formalism.
The TMD factorization theorem describes the transversely differential cross section 
in terms of three distinct kinematical regions:  The lowest transverse momenta are of order 
$\lesssim \Lambda_{\rm QCD}$, and transverse momentum 
dependence is understood here to be intrinsic and nonperturbative.  The second 
relevant region is where transverse momenta 
are large, of order $P_T \sim Q$, where transverse momentum dependence is 
described purely in terms of higher order collinear factorization (i.e., in terms of the $Y$-term alone). 
Finally there is an intermediate third region where $\Lambda_{\rm QCD} \ll P_T \ll Q$.  
Here, the TMD-factorization description applies, but the TMD PDFs are expressible in terms of 
collinear parton distributions.  A reliable description of the cross section in this last region  
requires both the TMD term and the $Y$-term to be present.  Note that 
the derivation of the TMD factorization theorem requires the use of approximations specific to 
each region separately.  

In the region approaching $Q \sim 1.0$~GeV, the distinction between the different regions of $P_T$ becomes less 
clear.  For instance, while the region of $P_T \sim 0.5$~GeV 
is generally expected to have a significant nonperturbative contribution, it is not clear 
that the approximation $P_T \ll Q$ is then reasonable for $Q$ of only $1.0$~GeV. 
Moreover, hadronic mass effects may become important when $Q$ is of order only a few GeV. 
Thus, it may be that the kinematics of the process need to be treated more exactly in an extended formalism, such as in 
the formalism proposed in Ref.~\cite{Collins:2007ph}.  
Finally, we recall again that flavor dependence will likely play an 
important role, as emphasized recently in Ref.~\cite{Signori:2013mda}.

However, the general trends that we observe are dramatic enough that we 
expect our main conclusions to be robust.
Even considering the issues related to the $Y$-term and possible limitations of TMD-factorization 
at very low $Q$ discussed above, it is difficult to reconcile the small values of $C_{\rm evol}$ found in  
Tables~\ref{table:g2valuespos},~\ref{table:g2valuesneg} with the much larger nonperturbative 
soft evolution found from direct extrapolations of global fits of Drell-Yan or 
large $Q$ processes to much lower $Q$,  if one limits the treatment of the nonperturbative large $b_T$ evolution 
factor to the quadratic form in Eq.~\eqref{eq:gaussform} for all $b_T$.  
In this regard, we confirm one of the main observations of Ref.~\cite{Sun:2013dya,Sun:2013hua}.

However, the meaning that we extract from these observations is 
very  different from Ref.~\cite{Sun:2013dya,Sun:2013hua}.  
In our analysis, performed within the TMD factorization theorem of Ref.~\cite{collins}, we find much greater sensitivity to the details of 
the nonperturbative large $b_T$ structure, rather than evidence that nonperturbative contributions to evolution are unnecessary. 
(Reference~\cite{Echevarria:2012pw} has also argued that nonperturbative TMD evolution is unnecessary, even at $Q \sim 1.0$ to $2.0$~GeV.)
Moreover, we find that the complications that arise from extrapolating from large 
to moderate $Q$ arise because of the greater care necessary in treating the nonperturbative 
contribution to evolution as larger $b_T$ values become relevant to evolution, not because such non-peturbative effects are less relavant.
The lack of such a detailed account of large $b_T$ evolution is a limitation of the TMD parametrizations produced in Ref.~\cite{Aybat:2011zv}.
By accounting for the nonperturbative behavior at 
very large $b_T$, as discussed in Sec.~\ref{sec:modlargebT}, we find that it is not difficult to reconcile past large $Q$ fits 
of nonperturbative evolution with the moderate $Q$ fits; see Fig.~\ref{fig:theorybdistlogmod}. 
Finally, our treatment of the perturbatively calculable parts 
of the evolution differs from that of Ref.~\cite{Sun:2013dya,Sun:2013hua}, as discussed in Sec.~\ref{sec:sunyuan}. 

In future fits, more detailed treatments of the functional form for $g_K(b_T;\bma)$ at large $b_T$ will be important both for extending 
TMD factorization to lower $Q$ where studies in hadronic structure are often performed, and for 
achieving the increasing demands for high precision at large $Q$.  
Reliable constraints on $g_K(b_T;\bma)$ might be obtained from global fitting that includes the $Y$-term and 
proper matching to collinear factorization at small $b_T$, such as is done in
Refs.~\cite{Nadolsky:1999kb,Nadolsky:2000ky}, but including newer low $Q$ data and alternative 
functional forms for $g_K(b_T;\bma)$.  
Another possibility is to repeat global fits to Drell-Yan type processes, 
but with the species of colliding hadrons held fixed. 
(See, also, the recent review of novel Drell-Yan phenomenology in Ref.~\cite{Peng:2014hta}.)
An important achievement would be to successfully identify differences in the transverse momentum dependence 
between different types of colliding hadrons, i.e. between valence and sea quark distributions, as discussed in the introduction.  
Addressing this and similar issues  within a complete TMD factorization formalism will help to unify
TMD evolution studies with hadronic structure phenomenology while also
addressing fundamental nonperturbative issues like those raised in Ref.~\cite{Schweitzer:2012hh,Schweitzer:2012dd}. 


\appendix

\section{TMD PDF \MSbar{} Anomalous Dimensions}
\label{sec:anondim}
The anomalous dimensions to order $\alpha_s(\mu)$ are the same for the TMD PDF and 
the TMD fragmentation function:
\begin{align}
\gamma_{\rm PDF}(\alpha_s(\mu),\zeta_{\rm PDF} / \mu^2) =  & 4 C_{\rm F} \left(\frac{3}{2} - \ln \left( \frac{\zeta_{\rm PDF}}{\mu^2} \right) \right) 
       \left( \frac{\alpha_s(\mu)}{4 \pi} \right) + \mathcal{O}(\alpha_s(\mu)^2), \, \label{eq:gammapdf} \\
\gamma_{\rm FF}(\alpha_s(\mu),\zeta_{\rm FF} / \mu^2) = & 4 C_{\rm F} \left(\frac{3}{2} - \ln \left( \frac{\zeta_{\rm FF}}{\mu^2} \right) \right) 
       \left( \frac{\alpha_s(\mu)}{4 \pi} \right) + \mathcal{O}(\alpha_s(\mu)^2).  \label{eq:gammaff}
\end{align}

The \MSbar{} anomalous dimension of the CS kernel to one loop is
\begin{equation}
\gamma_K(\alpha_s(\mu)) = 8 C_F \left( \frac{\alpha_s(\mu)}{4 \pi} \right) + \mathcal{O}(\alpha_s(\mu)^2) \, . \label{eq:gammaK}
\end{equation} 
$C_F = 4/3$ for QCD.

\section{Running Coupling}
\label{sec:runningcoupling}

For running coupling, we use the form
\begin{equation}
\alpha_s(\mu)  = \frac{A}{2 \ln \left(\mu / \Lambda_{\rm QCD} \right)} \, , \label{eq:alphas}
\end{equation}
which will allow us to easily obtain low order analytic expressions for perturbative quantities.
Here,
\begin{equation}
A = \frac{1}{\beta_0} = \frac{12 \pi}{33 - 2 n_f} = \frac{4 \pi}{9} \, .
\end{equation}
Since are interested in the behavior in the neighborhood of $\mu \sim 1.0$~GeV, we use $n_f = 3$. 
To determine a value for $\Lambda_{\rm QCD}$, we fit to the three loop $\beta$ function 2009 world average 
for $\alpha_s(\mu)$ in Ref.~\cite{Bethke:2009jm} in the region of $1.1$~GeV to $3$~GeV.  
We find $\Lambda_{\rm QCD} = 0.2123$~GeV.  Equation~\ref{eq:alphas} then closely 
matches the three loop behavior in the region of $\sim 1.0$~GeV, though the three loop $\beta$ function 
rises more steeply at small $\mu$.  
Thus, our calculations are consistent with a slight underestimate
of the approach to the nonperturbative region.  

\section{Evolution Exponents To One Loop}
\label{sec:evofacts}

Using Eq.~\eqref{eq:alphas} and Eqs.~(\ref{eq:gammapdf})-(\ref{eq:gammaK}) inside Eq.~\eqref{eq:evolution2}, 
we may calculate the perturbative part of the evolution factor in the 
Collins TMD factorization formalism analytically.  The result is,
\begin{align} 
          & \int_{\mu_b}^Q \frac{d \mu^\prime}{\mu^\prime} \left[ \gamma_{\rm PDF}(\alpha_s(\mu^\prime);1)  
         + \gamma_{\rm FF}(\alpha_s(\mu^\prime);1) 
         - 2 \ln \left( \frac{Q}{\mu^\prime} \right) \gamma_K(\alpha_s(\mu^\prime)) \right] \nonumber  \\ 
& \qquad = \frac{2 A}{\pi} \left[ \ln \left( \frac{\ln(Q/\Lambda_{\rm QCD})}{\ln(\mu_b/\Lambda_{\rm QCD})} \right) - \frac{4}{3} \ln(Q/\Lambda_{\rm QCD}) \ln \left( \frac{\ln(Q/\Lambda_{\rm QCD})}{\ln(\mu_b/\Lambda_{\rm QCD})} \right)  + \frac{4}{3} \ln(Q/\mu_b) \right] \, . \label{eq:collinsform}
\end{align}
Note that we have dropped the $\tilde{K}(b_{\ast};\mu_b)$ that appears in Eq.~\eqref{eq:evolution2} on the first line of Eq.~\eqref{eq:collinsform}.
This is because the order-$\alpha_s$ $\tilde{K}(b_{\ast};\mu_b)$ vanishes exactly when a choice of $C_1 = 2 e^{-\gamma_{\rm E}}$ is made.  
Note that 
the $b_T$-dependent part is linear in $\ln Q$.

The result for the Sun-Yuan formalism, Eq.~\eqref{eq:evolution3}, is
\begin{align} 
          &  -2 C_F \int_{Q_0}^Q \frac{d \mu^\prime}{\mu^\prime} \frac{\alpha_s(\mu^\prime)}{\pi} \left[ 
    \ln \left( \frac{Q^2}{{\mu^\prime}^2} \right) + \ln \left( \frac{Q_0^2 b_T^2}{C_1^2} \right) - \frac{3}{2} \right] \nonumber  \\ 
& \qquad = -\frac{4 A}{3 \pi} \left[ 2  \ln(Q/\Lambda_{\rm QCD}) \ln \left( \frac{\ln(Q/\Lambda_{\rm QCD})}{\ln(Q_0/\Lambda_{\rm QCD})} \right) - 2 \ln(Q/Q_0) + \left( \ln \left( \frac{Q_0^2 b_T^2}{C_1^2} \right) -\frac{3}{2} \right) 
\ln \left( \frac{\ln(Q/\Lambda_{\rm QCD})}{\ln(Q_0/\Lambda_{\rm QCD})} \right) \right] \, . \label{eq:sunyuanform}
\end{align}

\begin{acknowledgments} 
T.~Rogers is supported by the National Science Foundation, grant PHY-0969739, 
and L. Gamberg is supported by the U.S. Department of Energy under grant No. DE-FG02-07ER41460.
T.~Rogers thanks Christian Weiss for discussions that helped lead to the formulation of this project.  We especially 
thank the COMPASS collaboration, particularly Andrea Bressan, for discussions. 
We thank John Collins, Mariaelena Boglione, Zhongbo Kang, Pavel Nadolsky, Alexei Prokudin, Andrea Signori, George Sterman, and Mark Strikman 
for helpful discussions.
\end{acknowledgments}


\bibliography{evolestimates}

\end{document}